\documentclass{jfm}
\usepackage{floatrow}
\usepackage{graphicx}
\usepackage[labelfont=rm,labelformat=parens]{subfig} % <-- changed
\usepackage{caption}
\usepackage[justification=centerlast]{caption}
\usepackage{epstopdf, epsfig}
\usepackage[section]{placeins}
\usepackage{float}
\usepackage{cancel}
\usepackage{color}
\usepackage{xcolor}
\usepackage{bm}
\usepackage{natbib}
\usepackage[normalem]{ulem}  %means that \uline is defined but \emph isn't redefined.
\usepackage{amsmath}
\usepackage{url}
\usepackage{makecell}
\usepackage[colorlinks,	
linkcolor   =blue,
anchorcolor =blue,
citecolor   =blue,
urlcolor    =blue]{hyperref}

\newcommand{\RomanNumeralCaps}[1]
\linenumbers

\shorttitle{Onset of vortex shedding around a short cylinder}
\shortauthor{Y. Yang, Z. Feng and M. Zhang}

\title{Onset of vortex shedding around a short cylinder}

\author{Yongliang Yang,\aff{1,2}
Zhe Feng\aff{2}
\and  Mengqi Zhang\aff{2} \corresp{\email{mpezmq@nus.edu.sg}}
}

\affiliation{\aff{1} School of Mechanical Engineering, Nanjing University of Science and Technology, Xiaolingwei 200, 210094, Nanjing, China 
\aff{2} Department of Mechanical Engineering, National University of Singapore, 9 Engineering Drive 1, 117575 Singapore}

\date{\today}

\graphicspath{{./}}
\newlength\savewidth

\begin{document}

\maketitle

\begin{abstract}

This paper presents results of three-dimensional direct numerical simulations (DNS) and global linear stability analyses (LSA) of a viscous incompressible flow past a finite-length cylinder with two free flat ends. The cylindrical axis is normal to the streamwise direction. The work focuses on the effects of aspect ratios (in the range of $0.5\leq\textsc{ar} \leq2$, cylinder length over diameter) and Reynolds numbers ($Re\leq1000$ based on cylinder diameter and uniform incoming velocity) on the onset of vortex shedding in this flow. All important flow patterns have been identified and studied, especially as $\textsc{ar}$ changes. The appearance of a steady wake pattern when $\textsc{ar} \leq1.75$ has not been discussed earlier in the literature for this flow. LSA based on the time-mean flow has been applied to understand the Hopf bifurcation past which vortex shedding happens. The nonlinear DNS results indicate that there are two vortex shedding patterns at different $\it Re$, one is transient and the other is nonlinearly saturated. The vortex-shedding frequencies of these two flow patterns correspond to the eigenfrequencies of the two global modes in the stability analysis of the time-mean flow. Wherever possible, we compare the results of our analyses to those of the flows past other short-$\textsc{ar}$ bluff bodies in order that our discussions bear more general meanings.

\end{abstract}


\section{Introduction}

 Finite-length cylinders and their variants are ubiquitous in human lives and engineering, including the cylinder with one free end, e.g. chimneys, cylindrical tall buildings, etc. and the cylinder with two free ends, e.g. submarine-like shape (Tezuka $\&$ Suzuki \citeyear{Tezuka2006}; Sheard $et~al.$ \citeyear{sheard2008}), torpedoes-like shape (Schouveiler $\&$ Provansal \citeyear{SCHOUVEILER2001}), wheels (Zdravkovich $et~al.$ \citeyear{Zdravkovich1989}) and the short cylindrical bluff bodies (Prosser $\&$ Smith \citeyear{Prosser2016JFM}; Yang $et~al.$ \citeyear{YANG2021}) etc. 
However, compared to the extensively-studied flow past an infinite cylinder, there are relatively fewer works on the flow past a finite-length cylinder with two free ends and thus the current understanding of the flow dynamics past a finite-length cylinder is insufficient. The works on the flows past finite-length cylinders in the literature also appear scattered, mainly because of the many different configurations that are possible for the cylinder. For example, the cylinders can be classified as (1) having two flat free ends versus two non-flat free ends; (2) having two free ends versus one free end and one fixed end (connected to a ground plane); (3) being axially perpendicular to the incoming flow versus axially parallel to the incoming flow, etc. Plus, important parameters, such as Reynolds number ($Re$), aspect ratio ($\textsc{ar}$), yaw angle, etc., are all different in each work, further enriching but at the same time complicating the current literature. In this work, we will focus on the flow past a cylinder with two free ends and with an axis perpendicular to the incoming flow with $0.5\leq \textsc{ar} \leq2$ and $Re\le1000$. In the following, we will summarise the relevant works in the literature on the studies of the flows past a finite-length cylinder and on stability analyses based on a time-mean flow. At the end of this section, we will clarify the position of the current work by identifying the research gap we are going to fill.

\subsection{Flows past a finite cylinder with two free ends}

In an oil shadow experiment to visualise the flow past a finite-length cylinder, Zdravkovich $et~al.$ (\citeyear{Zdravkovich1989}) found that periodical vortices can be observed at the aspect ratio of $2<\textsc{ar}<8 $ and $6\times10^4<Re<2.6\times10^5$ (where $\textsc{ar}=L/D$ with $L$ being the length or height and $D$ the diameter of the cylinder), and when $\textsc{ar}$ is reduced to 3, the eye-like low pressure area near the free ends gradually disappears. Moreover, when $\textsc{ar}<3$, the pressure distribution on the curved surface is no longer symmetric about the center plane, and this asymmetrical flow produces yaw and roll moments. In general, the drag coefficient decreases as the aspect ratio decreases. 
However, Zdravkovich $et~al.$ (\citeyear{Zdravkovich1989}) later conducted experiments on the flow around a coin-like cylinder with $0< \textsc{ar}\leq 1$ and showed that in the range of $2\times10^5<Re<6\times10^5$, the drag coefficient and the aspect ratio satisfy the relationship: $\overline{C}_d=0.024/\textsc{ar}+0.633, (2\times10^5<Re<6\times10^5)$, that is when $\textsc{ar}$ decreases, $\overline{C}_d$ increases. This trend contradicted their previous measurements (in Zdravkovich $et~al.$ \citeyear{Zdravkovich1989}) , which the authors attributed to the inappropriate reference area used in this range of $\textsc{ar}$ (once the side area $\frac{D^2}{4}\pi$ was used as the reference area, instead of the projected area $LD$, the previous trend that $\overline{C}_d$ decreases as $\textsc{ar}$ decreases held valid for small values of $\textsc{ar}$). Furthermore, the topology of the flow field has also been depicted: there are two horseshoe-shaped vortices on the two free ends, which are separated at an angular position of approximately $90^{\circ}$, and the free vortex forms two counter-rotating vortex pairs in the leeward zone. Based on this observation, Zdravkovich analysed and proposed a drag-reducing strategy of rounded sharp edges with a drag reduction effect up to 33.6\%. 

More recently, direct numerical simulations have been adopted in this field of research; for example, Inoue $\&$ Sakuragi (\citeyear{Inoue2008}) studied the flow around a finite cylinder with two free ends with $0.5\leq \textsc{ar}\leq100$ and $40\leq Re\leq 300$. Their results showed that the $\textsc{ar}$ and $Re$ have a great influence on the vortex shedding pattern, and five types of vortex shedding patterns were identified as: (I) Periodic oblique vortex shedding (large $\textsc{ar}$, $Re>Re_c$); (II) Quasi-periodic oblique vortex shedding (large $\textsc{ar}$, $Re<Re_c$); (III) Hairpin vortex that periodically falls off (moderate $\textsc{ar}$); (IV) Two stable counter-rotating vortex pairs (small $\textsc{ar}$ and $Re$); (V) The counter-rotating vortex pairs alternately shed from the free ends (small $\textsc{ar}$, high $Re$). Prosser $\&$ Smith (\citeyear{Prosser2016JFM}) conducted large-eddy simulations coupled with the unsteady Reynolds-averaged Navier–Stokes equations using a finite-volume method to study the flow around bluff bodies (finite cylinder or prism) with much higher $Re$ in the range of $[10^5, 10^6]$ and $\textsc{ar}=1$ and $2$. The influences of the geometric features and attitude of the bluff body on the reattachment distance, pressure coefficient and stagnation point position were investigated and modelled empirically. For example, they found that through proper normalisation, on the flat ends of the bluff body, the pressure coefficient and the stagnation point position are single-valued functions of the incident angle; but on the curved face of the bluff body, these two variables are affected by the aspect ratio.
 Gao $et~al.$ (\citeyear{Gao2018}) numerically studied the flow pattern of a cylinder with two free ends. Unlike the flow field of cylinder with one free end (where there is a large horseshoe vortex surrounding the fixed end of the cylinder), there is no horseshoe vortex appears in the flow past the cylinder with two free ends, and a new relationship between $\overline{C}_d$ and $Re$ has been proposed within the range of $5\leq Re \leq5\times10^6$, $1\leq \textsc{ar} \leq6$. 
In addition, Pierson $et~al.$ (\citeyear{Pierson2019}) used the finite-volume–fictitious-domain method to study the axial flow (i.e. the cylinder axis is parallel to the streamwise direction) around a finite-length three-dimensional cylinder ($\textsc{ar}=1, 20<Re<460$) with two free ends. They identified two bifurcation points both without hysteresis, one is the regular bifurcation at $Re\approx278$ and the other is the Hopf bifurcation at $Re\approx355$. Unfortunately, they did not explain how the aspect ratio affects these two bifurcation points.

In addition, finite-length cylinders with non-flat ends have also been studied by some groups. These works are reviewed here because some of them are inspiring and comparisons have been made to these works in our result section. Zdravkovich $et~al.$ (\citeyear{Zdravkovich1989}), Schouveiler $\&$ Provansal (\citeyear{SCHOUVEILER2001}) and Sheard $et~al.$ (\citeyear{Sheard2005}) used wind tunnel experiment and numerical simulations respectively to study flow around a finite cylinder with two hemispherical ends. The results show that the hemispherical ends can effectively reduce drag coefficient. Since the flow field past this type of cylinder has no fixed boundary layer (due to the rounded ends), there is no horseshoe vortex in flow field either. When the two hemispherical ends connect each other, a sphere is formed. The wake pattern, transition and instability of the flow around a sphere have been studied for a long time, especially, regarding the critical Reynolds number, beyond which the steady flow becomes unsteady. This critical Reynolds number has been reported as $Re_c=130$ in early experimental research (Taneda \citeyear{Taneda1956}), as $Re_c=175$ in LSA (Kim $\&$ Pearlstein \citeyear{kim1990}), as $Re_c=270$ in experiments of \cite{magarvey1965vortices,wu1993sphere,Johnson1999}, as $Re_c=277.5$ in LSA of \cite{natarajan1993} and finally as $270<Re_c<285$ in the DNS using the spectral element method by \cite{TOMBOULIDES2000}. In summary, the value of Hopf bifurcation $Re_c=277.5$ obtained by LSA has been generally accepted. This is the second transition in the sphere flow; and the first transition occurs around $Re=210$ \citep{natarajan1993} or $Re=212$ \citep{TOMBOULIDES2000}. In this work, we will focus on the wake patterns and transition of the flow around a short finite cylinder ($\textsc{ar}\leq 2$). As we will later show, the flow past a short finite-length cylinder ($\textsc{ar} \leq 1.7$) shows a transition process similar to that of a sphere, but also exhibits some unique characteristics.

\subsection{Stability analyses of a mean flow}
As we will conduct global stability analysis of the flow around the finite cylinder, it is pertinent to review works on the application of this analysis to the infinite-length cylinder, pioneered by \cite{Jackson1987,Zebib1987,Noack1994}. Especially, we will review the stability analyses applied to the 2D wake flows based on a time-mean flow.

For the wake flow behind an infinite-length (2D) cylinder, \cite{hammond1997} and \cite{pier2002} noted that applying the linear stability analysis to the time-averaged flow field (which will be called mean flow for simplicity in the following) instead of the steady base flow, that is an unstable solution to the Navier--Stokes equations, can better predict the shedding frequency of the unsteady flow. \cite{barkley2006} conducted a global LSA  of the 2D cylinder flow based on its mean flow and observed that the eigenfrequency is consistent with the nonlinear vortex shedding frequency $St$ found by \cite{Williamson1988} in his famous experiments of the cylinder flow. In addition, \cite{barkley2006} also showed that the mean flow is marginally stable.
This striking result was explained in the asymptotic analysis by \cite{sipp2007} who demonstrated that the two-dimensional cylinder flow can be well represented by the mean flow value and a single eigenmode, and there is almost no nonlinear interaction between them, validating the usage of the mean flow in the stability analysis. Therefore, the premise of \cite{barkley2006} for the global stability analysis of using the time-averaged flow was verified, that is, the Reynolds stress caused by the pulsating wake is not disturbed at the linear order. \cite{sipp2007} also provided a theoretical proof for the validity of the global weakly nonlinear analysis of the 2D cylinder flow near the Hopf bifurcation. More specifically, the theoretical conditions were given on how to keep the mean flow linearly stable and when the eigenfrequency obtained using the mean flow matches the experimental frequency. This condition can be qualitatively described as a situation where the zeroth harmonic (mean flow) is much stronger than the second harmonic. Meanwhile, they also pointed out that the effective results obtained in the case of the 2D cylinder based on the mean flow are by no means general (in the context of the finite-length cylinder whose axis is perpendicular to the flow direction, the stability analysis based on its mean flow has not been performed and will be conducted here). Besides, because of the weakly nonlinear expansion employed in the theoretical development, the analysis of \cite{sipp2007} is strictly valid only very close to the bifurcation point, and cannot fully explains the success of \cite{barkley2006}'s global LSA performed on the mean flow across a wide range of Reynolds numbers. 

Later, \cite{leontini2010mean_flow} applied the saddle-point criterion (that the zero group velocity is found at the saddle point in the complex wavenumber plane or the cusp point in the complex frequency plane when local mean flows are analysed in the streamwise direction) to spanwise-averaged three-dimensional infinite cylinder flow. Their results showed that if the local curvature is not too large and the assumption of the flow changing slowly is valid, even if the flow is three-dimensional, the saddle-point criterion can work well. Their global LSA showed that even for the three-dimensional flow, the spanwise-averaged mean flow remains marginally stable, which supports the following hypothesis: the dynamics of the cylinder wake is mainly dominated by the first linear eigenmode generated on the nonlinear modified mean flow. \cite{turton2015mean_flow} proved in a more general manner that if the flow exhibits quasi-monochromatic oscillations, the eigenfrequency of the linearisation operator based on the temporally mean flow is indeed equal to the true nonlinear flow frequency. They further looked into the amplitudes of the waves in the spectral space of the signal and demonstrated that if the amplitude of the first harmonics is sufficiently larger than those of higher-order harmonics, the stability analysis based on the mean flow can represent fairly well the dynamics of the nonlinear flow. 

For a weakly non-parallel and strongly convectively unstable flow, whose first singular value largely dominates the others, a strong link between the mean flow and the fully nonlinear dynamics of a turbulent flow has been explained and verified by \cite{Beneddine2016,Beneddine2017} from both theoretical and experimental perspectives.
According to the RZIF property (real zero imaginary frequency named by Turton $et~al.$ \citeyear{turton2015mean_flow}), Bengana $et ~al.$ (\citeyear{bengana2019}) identified two successive Hopf bifurcations and classified three situations by using mean flow and base flow LSA in the shear-driven cavity flow and summarizing previous research. Another interesting finding is that although no mathematical proof is given, the cross-eigenvalues would then correspond to the relative stability of first two eigenmodes, which can good qualitatively explain the hysteresis in the nonlinear flow system.

In summary, it can be concluded that the global LSA based on the mean flow can predict accurately the nonlinear frequency in a wider range of parameters for the quasi-monochromatic oscillation flows. It remains to be seen if this method is applicable to the flow to be studied here.

\subsection{The current work}
As we can see from the previous sections, there is no regular bifurcation (so-called pitchfork) being reported in the previous research on 3D finite-length cylinder flows with $\textsc{ar}>2$. It has not been studied systematically how this flow bifurcates when $\textsc{ar}<2$. On the other hand, in the flow around other bluff bodies, many works have confirmed the existence of a regular bifurcation, such as sphere \citep{Johnson1999, TOMBOULIDES2000, thompson2001kinematics, sheard2004spheres}, axial flow around short cylinders with $\textsc{ar}=1$ \citep{Pierson2019} and ellipsoids \citep{sheard2008, Tezuka2006}. Thus, one can infer that the aspect ratio may be the decisive factor for educing the regular bifurcation in the finite-length cylinder flows and the regime $\textsc{ar}\le2$ should be explored to see if this bifurcation exists therein. In this work, we will study systematically the influence of $Re(\leq 1000$) and $\textsc{ar}\in[0.5,2]$ on the flow stability/instability of the flow past a 3D finite-length cylinder by nonlinear DNS and linear global LSA (mainly based on temporal mean flow). We will examine whether the LSA results obtained based on the mean flow can be compared to the nonlinear results in the DNS (e.g. in terms of the shedding frequency in the flow). Besides, in order to complete the discussions, we will also compare the LSA results based on the mean flow with the LSA results based on a base flow obtained using the selective frequency damping (SFD) method \citep{SFD2006}. The most important result will be summarised in {\color{blue}figure~}\ref{fig:AR_Rec} on the $Re_c$-$\textsc{ar}$ relation.

The paper is organised as follows. 
\S~\ref{sec:problemformulation} introduces the configuration of 3D finite cylinder flow, the boundary conditions, the governing equations, i.e. nonlinear Nav\-ier--Sto\-kes equations and their corresponding linearised equations about the base state (mean flow or base flow) and the numerical method. \S~\ref{sec:Validation} provides a detailed verification step of nonlinear and linear numerical methods. In \S~\ref{sec:results}, we show the results and discuss the base states, cylinder wake patterns, critical $Re$, nonlinear DNS bifurcation scenario, global eigenmodes and effects of $\textsc{ar}$ and $Re$ on this flow. Finally, the results are summarised in \S~\ref{sec:Conclusions} and conclusions are provided.

\section{Problem formulation and numerical methods}\label{sec:problemformulation}

\subsection{Governing equations and geometry}\label{subsec:NSEq}
\begin{figure}[tb]\centering % R2 = revision 2, R1 = revision 1
	\hfil
	\includegraphics[trim=1.5cm 9.2cm 1cm 9.0cm,clip,width=0.47\textwidth]{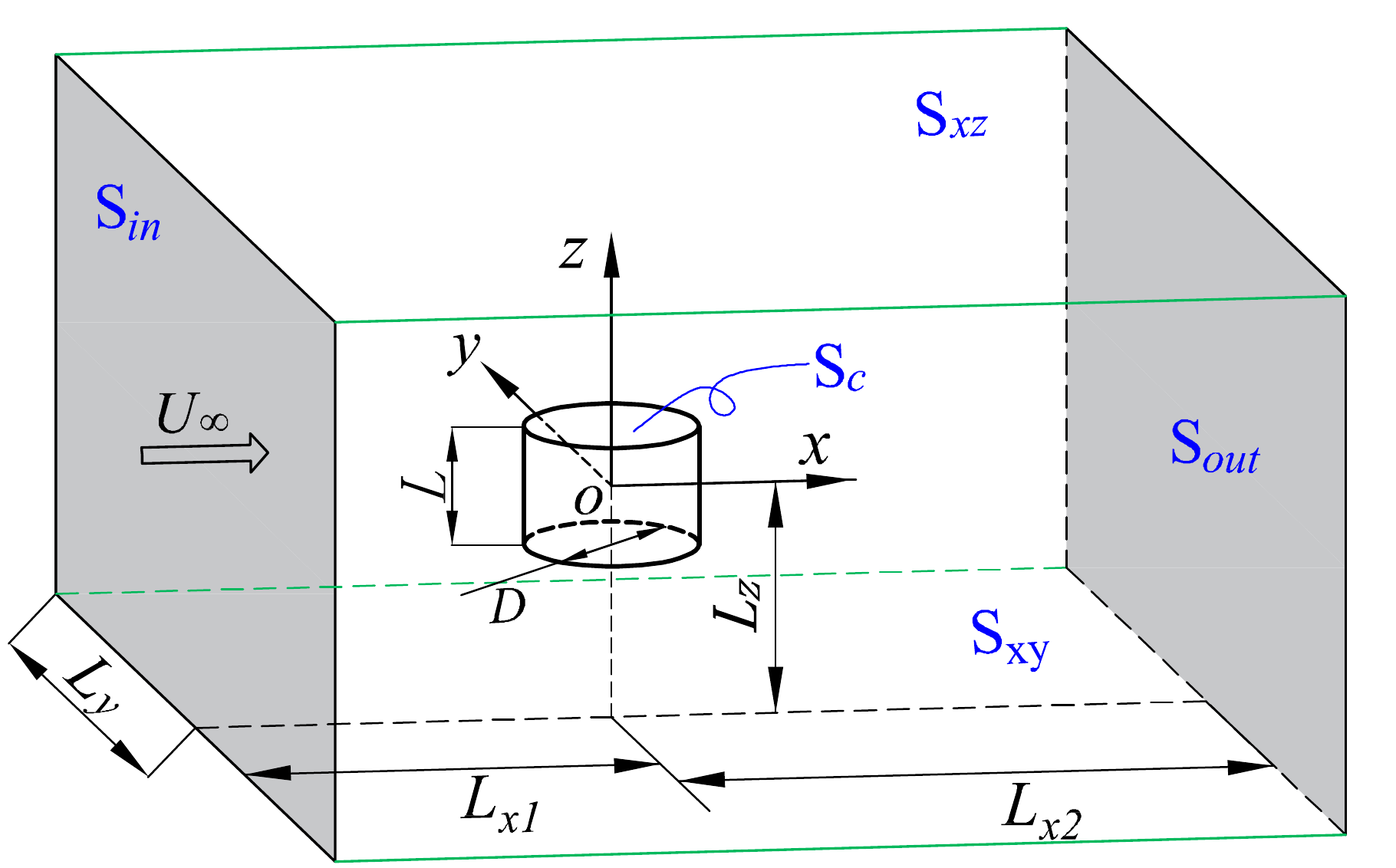} \label{fig:a1} \put(-195,102){($a$)} % in R1, figname is {CFDdomain_BC.pdf} 
  ~\includegraphics[trim=0.0cm 0cm 0.1cm 0.0cm,clip,width=0.49\textwidth]{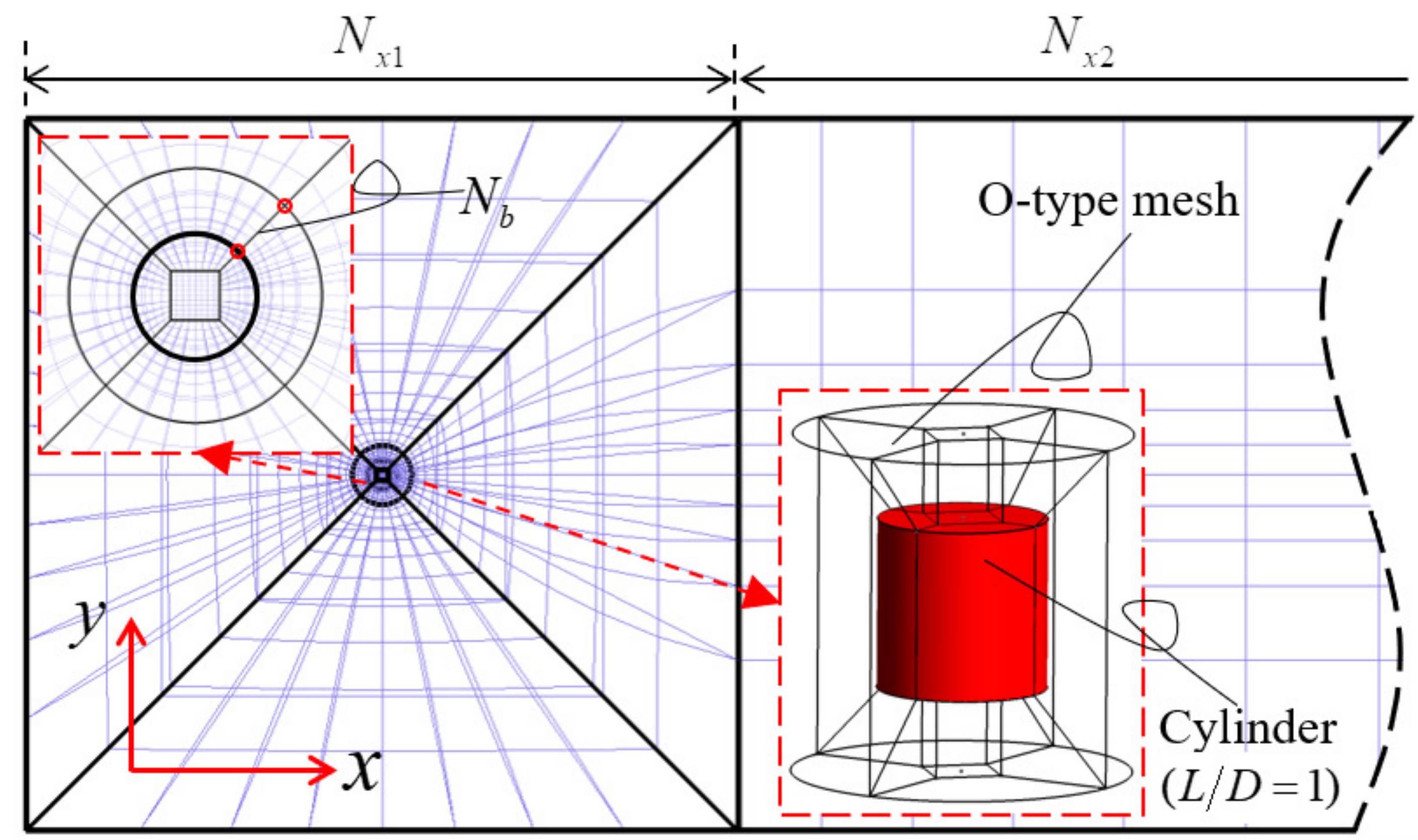}\label{fig:b1}    
        \put(-200,102){($b$)} % in R1, figname is {mesh.png} 
	\hfil
	\caption{($a$) The computational domain and boundary conditions (Not to scale). ($b$) The projection of the mesh topology on the $x$-$y$ plane.}
	\label{fig:CDBC}
\end{figure}

We study the three-dimensional stability of the flow around a finite-length cylinder (of a length $L$ and a diameter $D$) subjected to a uniform incoming flow in a Cartesian coordinate system. The computational domain size,  boundary conditions and geometry of the finite cylinder are shown in {\color{blue}figure~}\ref{fig:CDBC}. The origin of the Cartesian coordinate system is located at the center of the cylinder, the $x$ axis points in the flow direction, the $z$ axis extends along the center line of the cylinder, and the $y$ axis is determined by the right-hand rule. The axis of the cylinder is normal to the incoming flow. The Nav\-ier--Sto\-kes (NS) equations for the unsteady incompressible flow read 
\begin{align}
 \frac{\p \boldsymbol U }{\p t}+
  (\boldsymbol U \bcdot\nabla)\boldsymbol U =  -\nabla P+ \frac{1}{Re} \nabla^2\boldsymbol U, \ \ \ \ \ \  \nabla\bcdot \boldsymbol U  =   0,
 \label{NS}
\end{align}
where $\boldsymbol U$ is the velocity vector, whose components corresponding to the three directions of $x$, $y$, $z$ are $\boldsymbol U=(U_x,U_y,U_z)$ and $P$ is the pressure. In the equations \ref{NS}, the cylinder diameter $D$ is used as the reference length, the velocity $U_\infty$ of the uniform incoming flow at infinity as the reference velocity, and $\rho U^2_\infty$ as the reference pressure. Therefore, the Reynolds number, Strouhal number, drag coefficient $C_d$, lift coefficient $C_{l}$ and aspect ratio are defined respectively as
\begin{align}
    Re  =  \frac{U_\infty D}{\nu}, \ \ \ \  St  =  \frac{fD}{U_\infty}, \ \ \ \  C_d  =  \frac{F_d}{\frac{1}{2} \rho U^2_\infty A},\ \ \ \ C_{l}  =  \frac{F_l}{\frac{1}{2} \rho U^2_\infty A}, \ \ \ \ \textsc{ar} = \frac{L}{D},
	\label{Re}
\end{align}
where $f$ is the frequency of vortex shedding (thus, when there is no vortex shedding in the wake, $St=0$),  $\nu$ is the kinematic viscosity of the fluid and $\rho$ is the density. $F_d$ is the drag force on the surface of the cylinder, whose direction is the same as the streamwise direction. $F_l$ is the cross-stream lift force acting in either the $y$ or $z$ direction (see $C_{ly}$ and $C_{lz}$ below). $A$ is the reference area, and for a finite cylinder $A=LD$. For the lift coefficient $C_l$, there are two possibilities in $y$ and $z$ directions. Among them, $C_{ly}$ (in $y$ direction) is mainly discussed in present work, because its value is directly related to the regular bifurcation. $C_{lz}$ (in $z$ direction) or its mean is almost zero in most of the cases due to the symmetry of the flow. Besides, we will also use the letters $\overline{C}_d$ and $\overline{C}_{l}$ to denote the time-averaged values of $C_d$ and $C_{l}$.

Next, we define the boundary conditions in the computational domain. As shown in {\color{blue}figure~}\ref{fig:CDBC}($a$), $S_c$ represents the surfaces of the finite cylinder, on which no-slip boundary conditions are applied
\begin{equation}
\boldsymbol U =(U_x,U_y,U_z)=0 \quad  \rm{on} \quad \it\boldsymbol{S_c}.
\end{equation}
$S_{in}$ and $S_{out}$ represent the inlet and outlet surfaces of the rectangular computation domain, whose normal is in the $x$ direction. The inlet boundary conditions for the velocity imposed on $S_{in}$ are
\begin{equation}
	U_x=1,\quad U_y=U_z=0  \quad  \rm{on} \quad \it\boldsymbol S_{in}. %\\[6pt]
%	&v_x=0,\quad v_y=v_z=0  \quad  \rm{on} \quad \it\boldsymbol S_{in}
	\label{BCIN}
\end{equation}
 The pressure and the streamwise derivative of the $\boldsymbol{U}$ components are set to zero on the outlet $S_{out}$
 
\begin{equation}
	\frac{\p \boldsymbol{U}}{\p x} =0,\quad P=0  \quad  \rm{on} \quad \it\boldsymbol S_{out}.
	\label{BCout}
\end{equation}
$S_{xy,t}$, $S_{xy,b}$, $S_{xz,f}$ and $S_{xz,b}$ represent the surfaces of the cuboid on the top, bottom, front and back side walls, which are parallel to the $xy$, $xy$, $xz$ and $xz$ planes, respectively. Previous research indicates that the computation domain has some influence on the results of the LSA. For example, \cite{luchini2007} conducted an eigenvalue sensitivity analysis on the size of the computational domain in the problem of the two-dimensional cylinder wake flow with symmetric boundaries in the cross-streamwise direction. Their results showed that when the size in the cross-streamwise direction is greater than $10D$, the drift of the eigenvalue and eigenmode is small. Following them, we will perform a similar analysis. The symmetry boundary condition is imposed on the surfaces $S_{xy,t}$, $S_{xy,b}$, $S_{xz,f}$ and $S_{xz,b}$ at position $\pm 10D$, that is $L_y=L_z=10$:
\begin{equation}
	\left\{
	\begin{array}{lr}
		\frac{\p U_x }{\p y}=U_y=\frac{\p U_z }{\p y}=\frac{\p P}{\p y}=0, \quad \rm{on} \quad \it\boldsymbol S_{xz,f},S_{xz,b} \\[6pt]
		\frac{\p U_x }{\p z}=\frac{\p U_y }{\p z}=U_z=\frac{\p P}{\p z}=0, \quad \rm{on} \quad \it\boldsymbol S_{xy,t},S_{xy,b}
	\end{array}
	\right.
	\label{BCsym}
\end{equation}
The values of $L_{x1},L_{x2}$ define the position of the cylinder in the streamwise direction, which, together with the vertical length $L_{y}$ and span length $L_{z}$, will be discussed in a convergence study of the computational domain size on the linear dynamics.

\subsection{Linearisation}
In the present work, we will study the global linear stability/instability of the flows around the finite-length cylinder. As we have reviewed in the introduction section, we will focus on studying two types of base states (both of which are denoted as $(\boldsymbol U_b,P_b)$ in the text to follow as long as there is no confusion) to analyse their global instability. The first base state is called a base flow, which is a steady solution to the Nav\-ier--Sto\-kes equations \ref{NS} obtained by the selective frequency damping (SFD) method proposed by \cite{SFD2006}. This method damps the unsteady temporal oscillations by adding a dissipative relaxation term in the NS equations to let the low-frequency component of the velocity pass. The second base state is the mean flow, which is obtained by time-averaging the periodic flow with vortex shedding. For the present global stability analysis, at least ten vortex shedding cycles will be used for the time-average procedure. The linearisation starts with Reynolds' decomposition $\boldsymbol U = \boldsymbol U_b+\boldsymbol u, P=P_b+p$, which will be substituted into the nonlinear governing equations. Then, we discard the nonlinear terms and the terms satisfying the Navier--Stokes solution for the base states and retain terms of the order of the perturbation, yielding the linearised equations for the infinitesimal perturbations  residing on these base states

\begin{align}
\frac{\p \boldsymbol u }{\p t}+(\boldsymbol U_b \bcdot\nabla)\boldsymbol u + (\boldsymbol u \bcdot\nabla)\boldsymbol U_b 
=  -\nabla p+ \frac{1}{Re} \nabla^2\boldsymbol u, \ \ \ \ \ \  \nabla\bcdot \boldsymbol u  =   0,
\label{LNS}
\end{align} 
where $\boldsymbol u$ is the three-dimensional perturbation velocity vector $\boldsymbol u=(u_x,u_y,u_z)$ and $p$ is the perturbation pressure. Homogeneous boundary conditions are applied for the perturbed variables:
\begin{subeqnarray}
u_x=u_y=u_z=0, \quad  &\rm{on} \quad \it\boldsymbol{S_c}, \\
u_x=u_y=u_z=0,  \quad  &\rm{on} \quad \it\boldsymbol S_{in} \label{BCINpert}, \\
\frac{\p \boldsymbol{u}}{\p x} =0,\quad p=0, \quad &\rm{on} \quad \it\boldsymbol S_{out},\\
\frac{\p u_x }{\p y}=u_y=\frac{\p u_z }{\p y}=\frac{\p p}{\p y}=0, \quad &\rm{on} \quad \it\boldsymbol S_{xz,f},S_{xz,b}, \\
\frac{\p u_x }{\p z}=\frac{\p u_y }{\p z}=u_z=\frac{\p p}{\p z}=0, \quad &\rm{on} \quad \it\boldsymbol S_{xy,t},S_{xy,b}.
\end{subeqnarray}
We found that the numerical results obtained by these boundary conditions in the eigen-solver (to be discussed shortly) did not manifest the non-physical oscillations at outlet boundary encountered by \cite{luchini2007}. So, we did not set the partial derivative of pressure in the streamwise to zero. 

Equations \ref{LNS} can be written in matrix form with $\boldsymbol q=(\boldsymbol u, p)$ as
 \begin{align}
\boldsymbol M \frac{\p \boldsymbol q }{\p t} = \boldsymbol A\boldsymbol q
 \label{eig1}
 \end{align}
where the mass matrix $\boldsymbol M$ and the Jacobian matrix $\boldsymbol A$ are given by
\begin{equation}
\boldsymbol M =\left(\begin{array}{cc}
	\boldsymbol I & 0\\
	  0           & 0
	\end{array} 
	\right), \ \ \ \
\boldsymbol A =
    \left(\begin{array}{cc}
    -\boldsymbol U_b \bcdot\nabla - \nabla\boldsymbol U_b + Re^{-1}\nabla^2 & \quad-\nabla\\ 
    \nabla\bcdot & \quad 0
    \end{array} 
    \right).
    \label{eig2}
\end{equation}
We expand the perturbation $\boldsymbol{q}$ in time $\boldsymbol{q}(x,y,z,t)=\hat{\boldsymbol{q}}(x,y,z)e^{\lambda t}$, where $\lambda=\sigma+\boldsymbol{i}2\pi \omega$.
The equation \ref{eig1} can be transformed into the following generalised eigenvalue problem by substituting the solution ansatz of $\boldsymbol{q}$:
\begin{align}
\lambda \boldsymbol M \hat{\boldsymbol{q}} = \boldsymbol A\hat{\boldsymbol{q}}
\label{eig}
\end{align}
In this formulation, the stability of the base state $\boldsymbol U_b$ is dictated by the eigenvalues $\lambda$ in the linearised problem with $\hat{\boldsymbol{q}}$ being the eigenmode, $\sigma$ temporal growth/decay rate of perturbations and $\omega$ the eigenfrequency. Because the flow problem is not spatially periodic or homogeneous in either $x,y,z$ directions, $\hat{\boldsymbol{q}}$ depends on all the three coordinates, giving rise to a global stability problem \citep{Theofilis2011} to be solved by the Arnoldi method. % in Sec.~\ref{subsec:numMethod}. 
In addition, the eigenfrequency $\omega$ of the first eigenvalue also determines whether the base state $\boldsymbol U_b$ experiences a regular bifurcation ($\omega=0$) or a Hopf bifurcation ($\omega>0$)\citep{bengana2019}. 

\subsection{Numerical methods: DNS and IRAM} \label{subsec:numMethod}
In order to obtain the accurate wake pattern and base state of the fully three-dimensional flow past a finite-length cylinder at medium and low Reynolds numbers, we conduct DNS of the flow past the cylinder. The high-order parallelised open-source code Nek5000 \citep{nek5000} (version 19.0) is used, which is based on the nodal spectral element method (SEM) originally proposed by \cite{patera1984SPE}. Hexahedral elements for a polynomial order N = 7 (the optimal polynomial order suggested by \citealt{nek5000}) are used to get a better performance of the code. 
The time step $\Delta t$ is determined by the Courant--Friedrichs--Lewy (CFL) condition with the target Courant number being in the range of $0.5$-$1.0$. In the following numerical simulations, the boundary layer elements have been refined by the O-type. To achieve the requirement of adequate resolution near the surface of the cylinder, the value of the smallest boundary thickness for each Reynolds number will be discussed in section \ref{subsec:valDNS}. 

For a non-parallel three-dimensional flow past the finite cylinder, the numerical discretisation of linearised NS equation \ref{LNS} will results in a large-scale Jacobian matrix $\boldsymbol{A}$ in the generalised eigenvalue problem \ref{eig}. It is often impractical to solve a large-scale eigenvalue problem for its whole eigenspectrum in a 3D flow. Based on Nek5000 solver and the ARPACK package \citep{lehoucq1998arpack}, the matrix-free time-stepper method \citep{Theofilis2011,doedel2012} is used in the present work. In this method, one does not need to explicitly construct the matrices, but instead power-iterate the linearised NS equations \ref{LNS} in the temporal direction from an initial condition $\boldsymbol{u}_0$. In the long-time limit, the power iteration will converge to the asymptotic state of the linearised system, that is, the least stable/most unstable eigenmode of Eq. \ref{eig}. More eigen-information cannot be extracted from the simple power method. To remedy this, the classical Implicitly Restarted Arnoldi Method (IRAM) \citep{radke1996MatIRAM,lehoucq1998arpack} based on Krylov subspaces will be used to reduce the order of the original matrix, forming the much smaller Hessenberg matrix, to obtain the Ritz eigenmodes approximating the leading eigenmodes of the full system. 

\section{Validation} \label{sec:Validation}

\subsection{Validation of nonlinear DNS} \label{subsec:valDNS}
\begin{table}
	\begin{tabular}{p{0.9cm} p{0.6cm} p{2.4cm} p{1.0cm} p{2.4cm} p{2.4cm} p{2.4cm}}
		\hline
		\centering	
		Mesh & $N_b$ & \makecell*[l]{\ \ $(N_{x1}+N_{x2})$ \\ \ \ \ $\times{N_y}\times{N_z}$} & $N_{total}$ & $\overline{C}_d$(Err\%) & $\overline{C}_{ly}$(Err\%) & $St$(Err\%)\\
		G1 & 2 & $(4+6)\ \times 4^2$ & 717  & 0.7775 (4.08) & -0.06956 (2.20) & 0.1386 (3.01) \\
		G2 & 2 & $(5+8)\ \times 5^2$ & 1568 & 0.7760 (3.88) & -0.06980 (2.56) & 0.1325 (7.28) \\
		G3 & 2 & $(6+10)\times 6^2$  & 2885 & 0.7764 (3.94) & -0.06851 (0.66) & 0.1375 (3.78) \\
		G4 & 6 & $(6+10)\times 6^2$  & 4125 & 0.7510 (0.54) & -0.06830 (0.35) & 0.1420 (0.63) \\
		G5 & 4 & $(8+10)\times 8^2$  & 6909 & 0.7480 (0.13) & -0.06833 (0.40) & 0.1416 (0.91) \\
		G6 & 6 & $(8+10)\times 8^2$  & 7945 & 0.7500 (0.40) & -0.06845 (0.57) & 0.1434 (0.35) \\
		G7 & 8 & $(8+10)\times 8^2$  & 8981 & 0.7493 (0.31) & -0.06846 (0.59) & 0.1434 (0.35) \\
		G8 & 6 & $(9+10)\times 11^2$ & 15804 & 0.7470 (/) & -0.06806 (/)  & 0.1429 (/)  \\
		\hline		
	\end{tabular}
	\caption{A grid sensitivity test for the case $Re=300$, $\textsc{ar}=1$, $\Delta t=10^{-3}$. $N_b, N_{x1}$ and $N_{x2}$ are shown in {\color{blue}figure~}\ref{fig:CDBC}($b$). $N_y/2$ and $N_{z}/2$ are the node numbers on the edges $L_y$ and $L_z$, respectively. $N_{total}$ is the total number of hexahedral elements in the calculation domain. The errors in the parentheses are computed using as the reference the results of mesh G8.}\label{tab:Re300L1}
\end{table}
\begin{table}
	\centering
	\begin{tabular}{ p{1cm} p{2cm} p{3cm} p{3cm} p{2.3cm}l}	
		\hline	
		Mesh & $N_{tol}$ & $\overline{C}_d$(Err\%) & $\overline{C}_{ly}$ \\
		\hline
		G4 & 4125  & 0.6008 (1.13)   & $-0.001845 $    \\
		G8 & 15804 & 0.5941 (/)    & $-0.001138$    \\
		\hline	
	\end{tabular}	
	\caption{Grid sensitivity test for a high $Re=1000$ with $\textsc{ar}=1$ and $\Delta t=10^{-3}$.}
	\label{tab:GridTestR1000L1}
\end{table}

In this section, we will first show the evidence of converged calculations using Nek5000 and explain the choice of the grid resolution. Eight sets of grids from G1 to G8 in table \ref{tab:Re300L1} have been used to analyse the influence of spatial resolution on the DNS results at $Re=300$ and $\textsc{ar}=1$. The  coefficients $\overline{C}_d$ and $\overline{C}_{ly}$ and the $St$ number are used as evaluation indicators. $N_{x1}$ and $N_{x2}$ are the numbers of the nodes in $x$ direction as shown in the right panel of {\color{blue}figure~}\ref{fig:CDBC}($b$); $N_y$ and $N_z$ are the node numbers for the sections related to $L_y$ and $L_z$, respectively. As previously mentioned, the polynomial order $N=7$  for all $Re$, which means that the number of the Gauss–Legendre–Lobatto nodes inside each hexahedral element is $8\times 8\times 8$ (in 3 directions). Since the flow in the boundary layer of cylinder is sheared most, we tested the layer number of the O-type mesh $N_b=2, 4, 6, 8$, while keeping the total thickness of the O-type mesh constant. We can see from table \ref{tab:Re300L1} that, comparing G5-G7, when $N_b\geq6$, the errors in $\overline{C}_d$ and $\overline{C}_{ly}$ are generally small. The error in $St$ (which appears more sensitive to the grid number) is also less than 1\%. The corresponding smallest boundary layer grid thickness is approximately 0.003 (a dimensionless value with the diameter $D$ as the reference length), which is smaller than the estimated value of 0.016 for $Re=1000$ in \cite{TOMBOULIDES2000} who also used the SEM. Next, keeping the parameters of the boundary layer mesh unchanged, we test the effect of the element size in the wake area on the results. Comparing cases G4, G6 and G8, we can see that if $(N_{x1}+N_{x2})\times{N_y}\times{N_z}\geq(6+10)\times{6}\times{6}$, the differences in $\overline{C}_d$, $\overline{C}_{ly}$, $St$ among G4, G6 \& G8 are also very small. Besides comparing cases G4 and \cite{Inoue2008}'s DNS at $\textsc{ar}=1, Re=300$, the difference of $\overline{C}_d$ is approximately 0.53\%; but the difference in $St$ is 10.5\% (our $St=0.142$ in table \ref{tab:CFDdomainsize} using mesh S2 and the DNS results in \cite{Inoue2008} is $St = 0.127$), and the reasons for this large difference will be discussed in Sec. \ref{subsec:globalmodes} to follow. Furthermore, as the $Re=300$ in Table \ref{tab:Re300L1} is relatively low, we have also tested the cases with $Re=1000$ using the meshes G4 and G8 for a turbulent wake flow. In this case, the time histories of $C_d$ and $C_{ly}$ are intermittent. We compare the averaged values of theirs within 400 dimensionless time units. As shown in table \ref{tab:GridTestR1000L1}, the difference in $\overline{C}_d$ between the two grids is small and the values of $\overline{C}_{ly}$ by themselves are also small, close to zero. The latter is because the flow is turbulent and statistically, there is not a preferred direction of the lift force in $y$. With these results, we decided to use mesh G4 for $Re\leq 500$, mesh G8 for $500\leq Re\leq 1000$ in our numerical simulations.

\begin{table}
	\begin{tabular}{p{0.7cm} p{2.8cm} p{0.8cm} p{1.4cm} p{1.5cm} p{1.4cm} p{2.8cm}}
		\hline	
		Mesh& \makecell{\ \ $(L_{x1}+L_{x2})$ \\ \ \ \ $\times L_y \times L_z$} & $N_{total}$ & 
		\makecell*[l]{\ \ $\overline{C}_d$ \\ (Err\%)}  & \makecell*[l]{\ \ $\overline{C}_{ly}$ \\ (Err\%)}    & 
		\makecell*[l]{\ \ $St$  \\ (Err\%)}  & \makecell*[l]{\ \ \ \ $\sigma$+$\boldsymbol{i}\omega$ \\ \ \ \ (Err\%)} \\
		S0& \makecell{$(2.5+12)$\\ $\times2.5^2$} & 4125 & 0.8303 (10.9)  & -0.07486 (9.60)  & 0.1542 (7.98)  & 8.63e-3+$\boldsymbol{i}$0.1525 \ \ \quad(7.32) \\
		S1& \makecell{$(5.0+25)$\\$\times5.0^2$} & 4125 & 0.7624 (1.83)  & -0.06925 (1.39)  & 0.1449 (1.47)  & 6.34e-3+$\boldsymbol{i}$0.1436 \ \ \quad(1.06) \\
		S2& \makecell{$(10+50)$\\ $\times 10^2$}  & 4125 & 0.7510 (0.31) & -0.06830 (0.01) & 0.1420 (0.56) & 6.07e-3+$\boldsymbol{i}$0.1420 \ \ \quad(0.07) \\
		S3& \makecell{$(16+80)$\\ $\times 16^2$}  & 5324 & 0.7487\quad(/)  & -0.06830\quad(/)  & 0.1428\quad(/)  & 7.07e-3+$\boldsymbol{i}$0.1421 \ \ \quad(/) \\
		\hline
	\end{tabular}
	\centering\caption{A sensitivity test for the computational domain size. $Re=300$, $\textsc{ar}=1$, $\Delta t=10^{-3}$. The errors are computed using as a reference the results of the mesh S3.}
	\label{tab:CFDdomainsize}
\end{table}

Next, we deal with the size of the computational domain for G4. As mentioned previously (\S~\ref{subsec:NSEq}), the numerical structural stability analysis \citep{luchini2007} of the flow around a two-dimensional cylinder indicates that the results of the stability analysis will change significantly if the computational domain is not sufficiently large and its boundaries are placed close to the flow core area. For a two-dimensional cylinder, the core area is the recirculation zone near the cylinder. In table \ref{tab:CFDdomainsize}, we tested four computational sizes for the G4 grid. The blockage ratios ($=A/S_{in}$) for S0, S1, S2 and S3 are 4\%, 1\%, 0.25\% and 0.098\%, respectively. In order to ensure that the spatial resolution remains relatively the same, the element number outside the O-type mesh zone of mesh S3 increases accordingly with the increase of computational domain size, so its $N_{total}$ is slightly larger. We set the results of the S3 case as the reference. It can be seen that when the blockage ratio is less than 1\%, the differences in time-averaged drag and lift coefficients are less than $2\%$, and $St$ is relatively more sensitive. For the blocking ratio less than 0.25\%, the differences in $\overline{C}_d$, $\overline{C}_{ly}$, $St$ are less than 0.6\%. Meanwhile, we also tested the influence of the computational domain size on the eigenvalue solution based on the mean flow. The effect of the computational domain size on the eigenfrequency $\omega$ is similar to that on the $St$ number. On the other hand, the effect of the computational domain size on the growth rate is stronger than the eigenfrequency. This is because when $Re$ is greater than the critical Reynolds number, the linear stability results based on the mean flow are marginally stable (meaning that the growth rate is nearly zero), so any influence of the computational domain size on the growth rate will appear relatively large. These tests are consistent with the numerical tests in \cite{luchini2007}, whose results indicated that if the boundaries are not placed close to the core zone of instability, one can get reasonable eigenvalues.

After the numerical verification step, in the result section, we will use a computational domain for the case of $\textsc{ar}=1$ with a blocking ratio of 0.25\%. For other $\textsc{ar}$ values of the cylinder flow, the computational domain size is increased accordingly to keep the blocking ratio no greater than 0.375\%. For the cases with the same $\textsc{ar}$, when $Re$ increases, the computational domain size remains unchanged. In this way, the effect of the computational domain size on the growth/decay rate $\sigma$ is fixed to analyse the influence of  $Re$ on $\sigma$.

\subsection{Global stability analysis and validation} \label{subsec:valIRAM}
\begin{figure}   % R1= revision 1, % R2= revision 2
	\hfil
	\centering\includegraphics[trim=0.0cm 0cm 0.01cm 0cm,clip,width=0.7\textwidth]{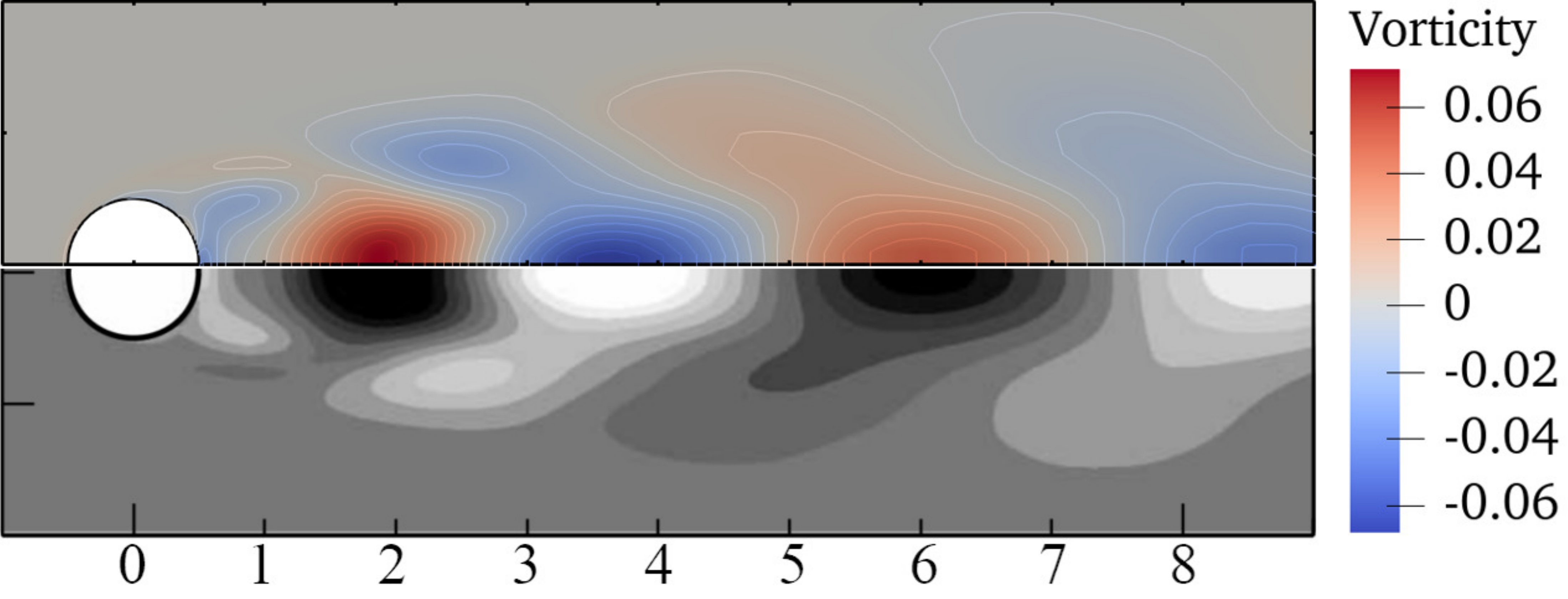}   
	\put(-283,95){($a$)}  % in R1, the figname is {IRAMmode2Dcyl2.png}
	\put(-283,47){($b$)}\\ 
	\centering\includegraphics[trim=0.0cm 0cm 0.0cm 0cm,clip,width=0.49\textwidth]{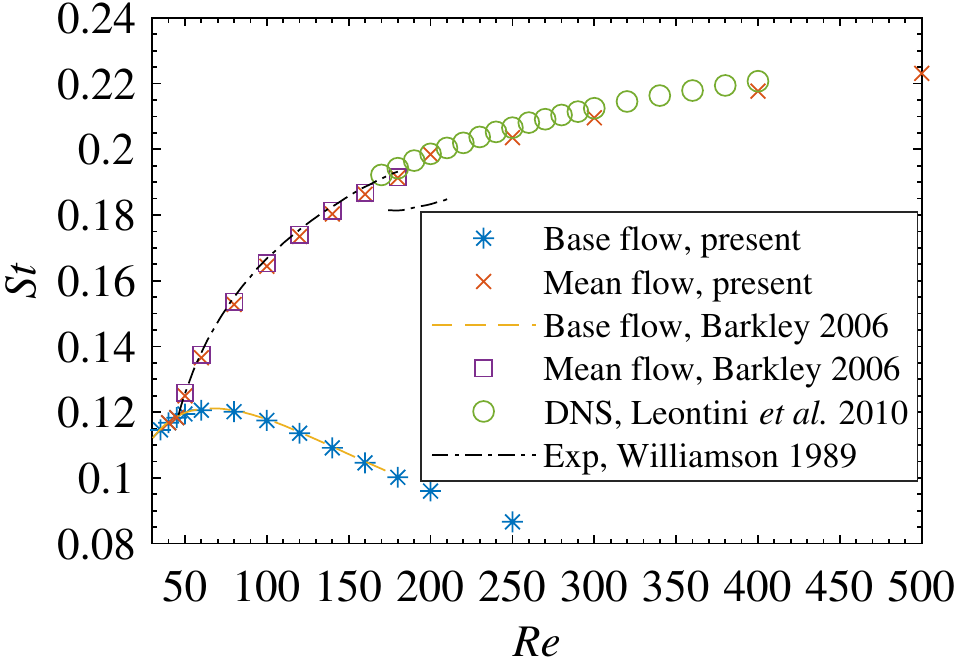}\put(-190,125){($c$)} % in R1, figname is {Re_St_2Dc1y.eps}
	\centering\includegraphics[trim=0.0cm 0cm 0.0cm 0cm,clip,width=0.48\textwidth]{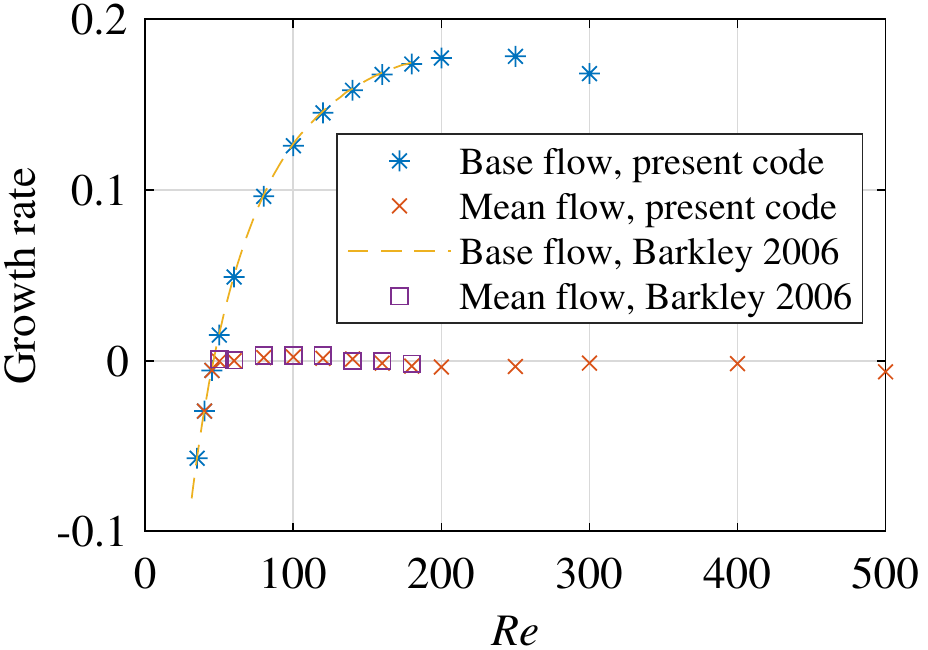}\put(-185,125){($d$)} % in R1, the figname is {GR-Re-validation2D-WithBarkley.eps}
	\caption{Comparison of eigenpairs between present IRAM code and the results in the literature for 2D cylinder wake flow. Panels $a$ and $b$ are the vorticity of leading eigenmode by present code and \cite{barkley2006} at $Re=100$, respectively. Panel $c$ is the eigenfrequency in IRAM and the $St$ in DNS \citep{leontini2010mean_flow} \& experiment \citep{Williamson1989}. Panel $d$ shows the growth rates calculated by the two methods.}
	\label{fig:eigenbyIRAM2Dcyl}
\end{figure}

In this section, we will show the global stability analyses of the flow around the finite-length cylinder. Since in the present literature, we are not able to find the global LSA results of the fully 3D flow around a short cylinder with two free ends, we will validate our implementation of the global stability code in the 2D cylinder flow. The Reynolds number range being explored is $35\leq Re\leq500$ ($Re$ here is also based on the cylinder's diameter $D$). In {\color{blue}figures~}\ref{fig:eigenbyIRAM2Dcyl}($a$, $b$), we compare the real part of the leading eigenmode (in the form of vorticity) by the present linear stability analysis based on the time-mean flow with the leading eigenmode calculated by \cite{barkley2006}. The two results are visually the same. In {\color{blue}figures~}\ref{fig:eigenbyIRAM2Dcyl}($c$, $d$), we further compare the leading eigenfrequency and the growth rate (by both base flow and mean flow) from our IRAM code with the $St$ results and growth rates obtained by previous stability analysis \citep{barkley2006}, DNS \citep{leontini2010mean_flow} and experiment \citep{Williamson1989} as function of $Re$. One can see a good agreement of our results with theirs.

In the infinite cylinder wake flow, there are two critical values of $Re$: one is around $46$, across which the flow undergoes a Hopf bifurcation giving rise to the 2D vortex shedding and the other is approximately $180$, across which the $St$-$Re$ curve shows discontinuity \citep{Williamson1989}, which is due to the three-dimensional nature of the wake flow. Comparing the results in {\color{blue}figure~}\ref{fig:eigenbyIRAM2Dcyl} regarding the critical Reynolds number, one can see that when $Re<46.1$, the eigenvalues obtained using the base flow and the mean flow are the same. This means that for the physical flows without oscillation or with decaying oscillation, stability analysis based on the base flow or the mean flow will yield the same result, which also verifies our implementation of the two methods (one is SFD-based LSA and the other is time-average DNS results). When $46.1<Re<180$, we can observe that stability analysis based on the mean flow can reproduce the experimental results but that based on the SFD base flow cannot, which has been observed and discussed in detail by \cite{pier2002,barkley2006} among many others. When $Re>180$, the result of LSA can no longer reflect the true vortex shedding frequency in the experiment \citep{Williamson1989}, but its eigenfrequency is still close to the vortex shedding frequency obtained by the two-dimensional DNS \citep{leontini2010mean_flow}. As previously mentioned, the discontinuous change of the $St$ number is caused by the three-dimensional nature of the wake, so two-dimensional DNS and LSA are no longer suitable for predicting the vortex shedding frequency in experiments (dot-dashed lines) when $Re>180$.

\begin{figure}[h]
	\centering
	\hfil
	\centering\includegraphics[trim=0.0cm 0cm 0.0cm 0cm,clip,width=0.45\textwidth]{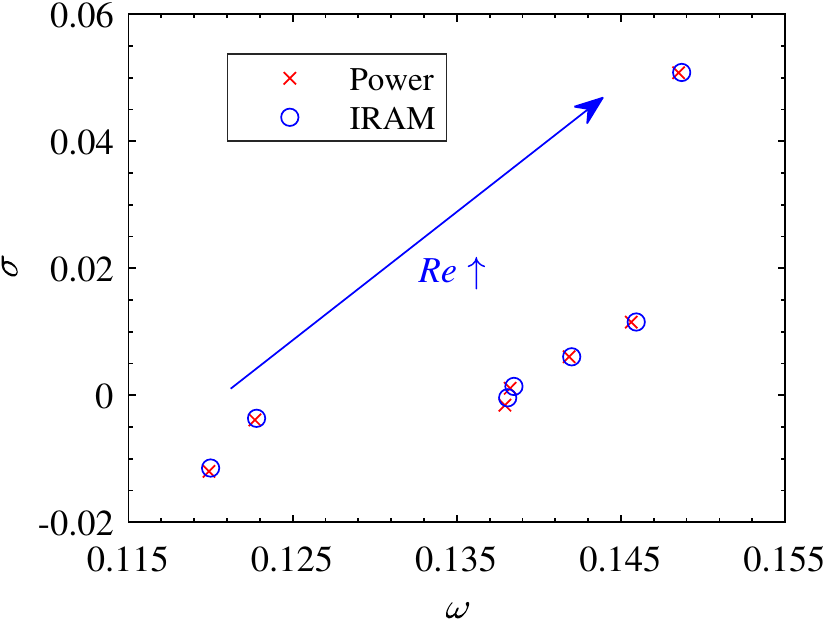}\label{fig:egva1}\put(-180,122){($a$)}    % in R1, the figmane is {GR_omega_power_iram.eps}
	~~~\centering\includegraphics[trim=0.0cm 0cm 0.08cm 0.0cm,clip,width=0.50\textwidth]{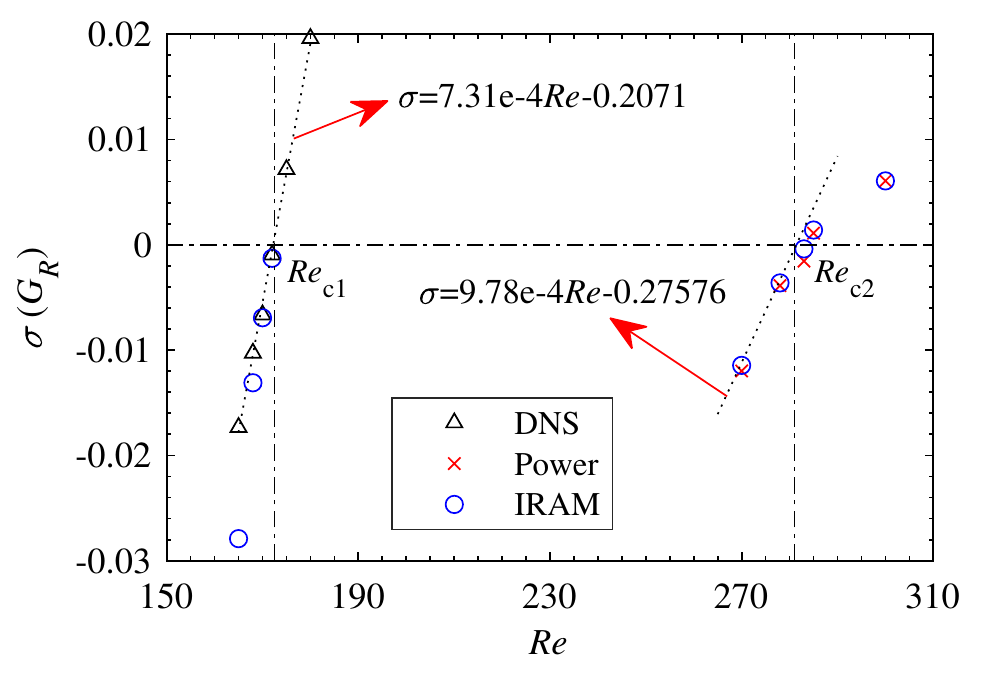}\label{fig:egvb1}\put(-190,122){($b$)}    % in R1, the figmane is {Re_GR_power_iram.eps}
	\hfil
	\caption{The leading eigenvalues obtained by various methods at $\textsc{ar}=1$, $270\leq Re\leq350$. ($a$) growth rates $\sigma$ (imaginary part) and eigenfrequencies $\omega$ (real part) for different $Re$. Each data point is at a certain $Re$. ($b$) growth rates as function of $Re$.}
	\label{fig:eigenbyPowerIRAM}
\end{figure}

As a further validation step of the three-dimensional results, we use the power method to solve the linear equation \ref{LNS} (linearised around the mean flow), then compare its growth rate and frequency with the leading eigenvalue obtained by IRAM. The power method can yield the eigenvalue of the largest absolute amplitude of a matrix, provided it is well separated from the second one. As shown in {\color{blue}figure~}\ref{fig:eigenbyPowerIRAM}($a$), the leading eigenvalues obtained by power method and IRAM are very close to each other at $\textsc{ar}=1$, $270<Re<350$. In {\color{blue}figure~}\ref{fig:eigenbyPowerIRAM}($b$), the growth rates at different $Re$ are obtained by the linear interpolation of the DNS results, the power method and the IRAM. The value of the critical $Re_{c1}$ obtained by the linear interpolation (of the DNS results) is almost the same as that determined by the IRAM, which are $172.2$ and $171.4$ respectively (the difference between them is $0.46\%$). For the cases far away from $Re_{c1}$, the growth rates obtained by these two methods appear different (due to the nonlinearity). The critical Reynolds numbers $Re_{c2}$ obtained by the power method and the IRAM are also close to each other.

\section{Result and discussions} \label{sec:results}

In this section, we will present detailed results of global stability analysis and DNS of the flows past a finite-length cylinder. 
We will show the base states in nonlinear system in the range of $Re\leq 1000$ and $\textsc{ar} \leq 2$, which is characterised by time-averaged aerodynamic coefficients (i.e. $C_d, C_l$ etc.).
We then study the flow transitions from steady wake pattern P1 to chaotic pattern P4 (see Sec.~\ref{subsec:nonlinearDNS}) for $\textsc{ar}=1$. This is followed by a discussion of the global modes and how they are connected to the monochromatic flow patterns P3. In the end, we will discuss the effect of $\textsc{ar}$ with the significance of our results summarised in {\color{blue}figure~}\ref{fig:AR_Rec}.

\subsection{Base states in nonlinear system} \label{subsec:base state}

\begin{figure}[h]
	\centering\includegraphics[trim=0cm 0cm 0cm 0cm,clip,width=0.99\textwidth]{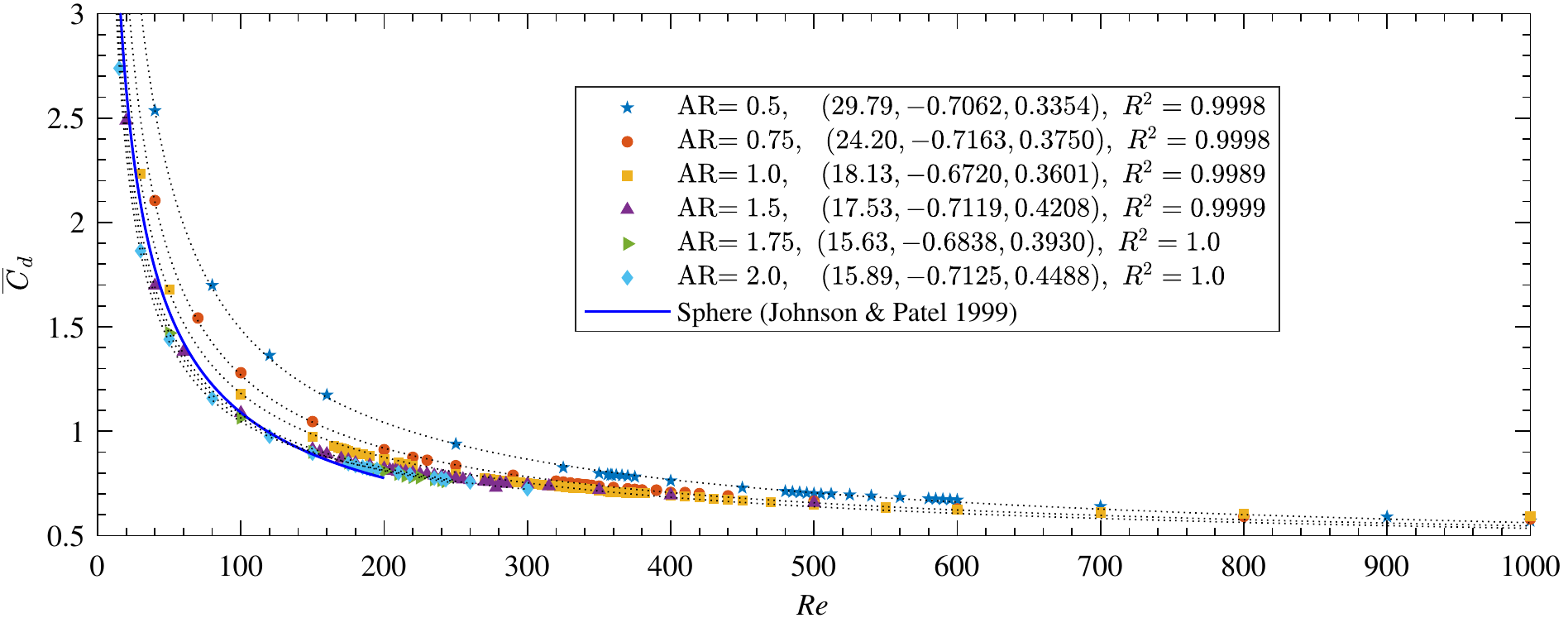}\put(-385,145){($a$)}\\ %  in R1, the figname is {Re_Cd}
	\centering\includegraphics[trim=0cm 0cm 0cm 0cm,clip,width=0.99\textwidth]{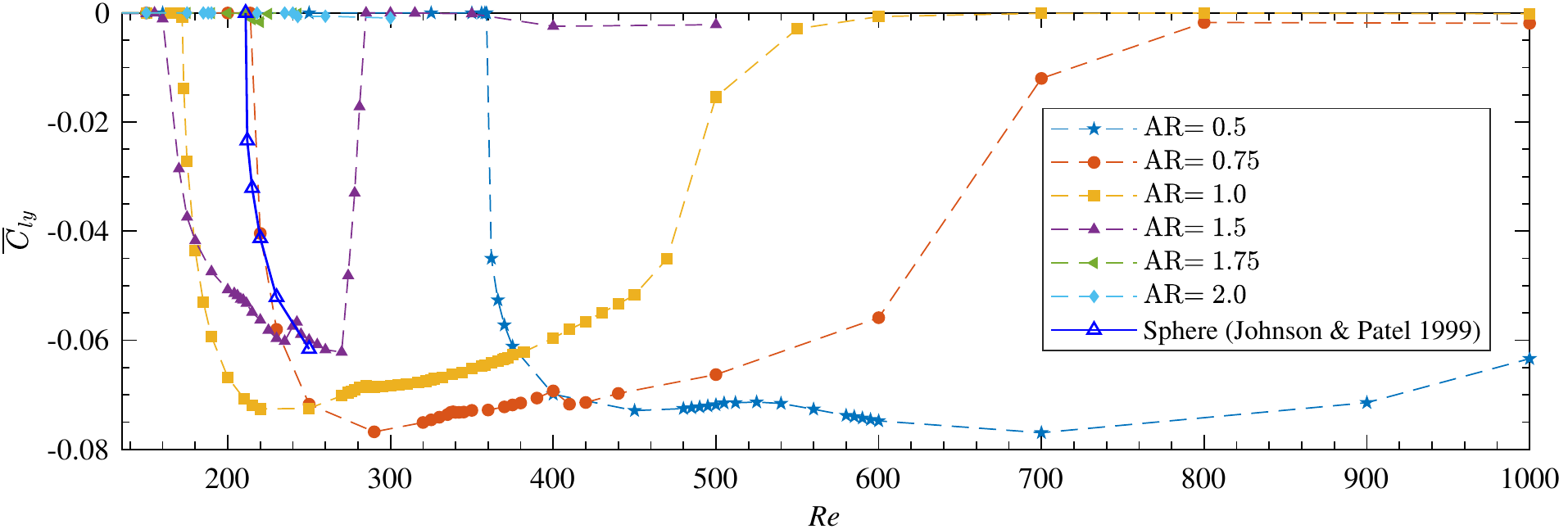}\put(-385,140){($b$)}   % in R1, the figname is {Re_cly}
	\caption{($a$) time-averaged drag coefficient $\overline{C}_d$ as function of Reynolds number. The dotted lines are fitted curves by the functions $\overline{C}_d =aRe^b+c$, and the fitting coefficients $(a, b, c)$ are shown in the parentheses of the legend. ($b$) time-averaged lift coefficient $\overline{C}_{ly}$ as function of Reynolds number. The sign of $\overline{C}_{ly}$ is not important because of the symmetry of the flow; here we choose to present it in negative values.}
	\label{fig:Re_cdcly}
\end{figure}

In this section, we will show the results of the nonlinear DNS for the flow around the finite-length cylinder. First, we observe that with the increase of $Re$, the wake pattern will change, which is reflected in the structure of the mean flow field. The drag and lift coefficients in the mean flow are shown in {\color{blue}figure~}\ref{fig:Re_cdcly}. We can fit the relation between $\overline{C}_d$ and the Reynolds number for the $\textsc{ar}$ values considered in {\color{blue}figure~}\ref{fig:Re_cdcly}($a$) using the power function $\overline{C}_d=aRe^b+c$. The fitting coefficients $(a, b, c)$ are shown in the legend of {\color{blue}figure~}\ref{fig:Re_cdcly}($a$). The goodness of fit $R^2$ for each $\textsc{ar}$ is greater than 0.99. On the other hand, for the lift coefficient, the change in wake pattern has a significant effect on $\overline{C}_{ly}$ as shown in figure \ref{fig:Re_cdcly}($b$). For example, at $\textsc{ar}=1$, each change in the wake pattern by increasing $Re$ will cause discontinuity or kinks in $\overline{C}_{ly}$, as shown at $Re=172,282$ and $550$. These discontinuous or non-smoothness positions correspond to bifurcation points, as we will explain in the following global linear stability analyses.

Comparing the values of $\overline{C}_{ly}$ of $\textsc{ar}=0.5, 0.75, 1$ \& $1.5$, we can observe that when $\textsc{ar}$ becomes larger, the value of $Re$ at which $\overline{C}_{ly}$ becomes non-zero decreases. For these $\textsc{ar}$ values, $\overline{C}_{ly}$ decreases to a negative value from zero and then increases and return to zero. This trend of the $\overline{C}_{ly}$ result with increasing $Re$ indicates that there is a spatial asymmetry generated in the flow evolving from the laminar solution when $Re$ becomes larger and this asymmetry disappears (so $C_{ly}\rightarrow 0$) when $Re$ becomes even larger. At the onset of non-zero $\overline{C}_{ly}$, the gradient of $\overline{C}_{ly}$ is quite steep, signalling a sudden transition of the flow states. On the other hand, when $\textsc{ar}$ becomes large, the $\overline{C}_{ly}$ results look very different: we can see from the panel $b$ that when $\textsc{ar}$ is large (e.g. $\textsc{ar}=2$), the value of $\overline{C}_{ly}$ is always approximately zero for $Re$ from 0 to 300 as we investigated here, in which region we can find non-zero values of $\overline{C}_{ly}$ for small-$\textsc{ar}$ flows. This means that the value of $\textsc{ar}$ is fundamentally important in determining the dynamics of the flow past a finite-length cylinder. We have also superposed the results of sphere \citep{Johnson1999} and interestingly, we find that the $\overline{C}_{ly}$ result of $\textsc{ar}=0.75$ in our flow is very close to that of the sphere. Besides, this similarity also exists in the characteristic size of the separation bubble as shown in {\color{blue}figure} \ref{fig:Re-cta-xs}($a$). Thus, it is reasonable to believe that the finite cylinder with a small $\textsc{ar}$ may behave dynamically similarly to the sphere. As a side note, \cite{Inoue2008} did not report the non-zero value of $\overline{C}_{ly}$ in their DNS results, but this phenomenon exists in our results and has also been confirmed in the flow past a sphere \citep{Johnson1999, TOMBOULIDES2000} and axial flow past a short cylinder \citep{Pierson2019}.

\subsection{Flows past a finite cylinder at $\textsc{ar}=1$} \label{subsec:nonlinearDNS}

The main parameters in this work are $Re$ and $\textsc{ar}$. In order to give a clear presentation of our results, in this section, we will mainly use the case $\textsc{ar}=1$ as a prototype. In subsection \ref{subsec:AReffect}, we will present the effect of changing $\textsc{ar}$.

\subsubsection{Steady flows} \label{subsec:P1}
%\yyl{With two symmetrical planes perpendicular to each other} \\

First, let us discuss the steady flows that one can obtain at different values of $Re$.
For the cylinder with $\textsc{ar}=1$ and $Re$ being larger than approximately $10$, the flow begins to separate on the ends of the cylinder, a closed recirculation zone begins to form, and a pair of counter-rotating vortices appear in the wake. At this time, there are two mutually perpendicular symmetric planes ($x$-$y$ and $x$-$z$ passing through the origin of the coordinate system) in the wake, and the centroid of the cylinder is on these two planes. We call this flow pattern P1. The corresponding Reynolds number at the onset of pattern P1 is $Re_{c0}$. 
{In general, the $\textsc{ar}$ is inversely correlated with $Re_{c0}$. The smaller $\textsc{ar}$ is, the closer $Re_{c0}$ is to 20 in the case of the sphere wake \citep{Johnson1999}, and the larger $\textsc{ar}$ is, the closer $Re_{c0}$ is to 4 in the case of infinite-length 2D cylinder flow. Specifically, the critical Reynolds numbers of the finite cylinder with $\textsc{ar}=0.5, 0.75, 1.0, 1.5$ and $2$ are $Re_{c0}\approx18, 15, 10, 8$ and $6$, respectively.} 

\begin{figure}
% \centering\includegraphics[trim=0.5cm 1.1cm 1.0cm 1.0cm,clip,height=0.245\textwidth]{modeM1_Re100.pdf}\put(-388,88){($a$)} \put(-190,88){($b$)} % R1
 \centering\includegraphics[trim=0.2cm 0.1cm 0.0cm 0.01cm,clip,height=0.265\textwidth]{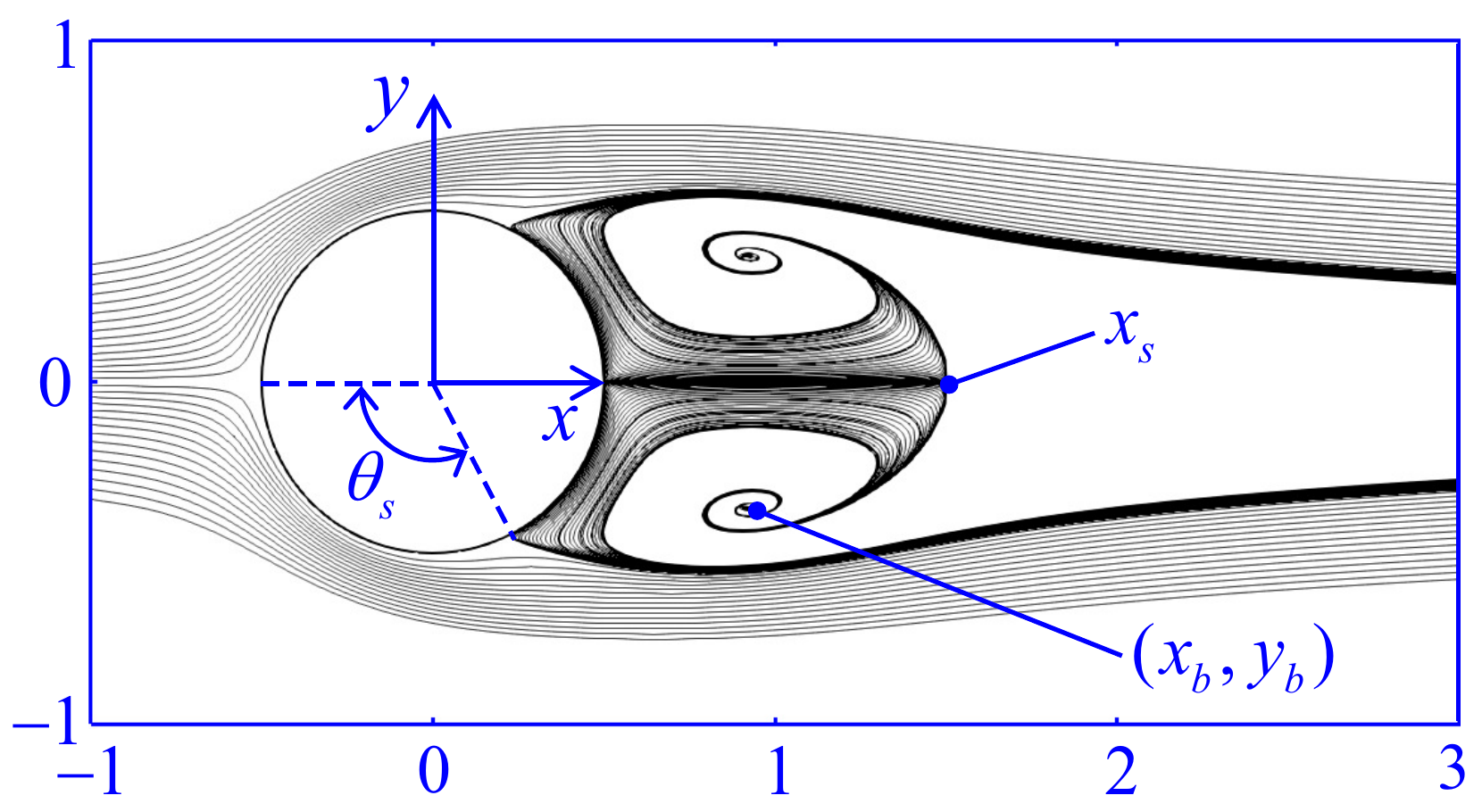}\put(-195,95){($a$)}\ \  % R2
 \ \centering\includegraphics[trim=0.0cm 0.1cm 0.0cm 0.01cm,clip,height=0.265\textwidth]{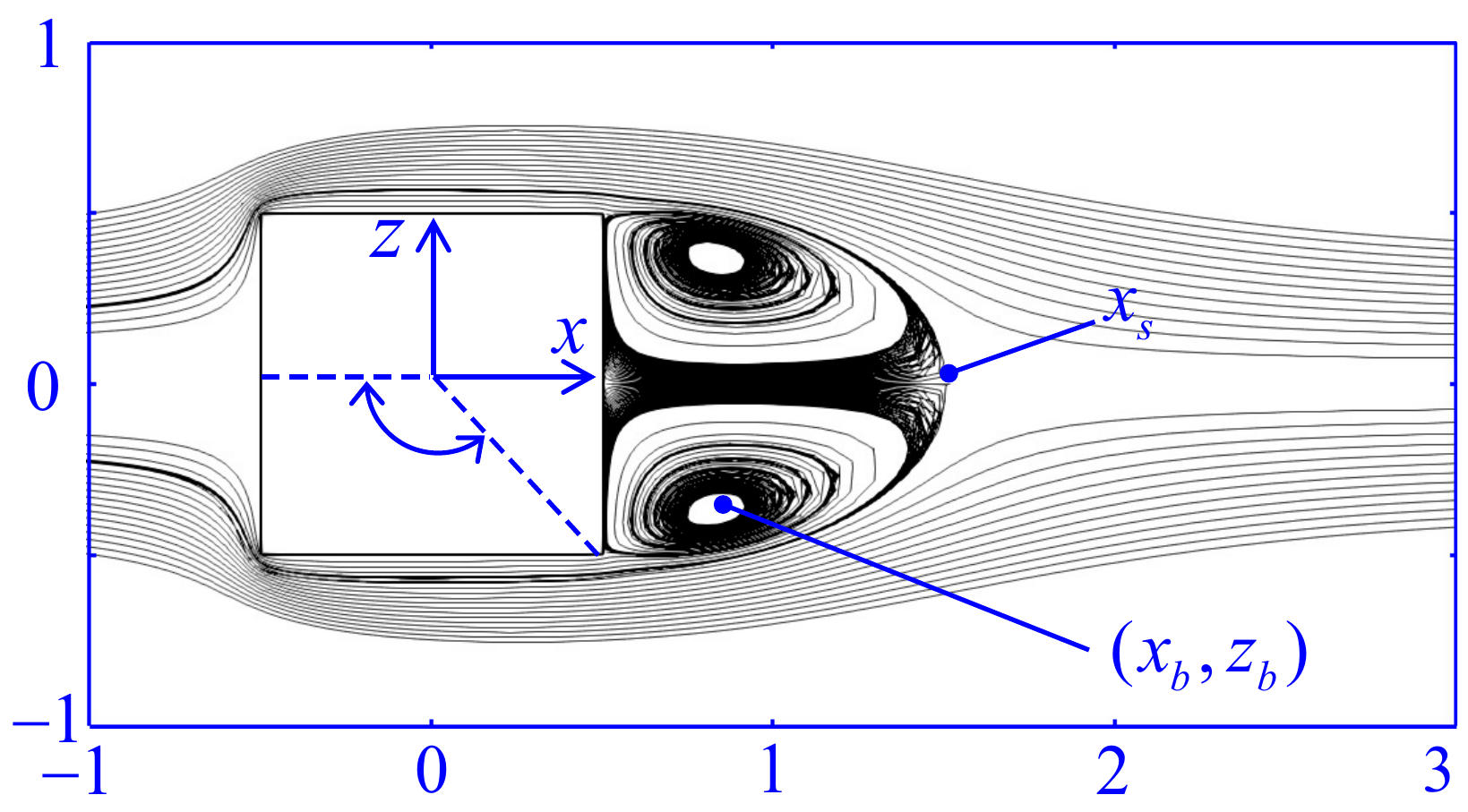} \put(-195,95){($b$)}     % R2
 \caption{A typical steady wake pattern P1 with two symmetrical planes perpendicular to each other, at $Re=100$ and $\textsc{ar}=1$. ($a$) streamlines on plane $x$-$y$, ($b$) streamlines on $x$-$z$ plane.}
 \label{fig:P1}
\end{figure}

 A streamline diagram is used to characterise the typical wake structure of the pattern P1 for $\textsc{ar}=1$ and $Re=100$ in {\color{blue}figure~}\ref{fig:P1}. The specific cross-section is the $x$-$y$ and $x$-$z$ planes of the flow. Unless stated otherwise, the streamwise direction of the flow field diagram in this article is from left to right. The topological traits of the pattern P1, such as the separation position on the cylinder's end ($\theta_s$), the vortex center point ($x_b,y_b$) or ($x_b,z_b$) and the length of the separation bubble ($x_s$), have also been shown in {\color{blue}figure~}\ref{fig:P1}. The fluid separates from the cylinder's end at the angle $\theta_s$ (the vertex of the angle is at the origin, one edge ends at the forward stagnation point, and the other at the separation point), and converges at a point $x_s$ on the $ox$ axis, forming a closed separation bubble in the $x$-$z$ plane. The coordinates of the center of the separation vortex pair in the $x$-$y$ and $x$-$z$ planes are $(x_b,y_b)$ and $(x_b,z_b)$, respectively, and the velocity at the vortex center point ($x_b,y_b,z_b$) is zero. All the lengths are non-dimensionalised by the cylinder diameter. With these parameters, we can quantify the effect of $Re$ on the pattern P1, as follows.

\begin{figure}[h]
	\centering\includegraphics[trim=0cm 0cm 0cm 0cm,clip,width=0.50\textwidth]{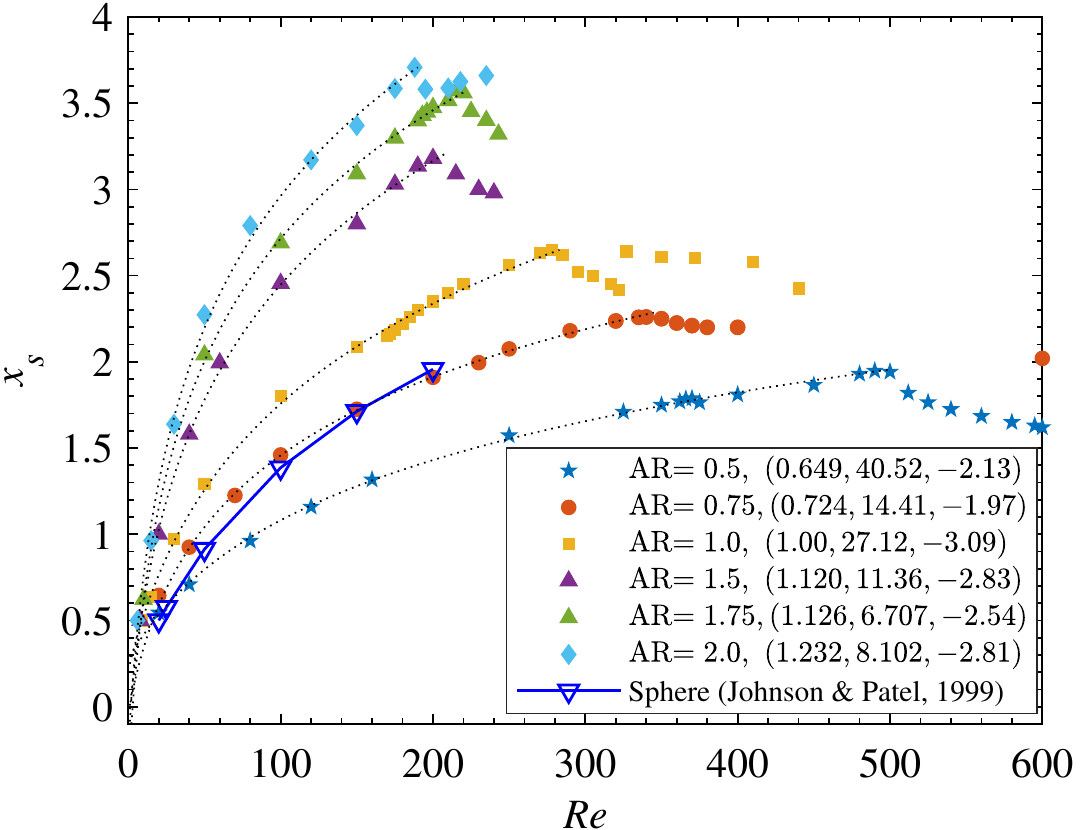}\put(-190,145){($a$)} % in R1, figname is Re_xs.eps
	\centering\includegraphics[trim=0cm 0cm 0cm 0.5cm,clip,width=0.5\textwidth]{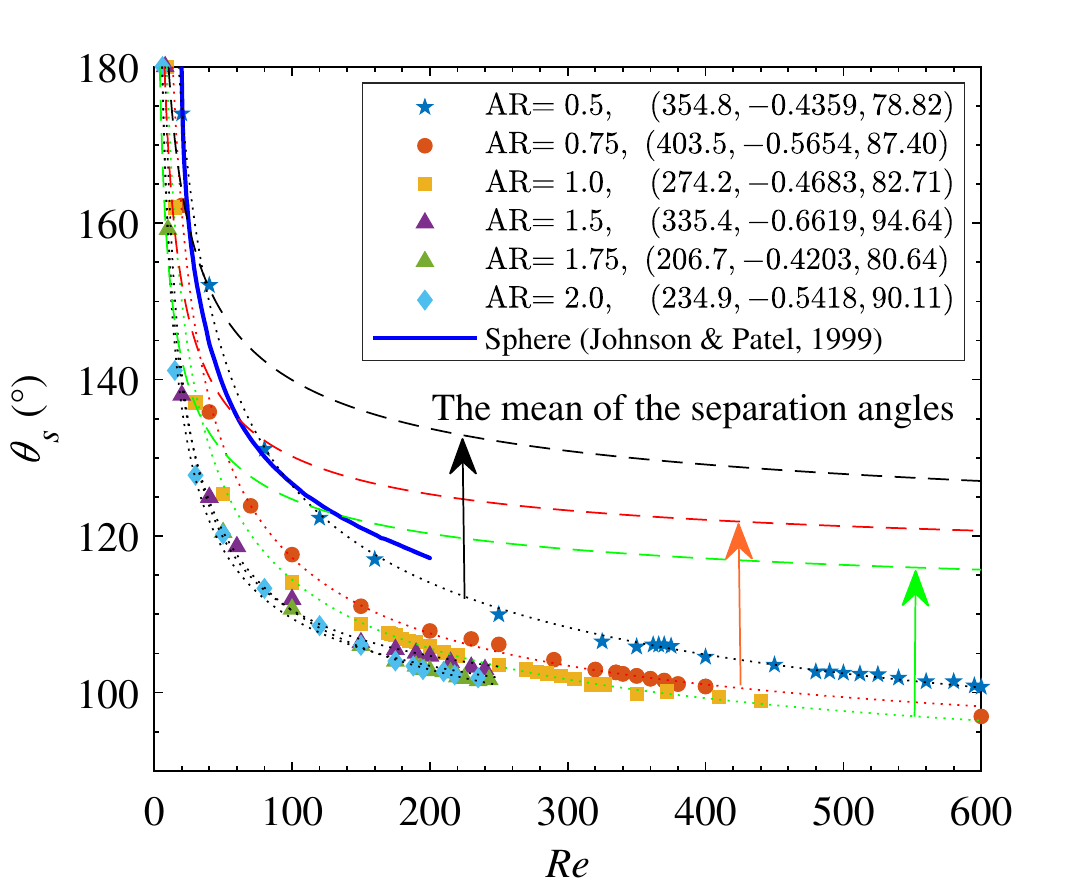}\put(-190,145){($b$)} % in R1, figname is Re_ctas.eps
	\caption{($a$) separation bubble length and ($b$) separation angle as function of Reynolds number. The dotted lines represent fitting functions $x_s=a\ln(Re+b)+c$ and $\theta_s =aRe^b+c$ (see figure \ref{fig:P1}($a$) for the schematic definitions). The fitting coefficients $(a, b, c)$ are shown in the legends. In panel ($b$), the black, red and green dashed lines are the averaged value of the separation angles in the $x$-$y$ and $x$-$z$ planes for the short cylinders with $\textsc{ar}=0.5,0.75$ and $1$, respectively. }
	\label{fig:Re-cta-xs}
\end{figure}

Previous works on flow past a sphere \citep{fornberg_1988,Johnson1999,TOMBOULIDES2000} found and confirmed that when $Re>75$, the separation angle and the separation bubble length are logarithmically related to the Reynolds number. Following this idea, we found that a logarithmic relation also holds in the flow past short cylinders, as illustrated in {\color{blue}figure~}\ref{fig:Re-cta-xs}. For the steady flow, the separation bubble length can be approximately related to $Re$ by $x_s=a\ln(Re+b)+c$ (see the dotted curves in panel ($a$) and for the unsteady flow, the separation angle and separation bubble length are measured in the mean flow). The goodness of fit $R^2$ for $\textsc{ar}=1$ is 0.9988, and exceeds $0.999$ for the other four cylinders. The turning points of curve $x_s$-$Re$ in {\color{blue}figure~}\ref{fig:Re-cta-xs}($a$) correspond to the transition of wake mode (from steady flows to time-periodic flows), which also can be observed in $\overline{C}_{ly}$-$Re$ ({\color{blue}figure~}\ref{fig:Re_cdcly}$b$). One can see that the $x_s$-$Re$ relation curve for the flow past a sphere is close to the $\textsc{ar}=0.75$ data. As pointed out by one of the referees, this aspect ratio is notable because, interestingly, it has a projected frontal area that is very close to that of a sphere (i.e. $\pi/4\approx 0.785$), which can also explains that the curve $\overline{C}_{ly}$-$Re$ ($\textsc{ar}=0.75$) in ({\color{blue}figure~}\ref{fig:Re_cdcly}$b$) almost coincides with that of sphere. 

In figure \ref{fig:Re-cta-xs}$(b)$, the relationship between the separation angle and the Reynolds number is approximately a power function $\theta_s=aRe^b+c$ within the range of $Re<600$. It turns out that the separation angle is less sensitive to the transition, and almost no visible kinks are found in these results. 
At a first glance, the results of the sphere from \cite{Johnson1999} deviate significantly from the result of $\textsc{ar}=0.75$ (note that these results in figure \ref{fig:Re-cta-xs}($b$) is only for $\theta_s$ in the $x$-$y$ plane). A more fair comparison entails an additional consideration of the separation angle in the $x$-$z$ plane. With reference to {\color{blue}figure~}\ref{fig:P1}($b$) (plane $x$-$z$), the separation angles at the sharp trailing corners of the short cylinder are fixed at 135$^\circ$ ($\textsc{ar}=1$), $143.13^\circ$ ($\textsc{ar}=0.75$) and $153.44^\circ$ ($\textsc{ar}=0.5$). Hence, we can simply average the two angles in the $x$-$z$ and $x$-$y$ planes and plot the new results as dashed lines in figure \ref{fig:Re-cta-xs}($b$). One can see that in this case, the results of short cylinders (for $\textsc{ar}=0.5,0.75,1$) indeed straddle that of the sphere, consistent with the previous simple argument as suggested by the referee.

\begin{figure}
  \centering\hfil\includegraphics[trim=0.05cm 0cm 0.02cm 0cm,clip,width=0.90\textwidth]{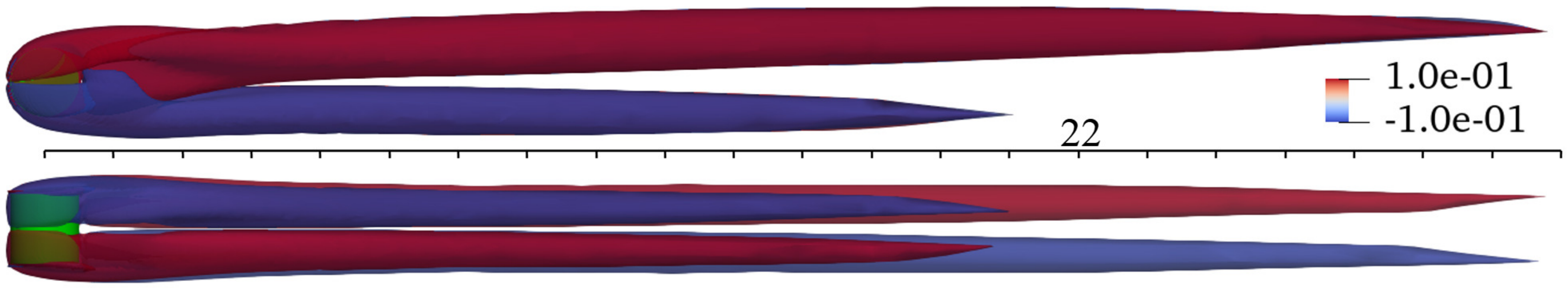}\label{fig:p2_xy}\put(-360,50){($a$)}\put(-360,20){($b$)} % in R1, the figname is {Re278Vorticity-xy-xz.png}
	\caption{A typical P2 type wake at $Re=278$ and $\textsc{ar}=1$, ($a$) $x$-$y$ plane, ($b$) $x$-$z$ plane. Vorticity ($x$-component) isosurface (60\% transparency) is colored by values $-0.1$ and $0.1$. The number 22 is length of the scale ruler. The green object is the cylinder.}
	\label{fig:P2}
\end{figure}

When $Re$ further increases, larger than a certain critical $Re_{c1}$, the current DNS results show that the pattern P1 will be unstable and transition to another steady wake pattern P2 ({\color{blue}figure~}\ref{fig:P2}) with only one symmetrical plane ($x$-$y$), simultaneously inducing non-zero values of $\overline{C}_{ly}$. The bifurcation of the short cylinder at $Re_{c1}$ is steady, time-independent without vortex shedding ($St=0$), so the flow experiences a pitchfork bifurcation. It is noted that the pattern P2 may not appear for certain values of $\textsc{ar}$. As we will show in the end of next section, $\textsc{ar}=1.75$ is such an example. We will present more systematic results of the effect of $\textsc{ar}$ in Sec. \ref{subsec:AReffect}.
This is similar to the flow past a sphere, where there is a regular bifurcation before the Hopf bifurcation. The difference is that the wake of the sphere retains a symmetric plane that is randomly oriented, which can be related to the numerical uncertainty or the perturbations in the experimental environment \citep{TOMBOULIDES2000}. In the literature, there is no discussion about the P2-type flow pattern (\citealt{Inoue2008} did not report this) or its associated regular bifurcation for the flow past a short cylinder with its axis being perpendicular to the streamwise direction (in the following, for the ease of discussion, we will call this flow a radial flow; likewise, we call the flow past a short cylinder with its axis parallel to the streamwise direction an axial flow).
Recently, \cite{Pierson2019} conducted a numerical simulation of an axial flow past a short cylinder at $\textsc{ar}=1, 20\leq Re\leq420$ and $\textsc{ar}=3, 25\leq Re\leq250$.
They found a regular bifurcation for the axial flow around the cylinder at $\textsc{ar}=1$ and $Re\approx278$ (based on the cylinder diameter) without hysteresis, but this bifurcation was not found for the cylinder with $\textsc{ar}=3$. 
\cite{sheard2008} found that there is a regular bifurcation for the radial flow past a short cylinder with hemispherical ends at $Re=350\pm2$.
Finally, three-dimensional transitions of the bluff-body cube have been numerically studied by \cite{saha2004}. The cube wake also undergoes a regular bifurcation (losing one symmetric plane but retaining time-independence) at $Re=216$ and a Hopf bifurcation at $Re=270$. 
 From these works, it can be hypothesised that the aspect ratio $\textsc{ar}$ of the bluff body (including the short cylinder with flat ends as we study here) is related to the existence of a regular bifurcation.

\begin{figure}
 	\centering\includegraphics[trim=0.1cm 0.05cm 0.2cm 0cm,clip,width=0.495\textwidth]{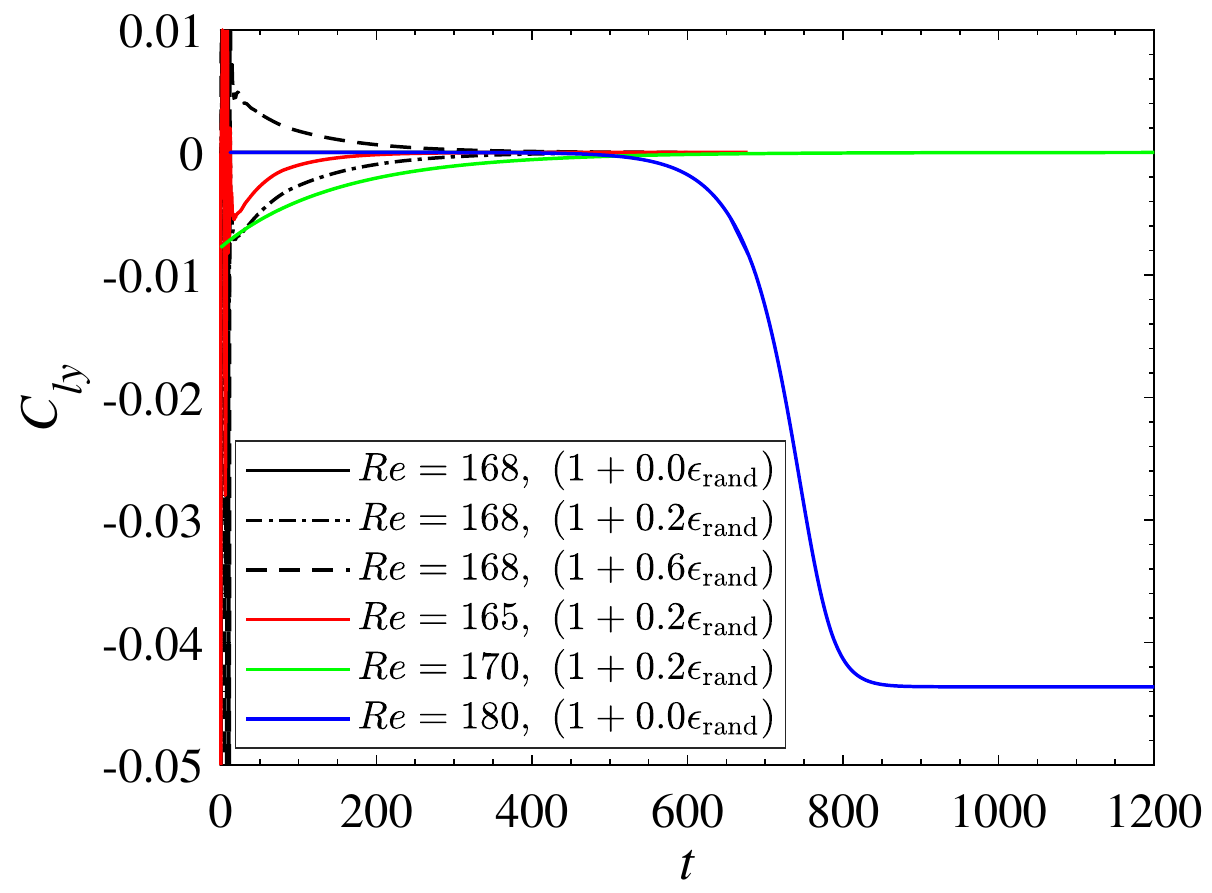} 
  % in R1, figname is clyRe165168170180.eps
	\centering\includegraphics[trim=0.1cm 0.0cm 0.1cm 0cm,clip,width=0.495\textwidth]{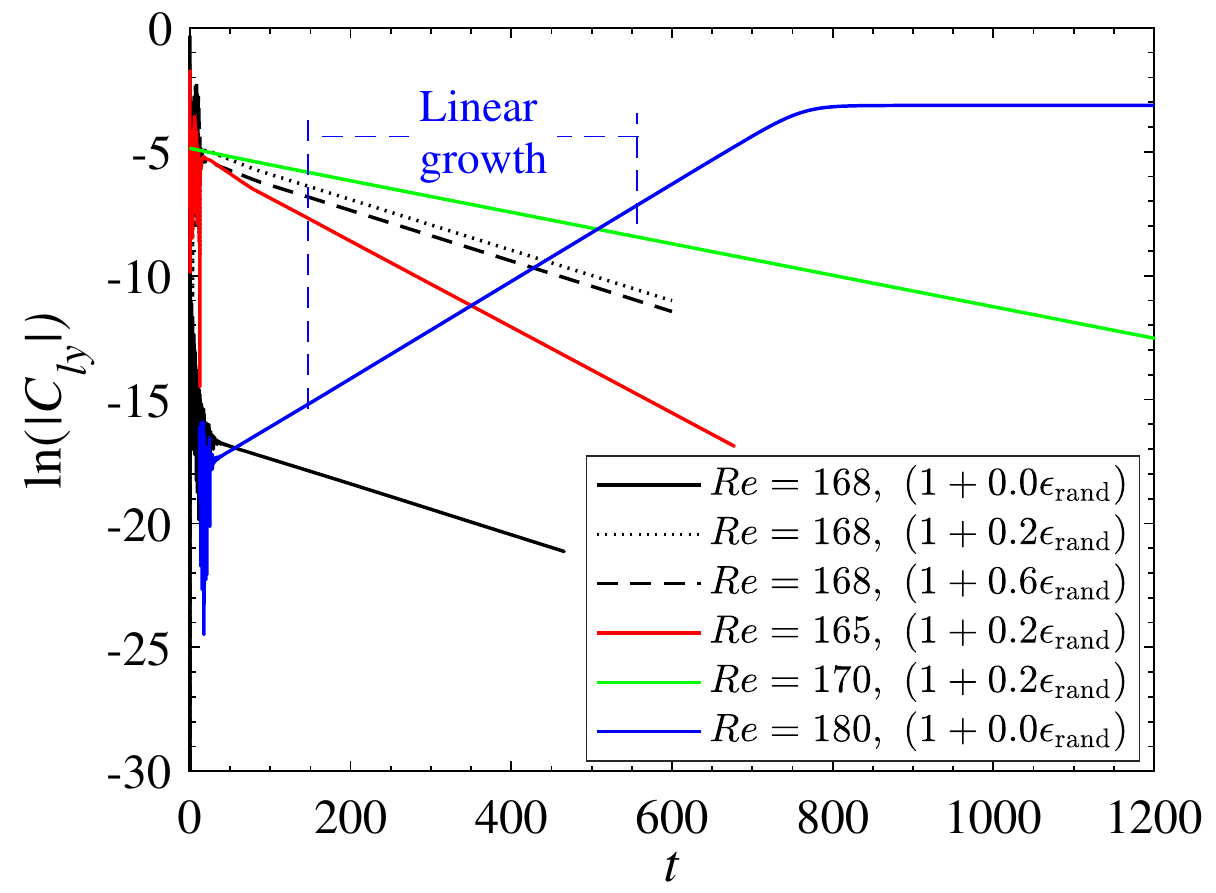}  
  % in R1, figname is lnclyRe165168170180.eps
 	\caption{Time histories of lift coefficient $\it C_{ly}$ and $\rm ln(\it |C_{ly}|\rm)$. $(1+0.2\epsilon_\textrm{rand})$ in the legend means that the initial condition is the base flow (1,0,0) plus a white noise of 0.2 times the base flow in the three velocity components. Likewise for the other initial conditions.}
 	\label{fig:lncly}
 \end{figure}

Now we present the way we determine $Re_{c1}$. We perform nonlinear DNS for $\textsc{ar}=1$, and the critical value $Re_{c1}$ is approached by gradually increasing the Reynolds number from a small value. For each $Re$, we can calculate the growth rate of disturbance development as in {\color{blue}figure~}\ref{fig:lncly}. In this paragraph, we focus on the blue curves for $Re=180$ without randomness in the initial condition and will discuss the other curves shortly. The time series of $C_{ly}$, when plotted in a logarithmic scale against the time in a linear scale (as in panel $b$), will display a period of a straight line, which represents the linear phase. The linear growth or decay rate can then be calculated and the critical $Re$ corresponding to the zero growth rate can be interpolated by fitting several data points using a linear function, which is shown in {\color{blue}figure~}\ref{fig:Re_GR_ln_cly}, to be discussed shortly. 
%For $\textsc{ar}$=1, we can see that $Re_{c1}=172.2$, which is consistent with the results in {\color{blue}figure~}\ref{fig:lncly}. 
For $\textsc{ar}$=1, we can see that $Re_{c1}=172.2$, which is consistent with the $Re$ of onset of non-zero $\overline{C}_{ly}$ shown in {\color{blue}figure~}\ref{fig:Re_cdcly}($b$).
The $Re_{c1}$ values of other $\textsc{ar}$ have also been calculated and it is found that when $\textsc{ar}$ is small or large, $Re_{c1}$ is relatively large. The $\textsc{ar}$ having the minimum $Re_{c1}$ is approximately located around $1.25\sim1.5$. Besides, the $\textsc{ar}$-dependence of the critical Reynolds number will be revealed in Sec.~\ref{subsec:AReffect}. In the analysis of \cite{TOMBOULIDES2000}, the authors also performed a logarithmic transformation of the azimuthal velocity at a point located in the wake and near the sphere, and found that the logarithm of the azimuthal velocity is a linear function of time during the initial evolution. The growth rate obtained by this method is consistent with the LSA result of \cite{natarajan1993}.

  	\begin{figure}[h]
%		\centering\includegraphics[trim=0.2cm 0.1cm 0.6cm 0.2cm,clip,width=0.95\textwidth]{dlogAdtA2_Re180.pdf} % R1
		\centering\includegraphics[trim=0.1cm 0.1cm 0.6cm 0.2cm,clip,width=0.46\textwidth]{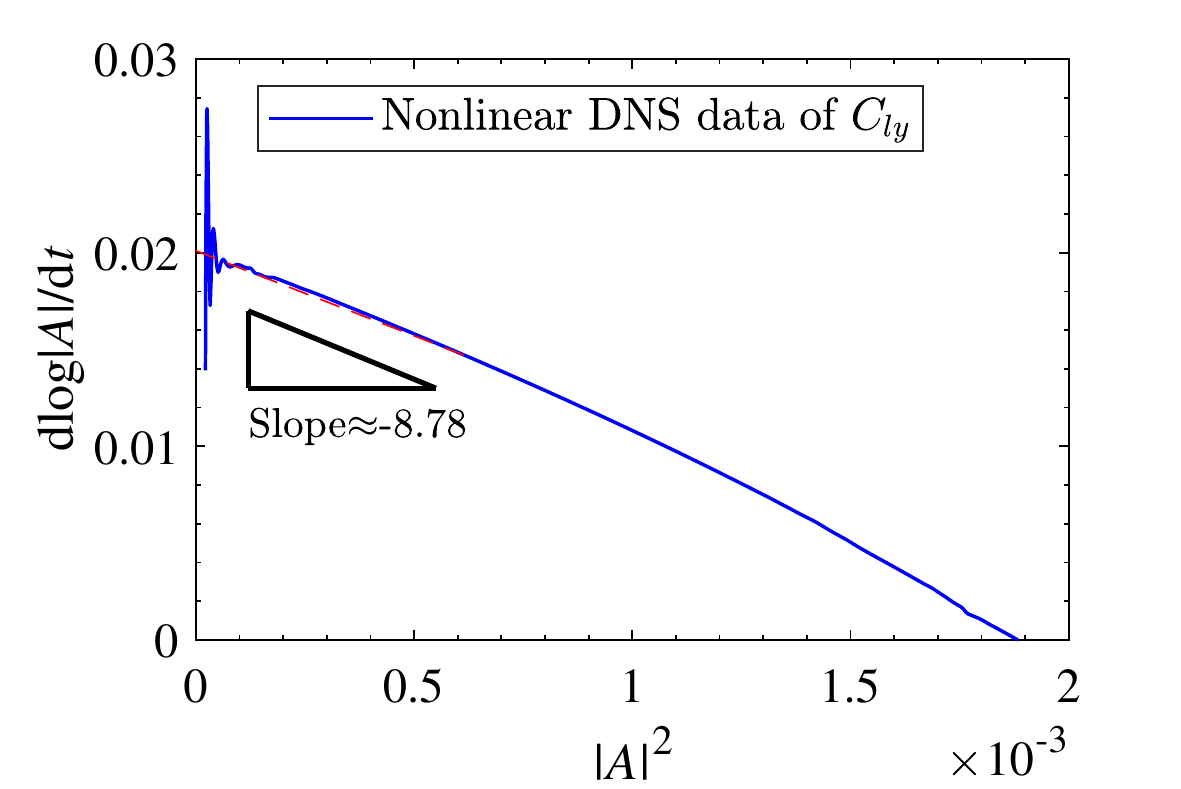}
		\centering\includegraphics[trim=0.0cm 0.1cm 0.6cm 0.2cm,clip,width=0.49\textwidth]{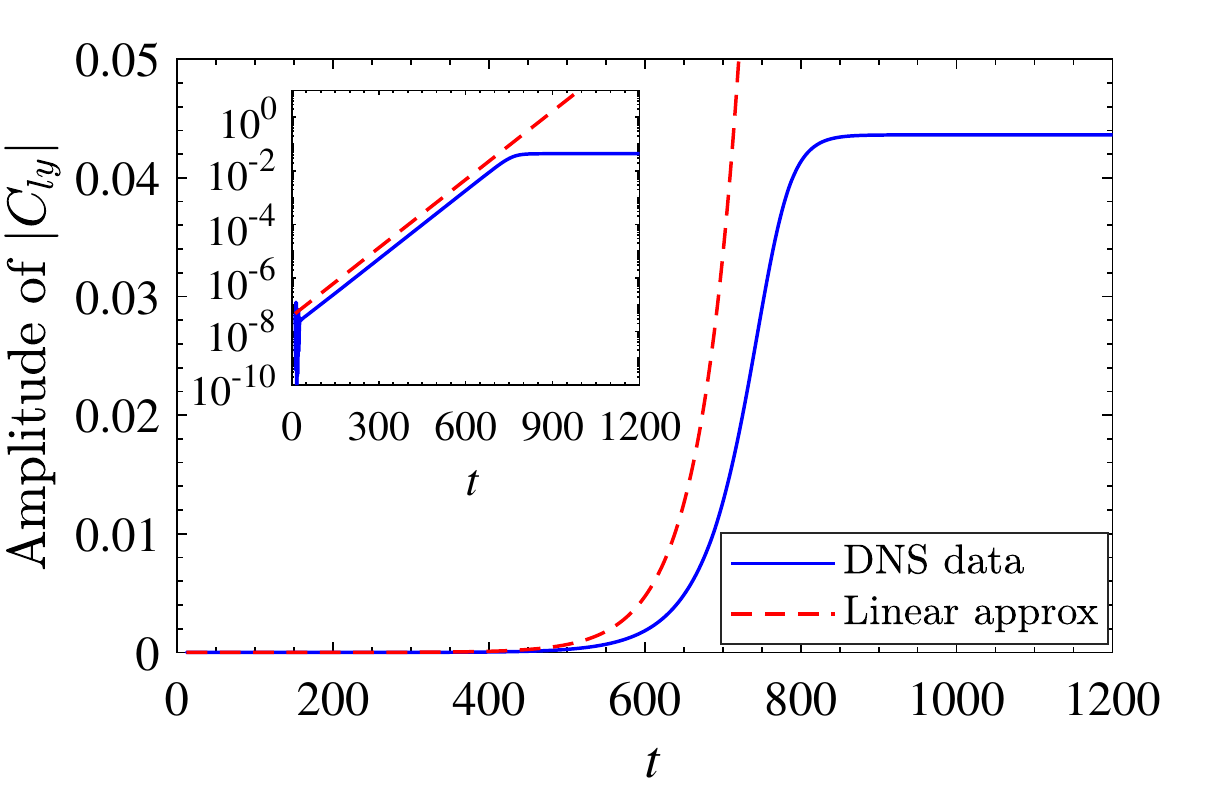}
		\caption{Time derivative of log$|A|$ versus $|A|^2$ with the vertical intercept giving the linear growth rate $c_1=-0.0201$ and the gradient near this intercept point giving the Landau coefficient $c_3=-8.78$, which indicates supercritical bifurcation behaviour. Nonlinear DNS data of case $\textsc{ar}=1$, $Re=180$.} 
		\label{fig:dlnClydt}
	\end{figure}
	
\begin{figure} % in R1, the figname is {Re_GR_ln_c1y.esp}
  \centering\includegraphics[trim=0cm 0cm 0cm 0cm, clip, width=0.95\textwidth] {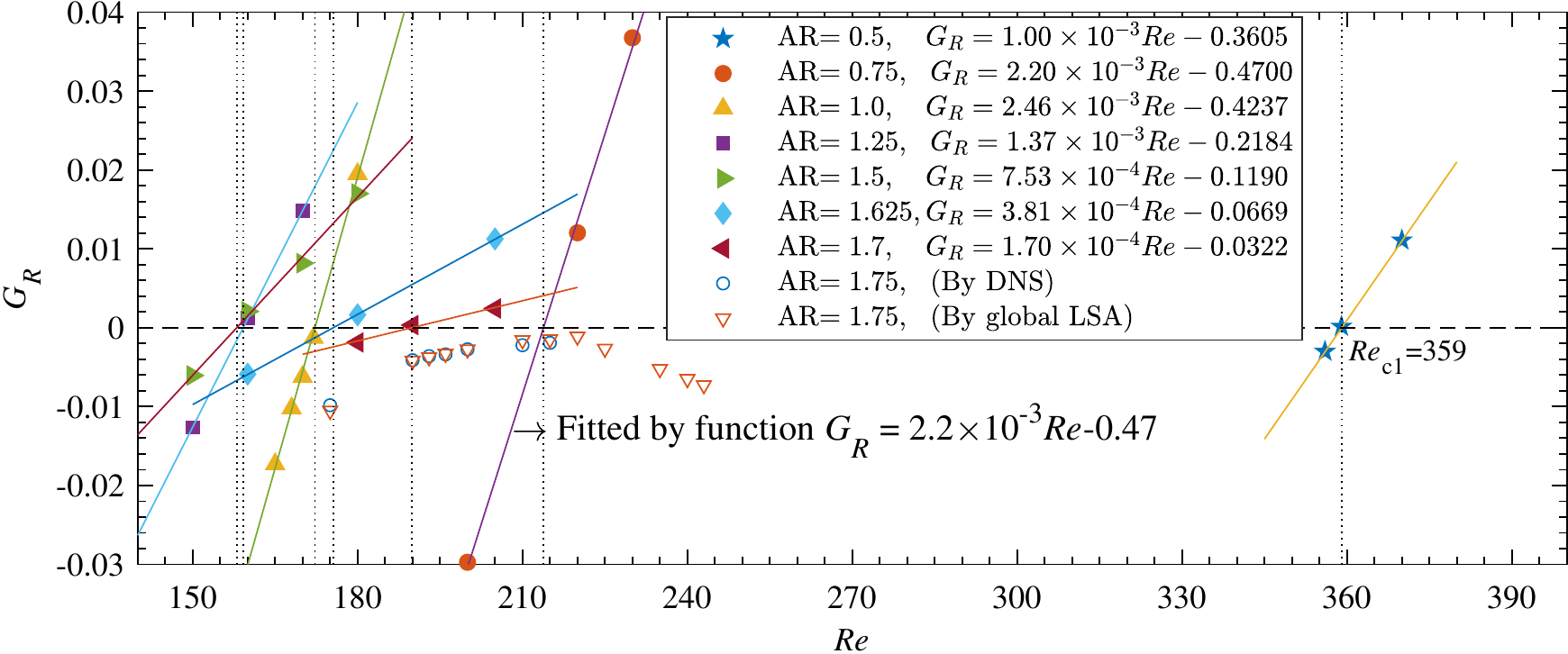} 
  \caption{The growth rate ($G_R$) calculated using the DNS results as function of Reynolds number. Unstable flows will transition from P1 pattern to P2 pattern via a regular bifurcation and the critical Reynolds number is $Re_{c1}$.}
  \label{fig:Re_GR_ln_cly}
\end{figure}

Some numerical experiments have also been conducted to perturb the flows in {\color{blue}figure~}\ref{fig:lncly}. We applied white noise (denoted as $\epsilon_\textrm{rand}$) to the initial conditions in the DNS at $\textsc{ar}=1, Re=165, 168, 170$ (subcritical, as $Re<Re_{c1}=172.2$). The random numbers are shifted to perturb the solution positively and negatively. The amplitude of the white noise is 0.2 times the inlet velocity for the three cases and additionally, for $Re=168$, we also examined the amplitudes of the (white-noise) disturbance being 0 and 0.6 times the inlet velocity. The white noise is applied to all the three velocity components through the following initial conditions (Eq. \ref{IC+Rand} and Eq. \ref{Rand}).\\
\begin{equation}
	\left\{
	\begin{array}{lr}
		U_x = 1.0 +a\epsilon_\textrm{rand},\quad (a_1=3.0\times10^4, a_2=-1.5\times10^3, a_3 =  0.5\times10^5)\\
		U_y = 0.0 +a\epsilon_\textrm{rand},\quad (a_1=2.3\times10^4, a_2=2.3\times10^3, a_3 = -2.0\times10^5)\\
		U_z = 0.0 +a\epsilon_\textrm{rand},\quad (a_1=2.0\times10^4, a_2=1.0\times10^3, a_3 =  1.0\times10^5)\\
	\end{array}
	\right.
	\label{IC+Rand}
\end{equation}	

\begin{equation}
	\left\{
	\begin{array}{lr}
		\epsilon_1 = a_1 (i_g+x \textrm{sin}y) + a_2 i_x i_y + a_3 i_x \\
		\epsilon_2 = a_1 (i_g + z \textrm{sin}\epsilon_1) + a_2 i_z i_x + a_3 i_z \\
		\epsilon_\textrm{rand} = \textrm{cos}(10^3\textrm{sin}(10^3\textrm{sin}\epsilon_2))\\
	\end{array}
	\right.
	\label{Rand}
\end{equation}
where, $a$ is the maximum amplitude of the perturbation. $\epsilon_\textrm{rand}$ denotes white noise with random numbers between -1 and 1 in Eq. \ref{Rand}. $x,y,z$ are the coordinate values of the element nodes.  $i_x,i_y,i_z,i_g$ are the numbers of the element nodes.
From {\color{blue}figure~}\ref{fig:lncly}, we can see that the lift coefficients $C_{ly}$ for all the cases aforementioned eventually converge to zero (because the flows are subcritical) and the linear phase survives for a long time. This demonstrates the robustness of the results.
A by-product of these numerical experiments is that they seem to indicate that the first bifurcation of the short cylinder around $Re_{c1}$ is probably supercritical for the parameters we consider here, because if the bifurcation was subcritical, the white noise with a large amplitude will probably bring the flow to a stable finite-amplitude nonlinear solution.  
In order to confirm the supercriticality of the first bifurcation in this flow, we estimate the coefficients of the Landau model from the DNS data (as is well known, the Landau coefficient calculated around the linear critical conditions determines the nature of the bifurcation, see \cite{Guckenheimer1983}). 
The global variable $C_{ly}$ that can directly reflect the transition from P1 to P2 is used to evaluate the Landau coefficient at $Re=180$ and $\textsc{ar}=1$. The original time history of $C_{ly}$ is shown in {\color{blue}figure} \ref{fig:lncly}($a$), where $|C_{ly}|$ grows and finally saturates. 
Now, based on Landau equation $ \frac{d A}{d t}=\it c_1 A+ c_3 |A|^2$ (where $\it A$ can be viewed as the amplitude of $C_{ly}$) or, equivalently, $\rm dlog|\it A\rm |/d\it t=c_1+ c_3|A|^2$, the Landau coefficients $c_1$ and $c_3$ can be calculated by plotting $\rm d(log|\it C_{ly}\rm |)/d\rm t$ versus $|C_{ly}|^2$ \citep{thompson2001kinematics}, as shown in {\color{blue}figure} \ref{fig:dlnClydt}. Therefore, the vertical intercept point gives an estimation of $\it c_1\rm=0.0201$ and the gradient near this point is an approximation of $\it c_3\rm=-8.78$. The value of growth rate $c_1=0.0201$ is close to the linear growth rate $0.0196$ (at $Re=180$) obtained by calculating the linear slope in {\color{blue}figure} \ref{fig:lncly}(b).
Furthermore, the negative $\it c_3\rm=-8.78$ indicates that the (lowest-order) nonlinearity stabilises the flow. In panel ($b$), one can also see that the nonlinearity stabilises the flow after the linear regime. Thus the bifurcation is supercritical transitioning from P1 to P2. We mention in passing that, theoretically, using this method to infer the bifurcation type of a flow is most accurate when the parameter is close to the critical condition. Nevertheless, when this is satisfied, the evolution time of the flow is very long in DNS. As long as the bifurcation type is concerned, the method can be applied slightly away from the critical condition \citep{Henderson1996,Gao2013,CARMO2010,Feng2021}. Besides, we arrive at the above conclusion using only one instance of $Re$ and admit that a more rigorous analysis would be to apply the weakly nonlinear stability analysis around the linear critical conditions, which can be pursued in a future work.

To sum up, the P2-type steady flow has been found for short cylinders with small $\textsc{ar}$. Most of the relevant research on the flow past a short bluff body have focused on the sphere or cylinder with a large $\textsc{ar}$, and no previous studies have systematically investigated this P2-type wake mode in the axial flow past a finite cylinder. Through our DNS results in the range of $0.5<\textsc{ar}<2$, it is found that the $\textsc{ar}$ of the short cylinder has a very significant and important influence on the critical Reynolds number $Re_{c1}$. Only when the $\textsc{ar}$ is relatively small ($\textsc{ar}< 1.75$) will there be a regular bifurcation. As can be seen in {\color{blue}figure~}\ref{fig:Re_GR_ln_cly}, when $\textsc{ar}=1.75$, we cannot find the regular transition by either DNS or LSA methods. In fact, when $\textsc{ar}\gtrsim1.75$, as $Re$ increases, the wake of the radial flow around the short cylinder will not experience the P2-type wake, and directly transition from the P1-type to a periodic shedding of hairpin-shaped vortices P3-0 with $\overline{C}_{ly}=\overline{C}_{lz}=0$, which has been reported many times in the flow past a bluff body with $\textsc{ar}>2$, for example, in the flow past a cylinder with two free hemispherical ends \citep{Sheard2005,sheard2008}, a cylinder with two free flat ends at moderate $2 \leq \textsc{ar} \leq 10$ (named type III by \citealt{Inoue2008}) and at $\textsc{ar}=3$ and $Re>125$ (\citealt{Pierson2019}'s figure 29), a prolate spheroid at $L/D=6$ and $Re=100$ \citep{el2012wakes} and a freely falling cylinder following rectilinear paths at Archimedes number $Ar=200$ and $L/D=5$ (\citealt{Toupoint2019}'s figure 17a).
Nevertheless, we will report in the next section that there are two unreported periodic vortex shedding wakes (P3-1 and P3-2) of radial flow around the short cylinder with $\overline{C}_{ly}\neq0$.

\subsubsection{Vortex shedding} \label{subsec:P3}

\begin{figure} % in R1, the figname is {Re_GR.eps}
	\centering\includegraphics[trim=0cm 0cm 0cm 0cm, clip, width=1.0\textwidth]{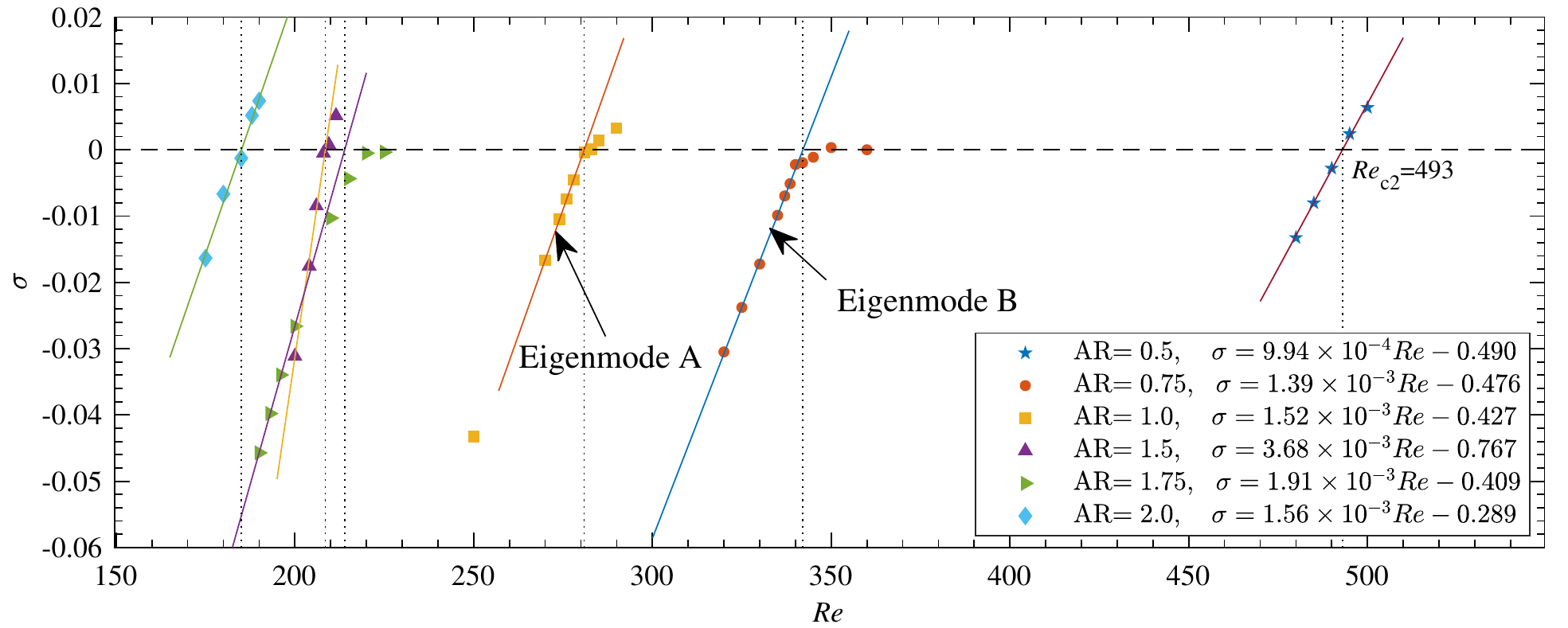} 
	\caption{The growth rate $\sigma$ of the leading eigenmode in the global LSA (of the time-mean flow) as function of $Re$. Unstable flows will transition from P2 pattern to P3 pattern via a Hopf bifurcation and the critical Reynolds number is $Re_{c2}$.}
%	For $\textsc{ar}=0.75$, we indicate two types of eigenmodes A (green dots) and B (red dots), which will be discussed in Sec. \ref{subsec:globalmodes} to follow.
	\label{fig:Re_GR}
\end{figure}

%\yyl{With one symmetrical plane}\\
When $Re$ increases further, a second bifurcation appears at $Re\approx282$ (for $\textsc{ar}=1$) with $St\neq0$, which belongs to a Hopf bifurcation. {\color{blue}Figure~}\ref{fig:Re_GR} shows the relationship between the linear growth rate and $Re$ for short cylinders of different $\textsc{ar}$ values from $0.5$ to 2 near the Hopf bifurcation point. The relationship between the growth rate and the Reynolds number is approximately linear when the flow is stable and near the critical condition. The solid lines in {\color{blue}figure~}\ref{fig:Re_GR} represent linear fitting results which are shown in the legend. The general trend is that the larger the value of $\textsc{ar}$ is, the smaller the critical value of $Re_{c2}$. The denoted eigenmodes A and B will be discussed in Sec. \ref{subsec:globalmodes}.

\begin{figure}
\centering
{\includegraphics[trim=2.8cm 4.0cm 1.5cm 3.0cm, clip, width=1\textwidth]{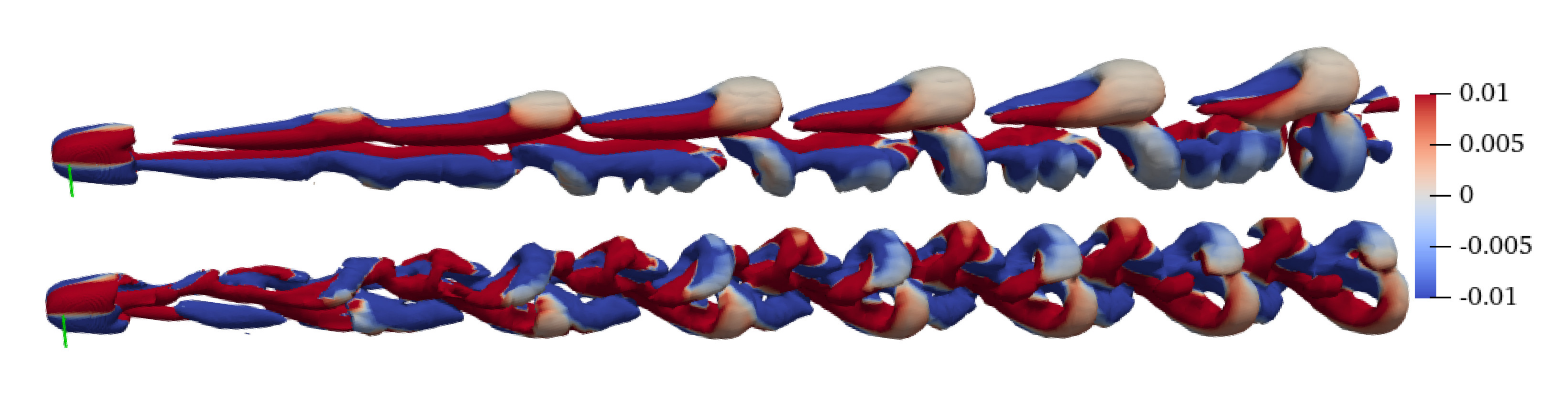} % in R1, the figname is {Re290P3_12_fig10.pns}
\put(-385,62){($a$)}
\put(-385,22){($b$)} } 
 \caption{Two typical periodic wake patterns for  the case $Re=290$, $\textsc{ar}=1$. ($a$) wake pattern P3-1 stimulated by the initial condition $1+0.0\epsilon_\textrm{rand}$; ($b$) wake pattern P3-2 stimulated by the initial condition $1+0.2\epsilon_\textrm{rand}$. The $Q$-criterion isosurfaces $Q=0$ are colored by the $x$-component of the vorticity ranging from $-10^{-2}$ to $10^{-2}$. The green ruler is collinear with the cylinder axis.}
 \label{fig:P3-12}
\end{figure}

\begin{figure}
	\centering\includegraphics[trim=1cm 0.8cm 7.0cm 0.5cm, clip, height=0.355\textwidth]{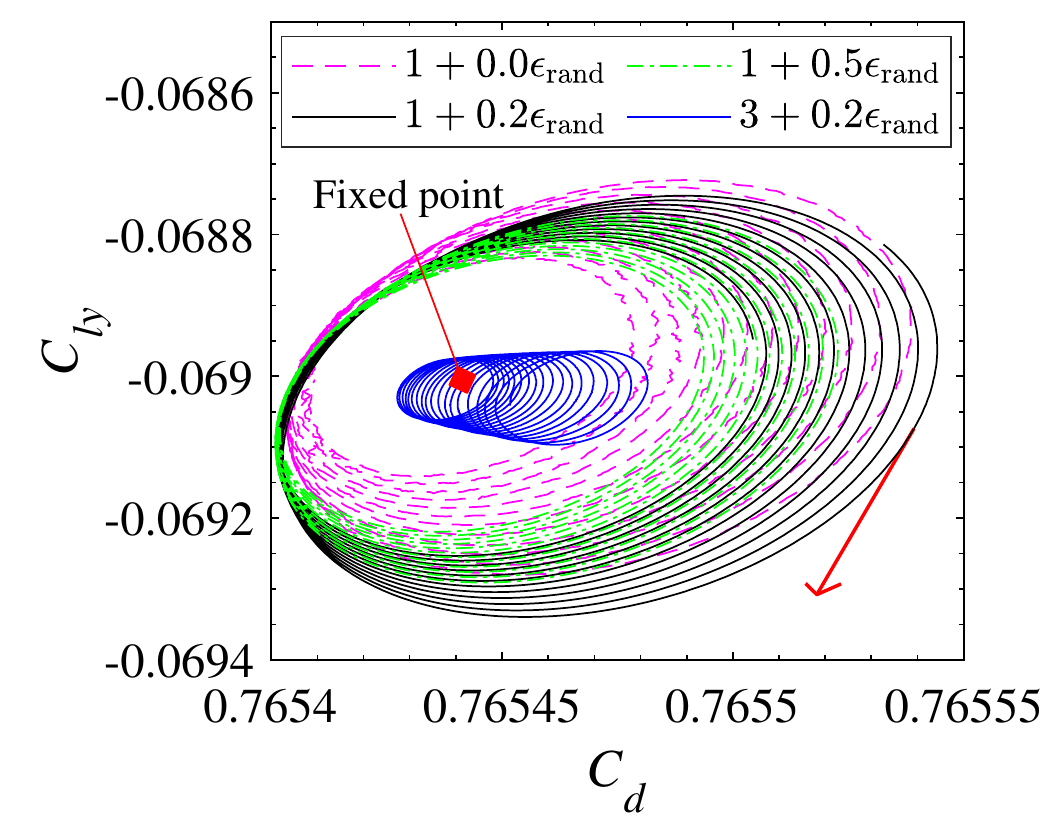}\put(-169,129){($a$)}  % in R1, the figname is {Re278cdclyA00_05.eps}
	~~~~~~~\centering\includegraphics[trim=0.0cm 0cm 0.1cm 0.05cm, clip, height=0.358\textwidth]{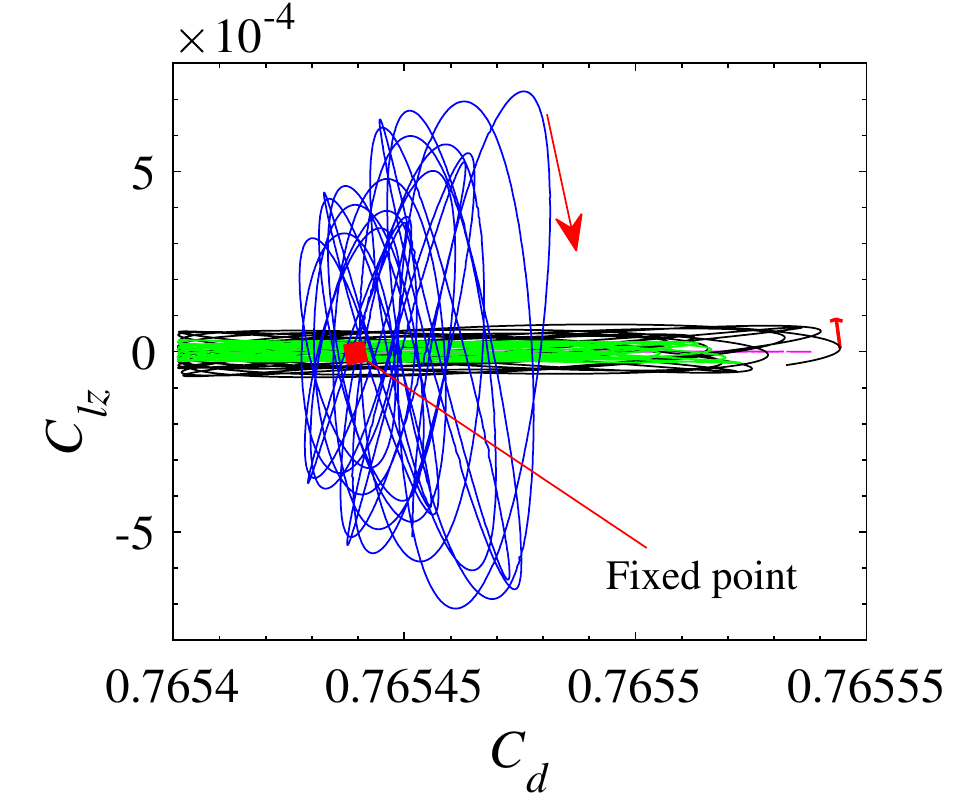}\put(-162,129){($b$)} % in R1, the figname is {Re278cdclzA00_06.eps}
	\caption{The diagram of $C_d$-$C_{l}$ at $Re=278$, $\textsc{ar}=1$. The expression $1+a \epsilon_\textrm{rand}$ in the legends indicates that the initial condition for the velocity is (1,0,0) plus white noise with an amplitude of $a$. The white noise is three-dimensional as seen in Eq. \ref{Rand}.}
	\label{fig:Re278cdcl}
\end{figure}

We found that near the Hopf bifurcation point, one of the two periodic wake patterns P3-1 (a transitional state) or P3-2 (a saturated state) may appear and can exist for a long time. The P3-1 wake pattern is a hairpin-like vortex structure shedding from the side surface of the cylinder, as shown in {\color{blue}figure~}\ref{fig:P3-12}($a$).
The P3-1 type wake is symmetric about the $x$-$y$ plane (note that the green ruler is collinear with the cylinder axis, that is $z$), and the lift coefficient $C_{lz}$ in the $z$ direction is non-oscillating and approximately zero. Fast Fourier transform (FFT) applied to the drag and the lift coefficients gives the same and single-periodic vortex shedding frequency $St_{Re=283}=0.125$. Interestingly, we notice that this value is comparable to the vortex frequency of the sphere $St_{Re=270}=0.1292$, $St_{Re=285}=0.1335$ obtained by \cite{TOMBOULIDES2000} through nonlinear DNS. The cross-section of the P3-1 hairpin vortex on the $x$-$y$ plane is asymmetric with respect to $x$ axis, and the wake is offset in the $y$ axis (similar to the P2 type wake).
This asymmetric periodic wake pattern is also similar to the asymmetric wakes of bluff bodies, for example, in the axial flow around a short cylinder studied by \cite{Pierson2019} (who found that the axial flow becomes unsteady at $Re\approx355, \textsc{ar}=1$, but it still maintains a symmetric plane in wake until $Re=420$), in the flow past a sphere by \cite{TOMBOULIDES2000, Johnson1999} (who showed that the regular bifurcation at $Re=211$, and the asymmetric steady flow becomes unsteady at $Re=275$), and in the flow past a short cylinder with hemispherical ends by \cite{sheard2008} (who found that the regular bifurcation at $Re=350\pm2$). The other three-dimensional wake pattern P3-2 is shown in {\color{blue}figure~}\ref{fig:P3-12}($b$).  
It can be seen that the vortex shedding from the ends begins to be dominant in the wake. The results of FFT applied on the aerodynamic coefficients show that the frequency of $C_d$ and $C_{ly}$ is the same, and the frequency of $C_{lz}$ is 1/2 that of $C_d$ and $C_{ly}$. 
Besides, the frequency of vortex shedding of the structure P3-2 is larger than that of the structure P3-1; for example at $Re=290$, we calculated that $St_{P3-2}=0.1395$ whereas $St_{P3-1}=0.1269$. More results of the $St$ data for other values of $Re$ are shown in {\color{blue}figure~}\ref{fig:eigenvalue} to be discussed shortly.

\begin{figure} %in R1, the figname is {Re290ar1CdCl.pdf}
~~~\centering\includegraphics[trim=4.8cm 0.3cm 0.0cm 2.0cm, clip, width=0.97\textwidth]{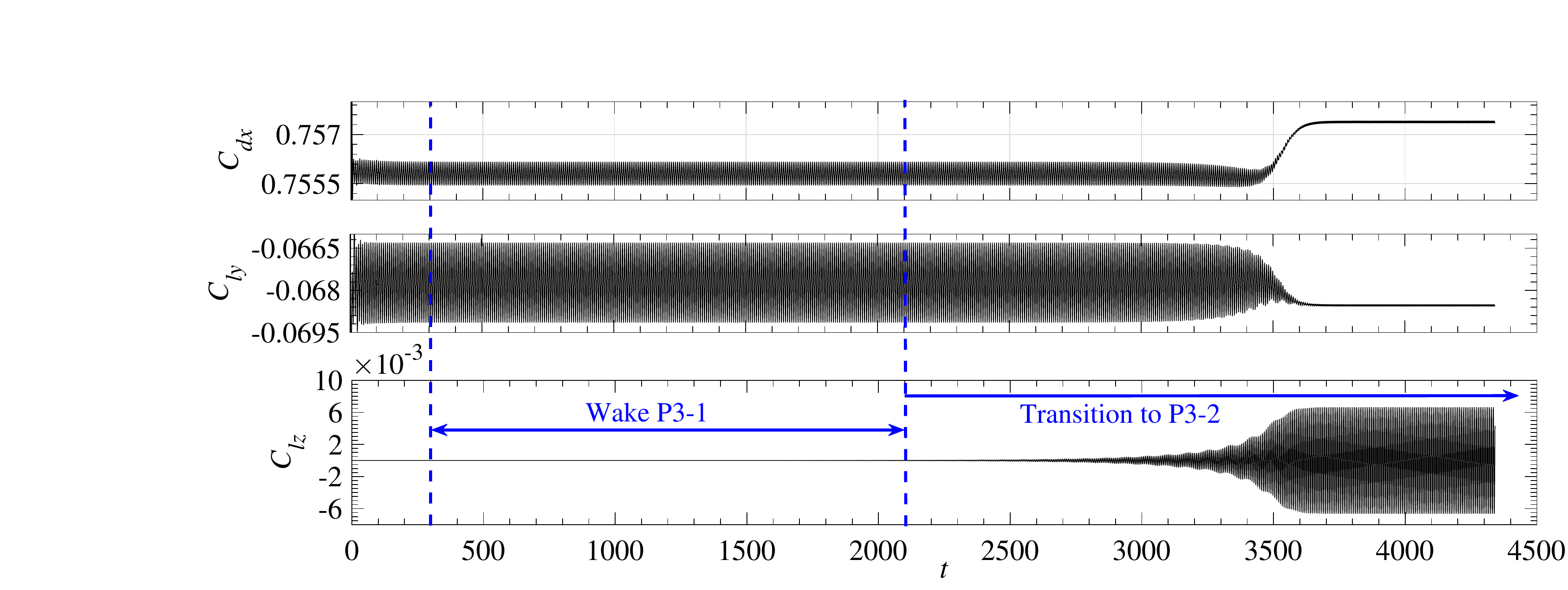}   
	\put(-386,122){($a$)}
	\put(-386,85){($b$)}
	\put(-386,49){($c$)}
	\caption{The time histories of aerodynamic coefficients for cylinder $\textsc{ar}=1$ at $Re=290$. The initial condition is the wake P3-1 of a lower-$Re$ case.}
	\label{fig:Re290CdClTimeHistory}
\end{figure}

\begin{figure}% in R1, the figname is {Re300ar1CdCl.pdf}
	~\centering\includegraphics[trim=0.1cm 0.1cm 0.5cm 0.2cm,clip,width=0.333\textwidth]{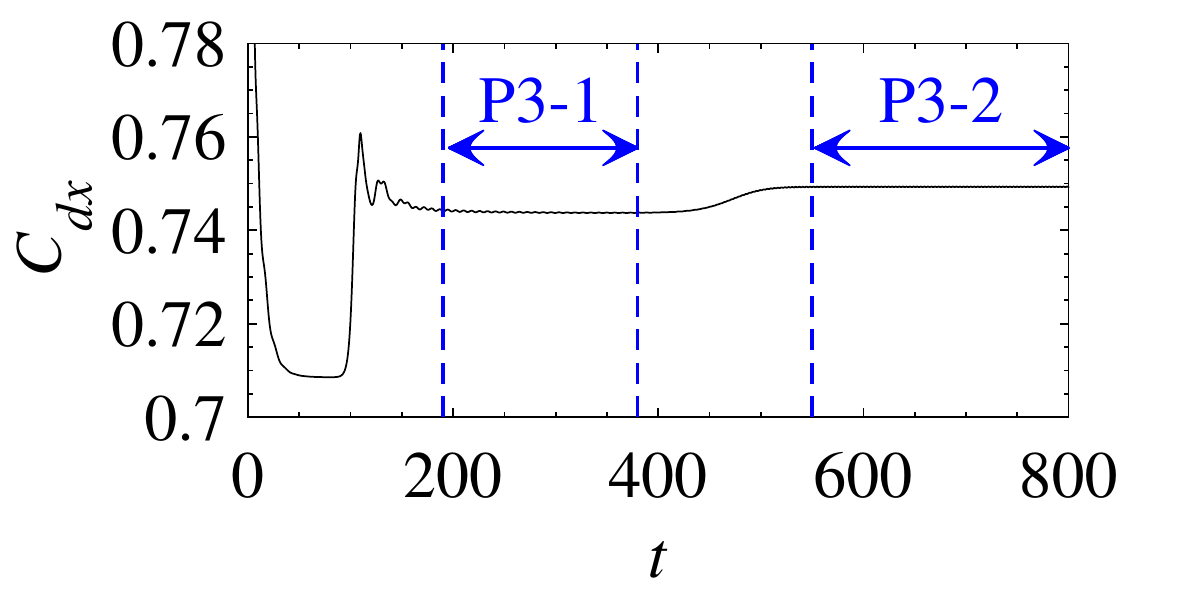}% R2
	\centering\includegraphics[trim=0.1cm 0.1cm 0.5cm 0.2cm,clip,width=0.333\textwidth]{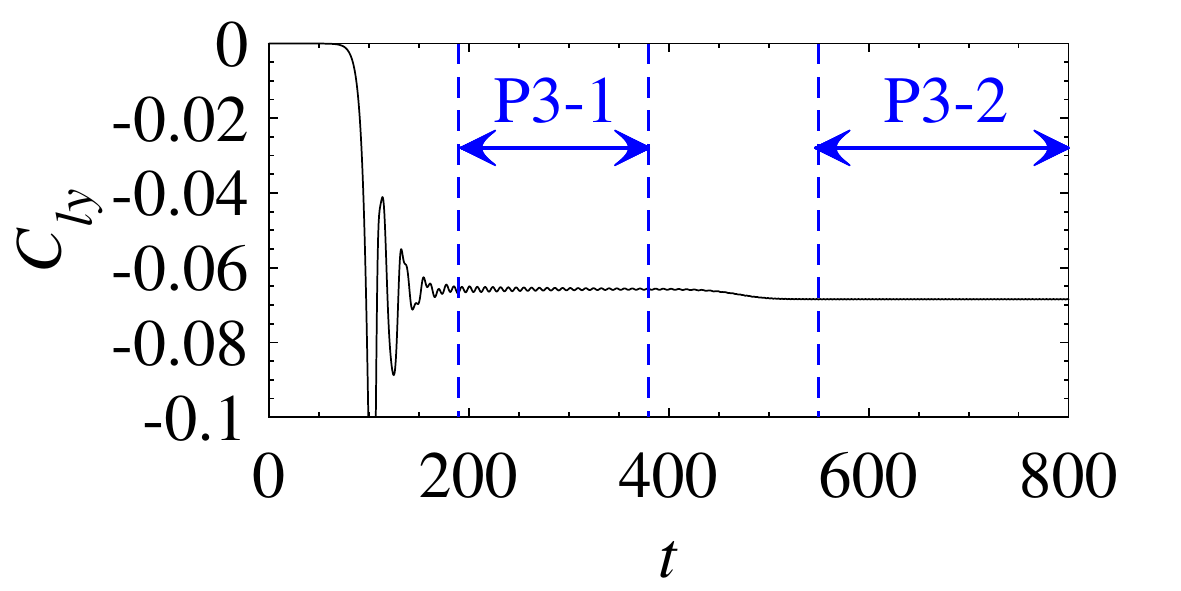}% R2
	\centering\includegraphics[trim=0.1cm 0.1cm 0.5cm 0.2cm,clip,width=0.333\textwidth]{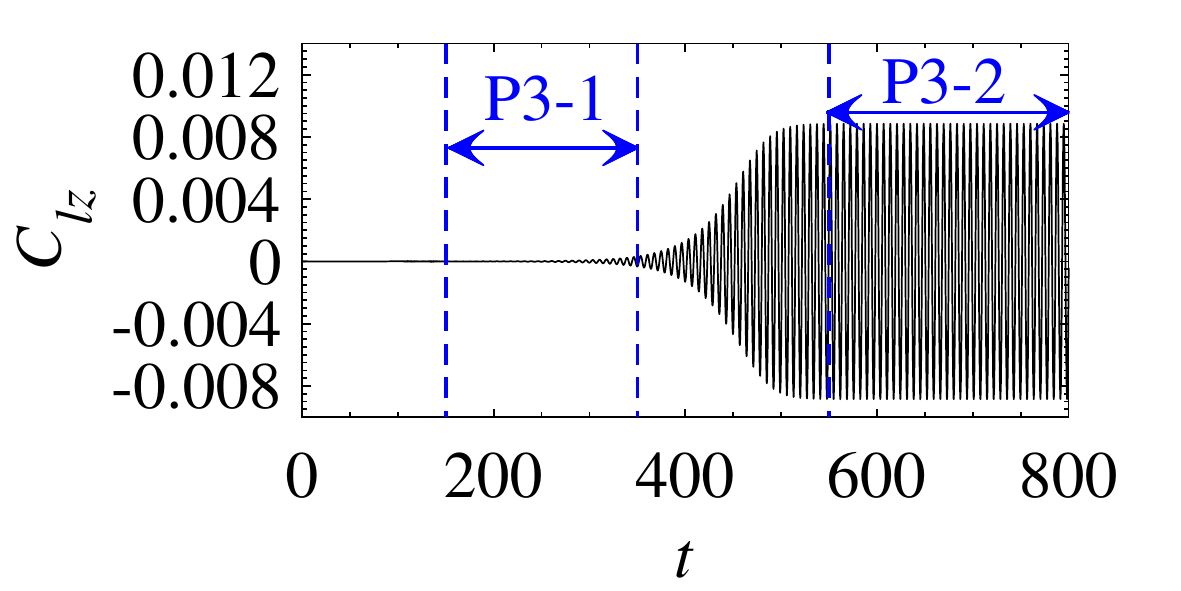}% R2
	\put(-387,58){($a$)}
	\put(-256,58){($b$)}
	\put(-129,58){($c$)}
	\caption{The time histories of aerodynamic coefficients for cylinder $\textsc{ar}=1$ at $Re=300$. The initial condition is $1+0.0\epsilon_\textrm{rand}$.}
	\label{fig:Re300CdClTimeHistory}
\end{figure}

In a subcritical flow (with $Re<282$ for $\textsc{ar}=1$), both P3-1 and P3-2 patterns are stable and will converge to the P2 pattern. This is demonstrated in figure \ref{fig:Re278cdcl} using DNS where we have used different initial conditions to trigger the appearance of the two P3 patterns. We found that if one uses initial perturbation with a smaller amplitude at $Re=278$, the flow will more likely develop to the P3-1 structure; whereas, if a large disturbance is adopted, P3-2 is more likely to appear. To explain, the initial condition is the base flow $(U_x, U_y, U_z)$, e.g. ($1, 0, 0)$, plus white noise with a maximum amplitude of $a=0.0, 0.2, 0.5$, denoted as $1+a\epsilon_\textrm{rand}$. 
Let us first claim the results that $1+a\epsilon_\textrm{rand}$ leads to the P3-1 structure, and $3+0.2\epsilon_\textrm{rand}$ gives rise to the P3-2 structure. The phase trajectory diagram of the lift-drag coefficient is shown in {\color{blue}figure~}\ref{fig:Re278cdcl}. For all the initial conditions considered, the $C_d$-$C_{ly}$ curves of P3-1 and P3-2 patterns, as converging spirals, shrink to the same fixed point, that is, the attractor of the subcritical cases is a fixed point, which is the steady flow pattern P2 as shown in {\color{blue}figure~}\ref{fig:P2}. On the other hand, the oscillation range of $C_{lz}$ corresponding to the initial conditions $1+a\epsilon_\textrm{rand}$ is much smaller than $C_{ly}$ and is close to zero, as shown in {\color{blue}figure~}\ref{fig:Re278cdcl}($b$). Based on this result, we think that the initial conditions $1+a\epsilon_\textrm{rand}$ leads to the P3-1 pattern. The phase trajectory of $C_d$-$C_{lz}$ corresponding to the initial condition $3+0.2\epsilon_\textrm{rand}$ is obviously different from that of $1+a\epsilon_\textrm{rand}$ (the P3-1 pattern) and corresponds to the P3-2 structure. 

In a supercritical flow (with $Re>282$ for $\textsc{ar}=1$), the P3-1 pattern is transient and will eventually converge to the saturated P3-2 pattern. However, one may be easily fooled to believe that the P3-1 pattern can be a stable mode because when $Re$ is close to the critical condition, the lingering time of the flow around the neighbourhood of P3-1 can be extremely long (before it saturates to the nonlinearly stable P3-2 state). 
This is shown in figures \ref{fig:Re290CdClTimeHistory} and \ref{fig:Re300CdClTimeHistory} for $Re=290$ and $300$, respectively. As one can see from figure \ref{fig:Re300CdClTimeHistory} for $Re=300$, the two wake patterns and their transition are clear in a relatively short time. When $Re=290$, however, the DNS method becomes quite difficult to differentiate the two because one needs to simulate for a very long time to see the transition. In the above two figures, because a high-$Re$ is more sensitive to numerical error, it makes the $Re=300$ flow transition to the saturated state more quickly. Another way to expedite the convergence to the saturated state is to add perturbation in the initial condition, resulting in a high possibility of throwing the flow out of the neighbourhood of P3-1 and converging to the final saturated state more quickly (results not shown).

\begin{figure}[h] % in R1 {dlogAdtA2_Re300.pdf}
  \centering\includegraphics[trim=0.4cm 0.0cm 0.4cm 0.4cm,clip,width=0.475\textwidth]{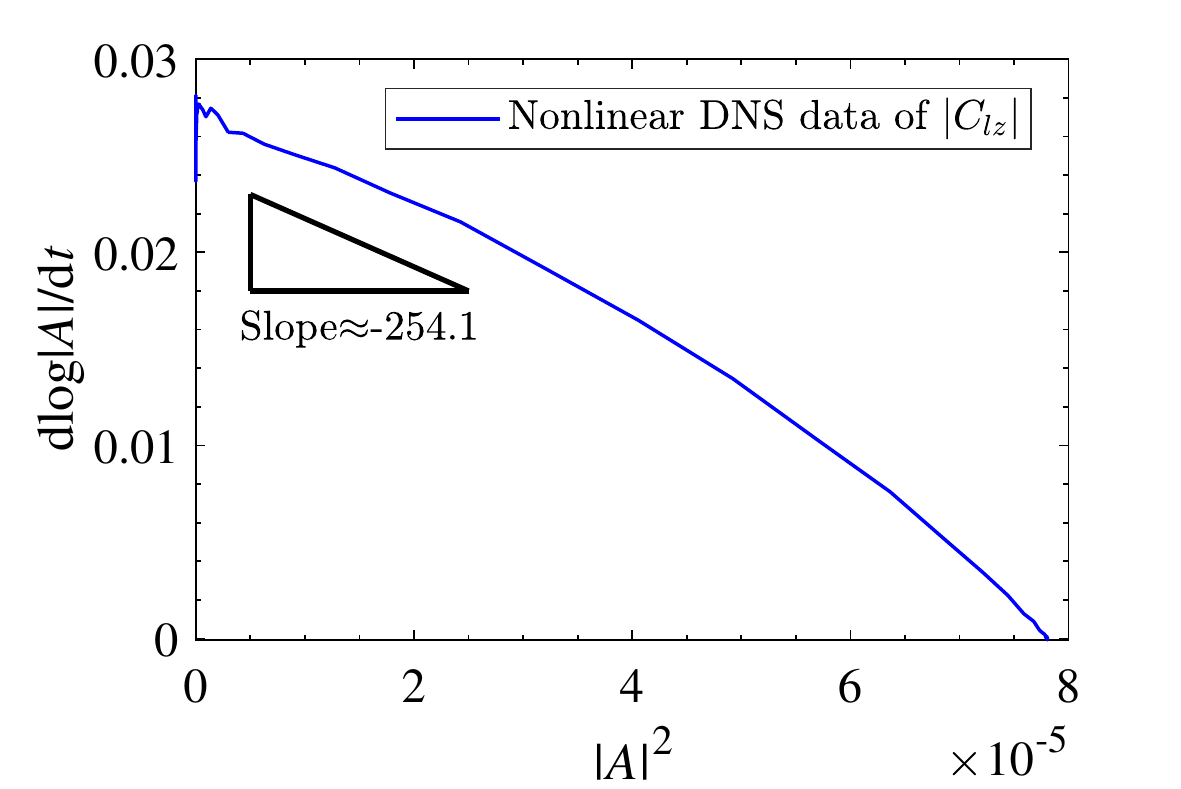} 
	\put(-190,118){$(a)$}   
  ~\centering\includegraphics[trim=0.0cm 0.0cm 0.9cm 0.4cm,clip,width=0.51\textwidth]{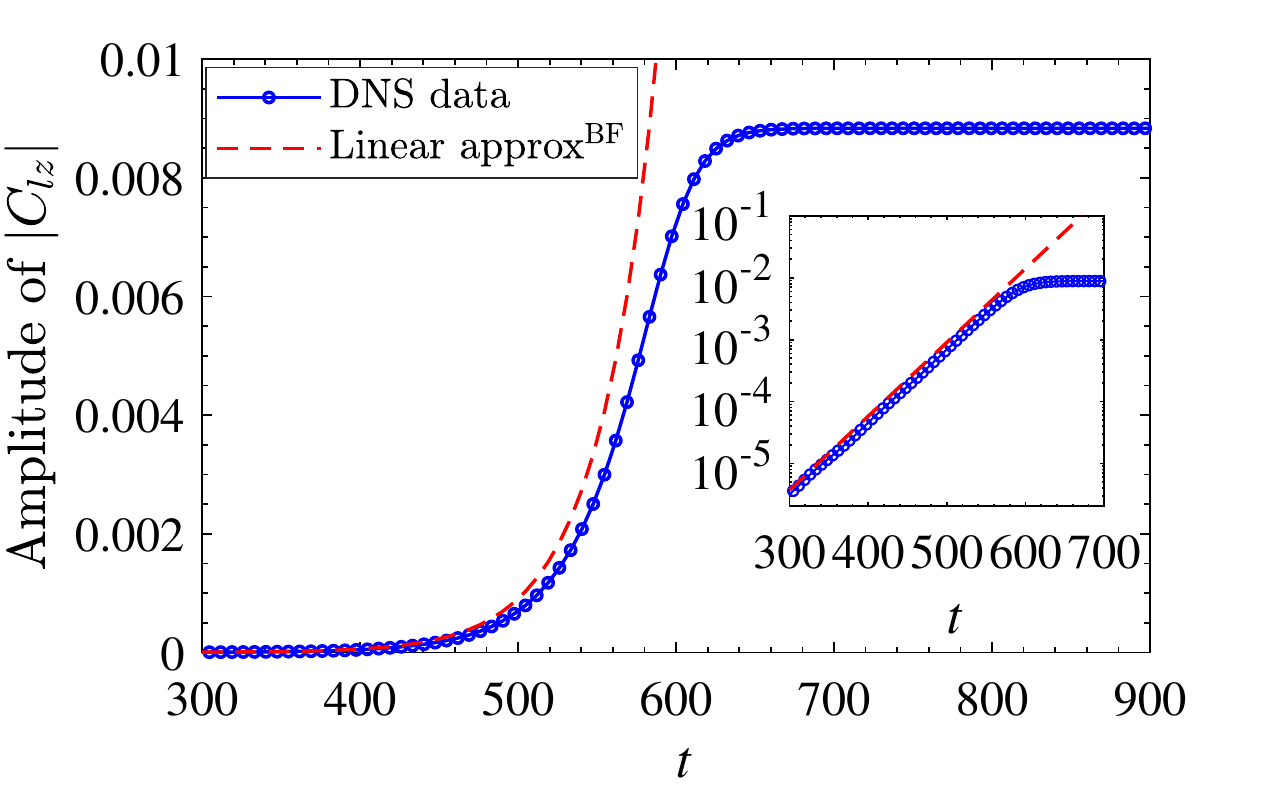}
	\put(-197,118){$(b)$}
  \caption{Time derivative of log$|A|$ versus $|A|^2$ with the vertical intercept giving the linear growth rate $c_1=0.0272$ and the gradient near this intercept point giving the Landau coefficient $c_3=-254.1$, which indicates supercritical bifurcation behaviour. Nonlinear DNS data of case $\textsc{ar}=1$, $Re=300$.} 
  \label{fig:dlnClydtRe300}
\end{figure}

Some discussions are in order regarding the possible bifurcation type around $Re_{c2}$. In the subcritical flows, our numerical simulations with different initial conditions cannot converge to a stable flow state other than P2 (which is a linearly stable state). In the supercritical flows, even though the lingering time around P3-1 can be very long, the final saturation of the flow is the P3-2 pattern; different initial conditions will only change the time spent by the flow around P3-1 before converging to the P3-2 wake. With these results, we infer that the nature of the bifurcation around $Re_{c2}$ should be supercritical. The Landau model has also been used to study the bifurcation type (following a similar analysis presented in figure \ref{fig:dlnClydt}). Again, $\it c_1\rm=0.0272$ obtained in the Landau model is close to the global stability analysis results based on the SFD base flow (whose linear growth rate is $0.02746$). The negative value of $\it c_3$ indicates that the flow bifurcation may be supercritical, which can also been seen in the stabilising effect of the (lowest-order) nonlinearity in  figure \ref{fig:dlnClydtRe300}($b$). More analyses of the P3-1 and P3-2 patterns and the flow bifurcation will be conducted in section \ref{subsec:globalmodes} on the global modes.

\subsubsection{Approach to chaos} \label{subsec:P4}
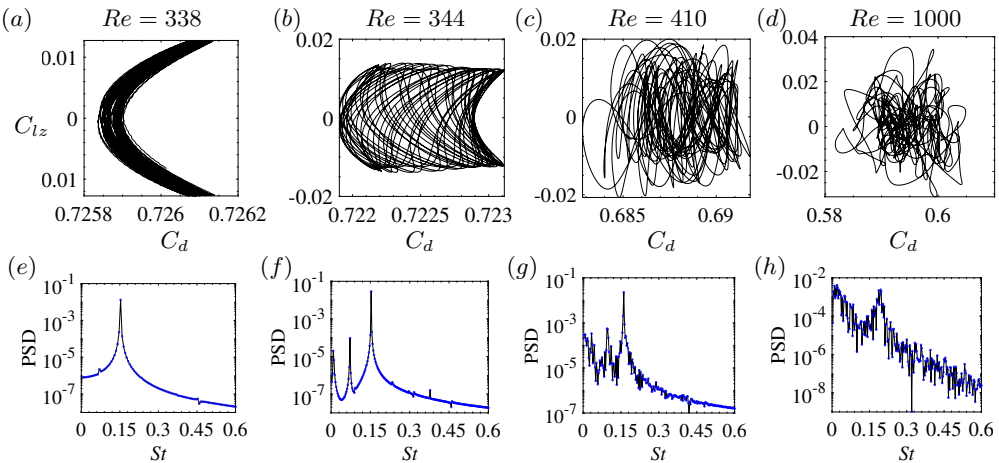
\begin{figure}[h]
\centering
\hfil
\includegraphics[trim=0.3cm -1.2cm 0.2cm 0.0cm,clip,width=0.22\textwidth]{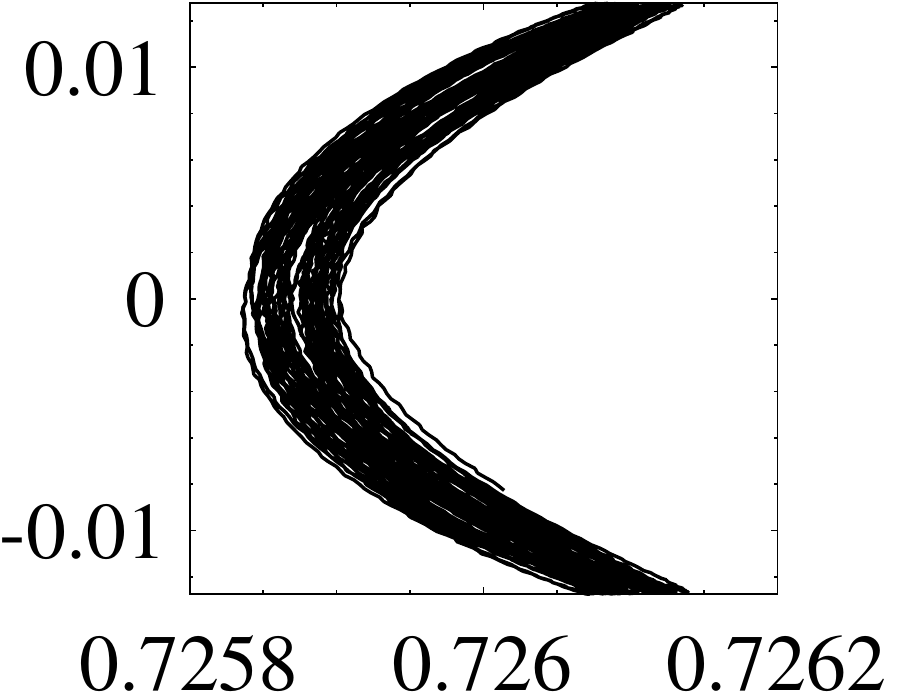}\label{fig:Re338cd_clz}\put(-95,45){\small{$C_{lz}$}}\put(-40,2){\small{$C_{d}$}}
\put(-62,85){\small{$Re=338$}}    \put(-100,85){\small{$(a)$}} % R1: Re338_cd_clznoXYlabel.eps
\hfil
\includegraphics[trim=0.0cm -1.2cm 0.2cm 0.0cm,clip,width=0.218\textwidth]{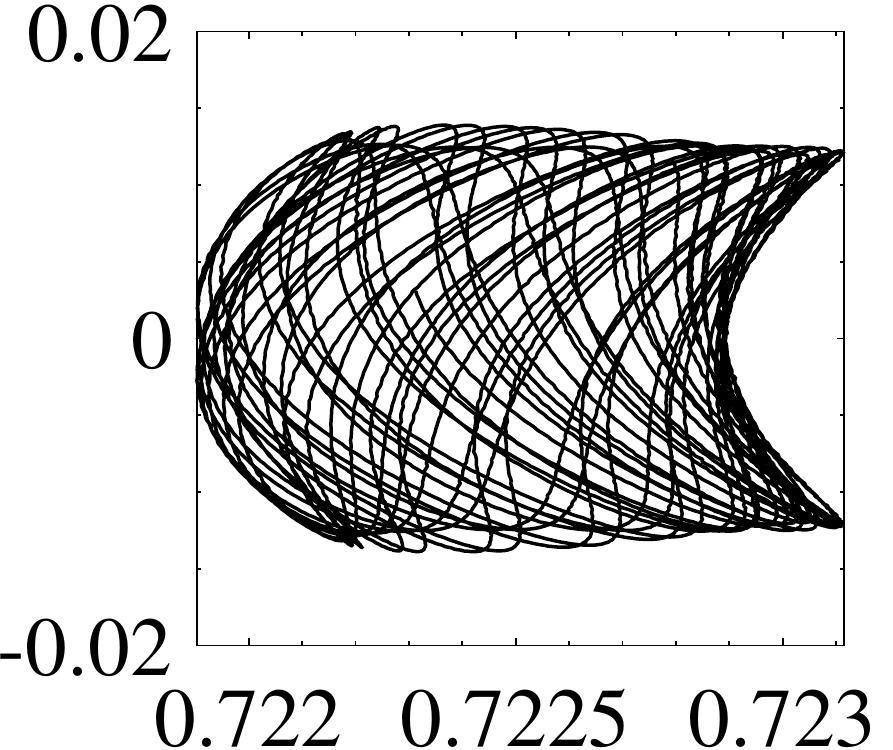}\label{fig:Re344cd_clz} \put(-40,2){\small{$C_{d}$}}
 \put(-55,85){\small{$Re=344$}}   \put(-90,85){\small{$(b)$}} % R1: Re344_cd_clznoXYlabel.eps
\hfil
\includegraphics[trim=0.0cm -1.2cm -0.1cm 0.0cm,clip,width=0.215\textwidth]{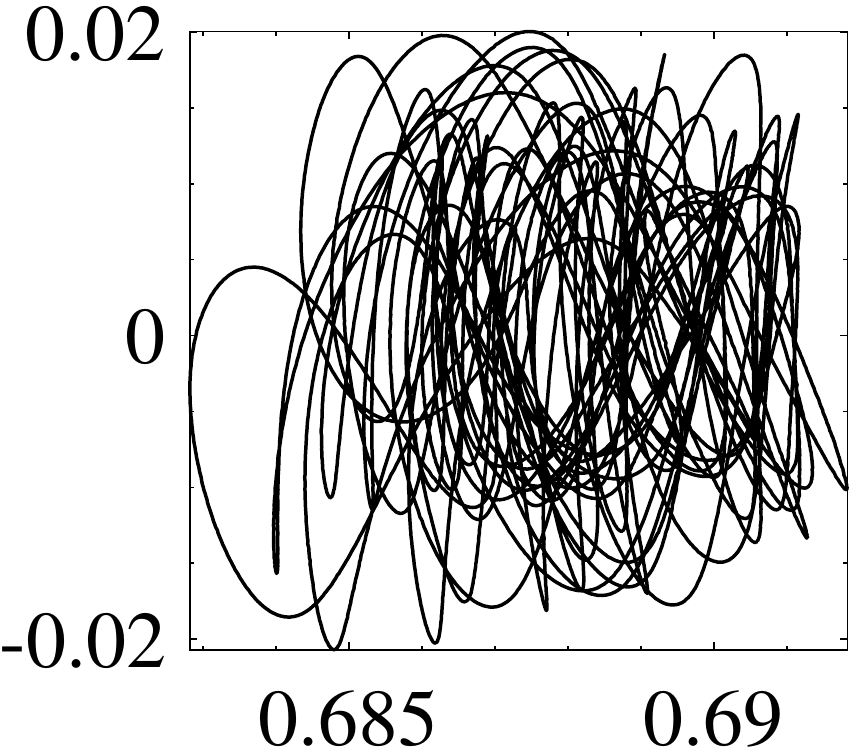}\label{fig:Re410cd_clz}\put(-40,2){\small{$C_{d}$}}
\put(-55,85){\small{$Re=410$}}   \put(-90,85){\small{$(c)$}}  % R1: Re410_cd_clznoXYlabel.eps
\hfil
\includegraphics[trim=0.0cm -1.2cm 0.0cm 0.0cm,clip,width=0.215\textwidth]{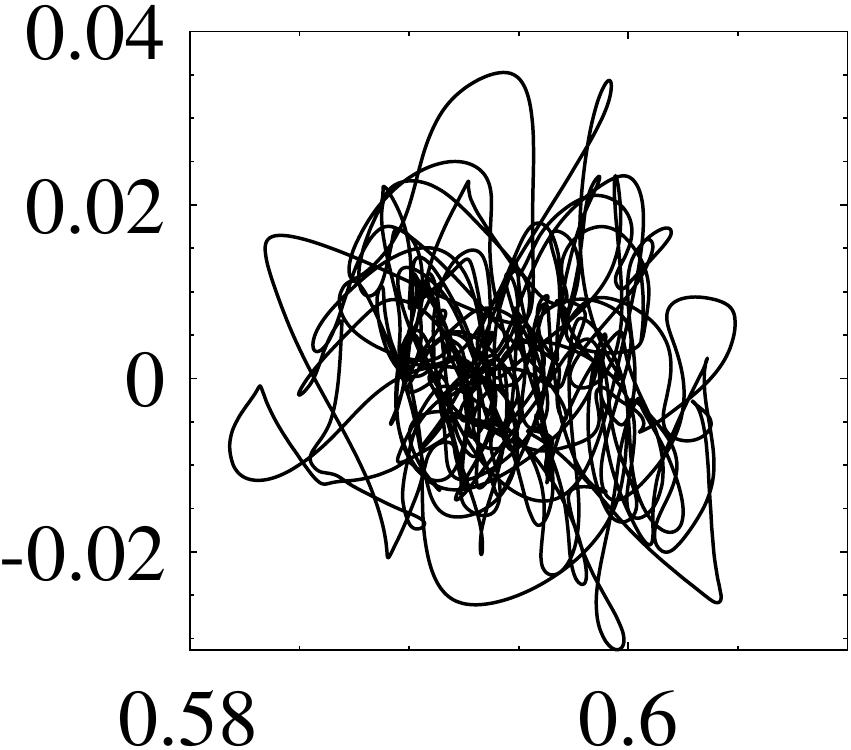}\label{fig:Re1000cd_clz}  \put(-40,2){\small{$C_{d}$}}
\put(-55,85){\small{$Re=1000$}}  \put(-90,85){\small{$(d)$}}  % R1: Re1000_cd_clznoXYlabel.eps
\hfil
\includegraphics[trim=0.1cm 0.0cm 0.15cm 0.0cm,clip,width=0.219\textwidth]{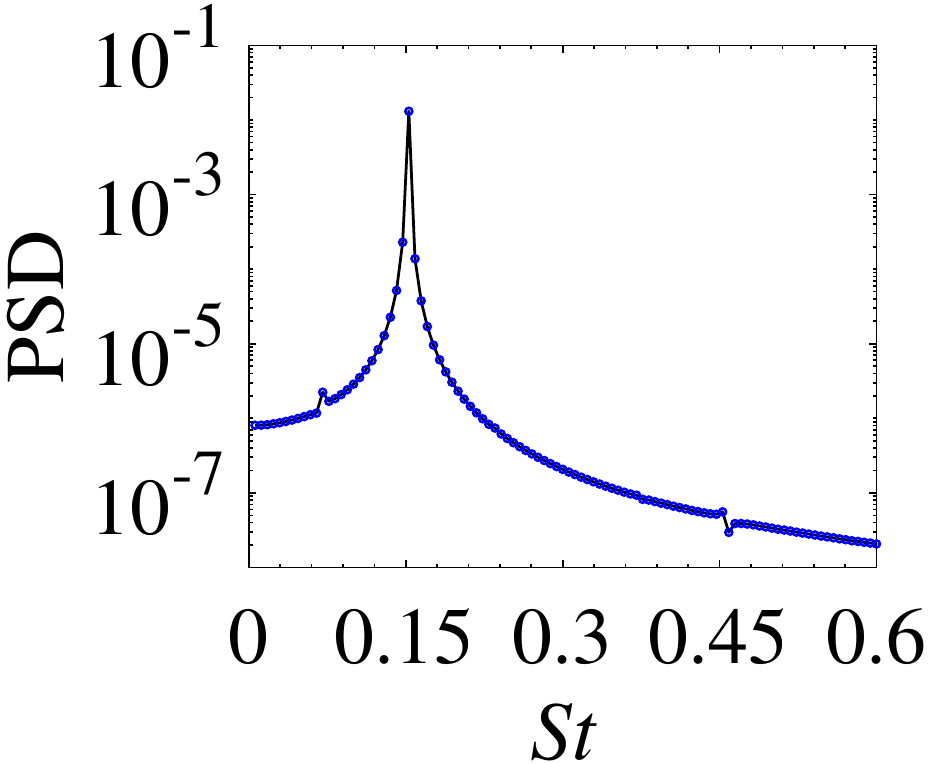}\label{fig:Re338PSD}\put(-90,70){\small{$(e)$}}    % R1: Re338PSD_clz.eps
\hfil
 \includegraphics[trim=0.05cm 0.0cm 0.15cm 0.0cm,clip,width=0.223\textwidth]{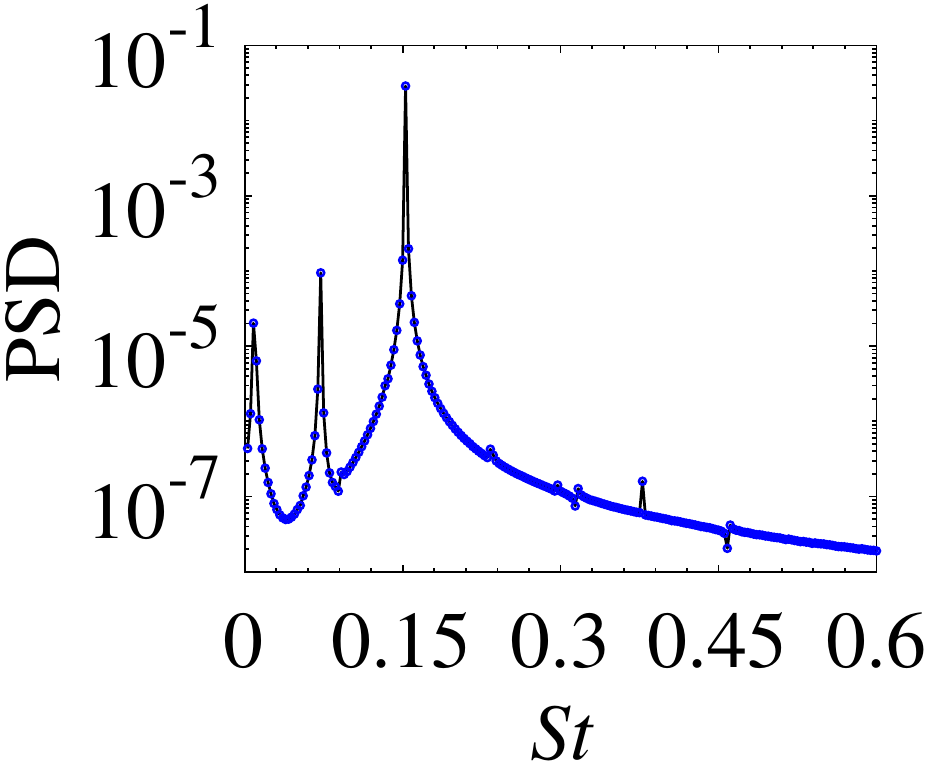}\label{fig:Re344PSD}\put(-90,70){\small{$(f)$}}   % R1: Re344PSD_clz.eps
\hfil
 \includegraphics[trim=0.05cm 0.0cm 0.15cm 0.0cm,clip,width=0.217\textwidth]{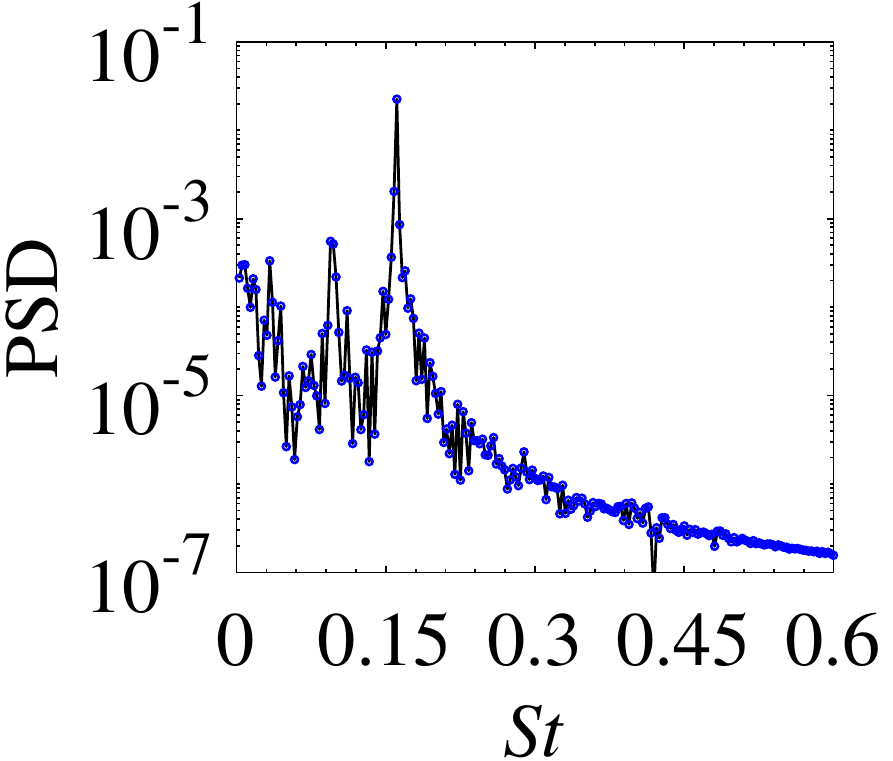}\label{fig:Re410PSD}\put(-90,70){\small{$(g)$}}   % R1: Re410PSD_clz.eps
\hfil
 \includegraphics[trim=0.05cm 0.0cm 0.15cm 0.0cm,clip,width=0.217\textwidth]{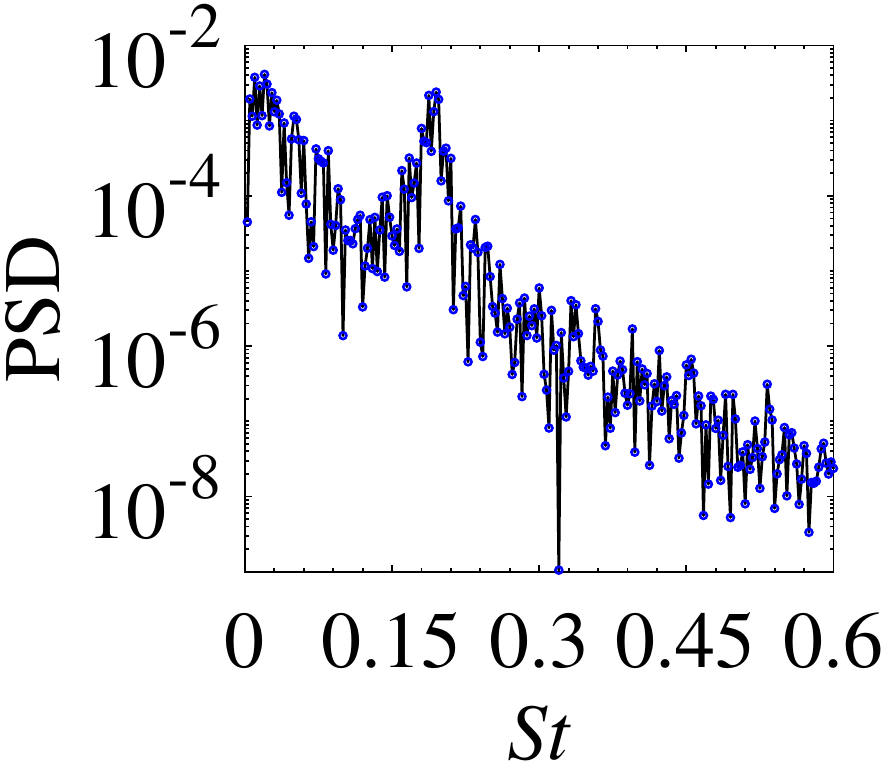}\label{fig:Re1000PSD}\put(-90,70){\small{$(h)$}}  % R1:Re1000PSD_clz.eps
	\caption{The diagrams of $C_{d}$-$C_{lz}$ in the first row and the power spectral density for the time history of $C_{lz}$ in the second row with $\textsc{ar}=1$, at $Re=338$ ($a$ \& $e$), $Re=344$ ($b$ \& $f$), $Re=410$ ($c$ \& $g$) and $Re=1000$ ($d$ \& $h$).}
	\label{fig:Re338_1000cdcly}
\end{figure}

When $Re$ is further increased, the flow becomes more chaotic. {\color{blue}Figure~}\ref{fig:Re338_1000cdcly} shows $C_d$-$C_{lz}$ phase diagrams and power spectral density (PSD) of the $C_{lz}$ data for four values of $Re$ in $338 \leq Re \leq 1000$. When $Re$ is larger than $332$, a new lower frequency $St=0.072$ starts to emerge in the flow field. This can be seen in panels ($a,e$) for $Re=338$ where a small peak begins to form around $St=0.072$ in the PSD figure. This frequency becomes incommensurable when $Re$ further increases and the wake will gradually become chaotic, which is called pattern P4-0 (with $\overline{C}_{ly}=\overline{C}_{lz}=0$) and P4-1 (with $\overline{C}_{ly}\neq 0$, occurs at $\textsc{ar}<1.75$) in this article. In \cite{TOMBOULIDES2000}'s DNS study on the sphere, a similar lower frequency value was detected as $St_{Re=500}=0.045$ at a relatively large $Re$, which was the second highest peak in the PSD of their DNS data. The frequency corresponding to the dominant peak was still $St_{Re=500}=0.167$, which was dominant at all positions of wake.
In the work of the axial flow  past a short cylinder, \cite{Pierson2019} also found a similar low frequency ($St_{Re=395}=0.03$) when $Re \gtrsim 395$ in their DNS study of the axial flow past a short cylinder. 

Back to {\color{blue}figure~}\ref{fig:Re338_1000cdcly}, at $Re=344$, a third new frequency signal $St=0.0083$ starts to appear. At this $Re$, the attractor of $C_d$-$C_{ly}$ is no longer a limit cycle, but becomes a limit torus. It means that certain two frequencies in the flow field are incommensurable, and the phase trajectory is no longer closed. The PSD of these two lower frequencies $St=0.0706$ and $St=0.0083$ is not negligible compared to the dominant frequency $St=0.1528$. 
At $Re=410$, the phase trajectory of $C_d$-$C_{ly}$ becomes highly disordered, and the PSD curve shows the characteristics of a chaotic signal. It can be seen from {\color{blue}figure~}\ref{fig:Re338_1000cdcly}($g$) that the dominant frequency is still around $St=0.1611$, and its PSD curve shows that there are many small peaks around $St=0.033$, $St=0.094$ and $St=0.1611$.
Finally, when $Re=1000$, the phase diagram is highly chaotic and a broad spectrum of frequencies appears signalling the disordered state of the flow.
In the literature, \cite{sakamoto1995}'s experimental work on the flow past a sphere showed that, when $Re>420$, the shedding direction of the hairpin vortex appears intermittent, the oscillation amplitude and waveform caused by the vortex shedding start to become irregular, and the flow field starts to transition from a single-frequency flow to a chaotic state. \cite{Pierson2019}'s DNS study (in its supplementary materials) on the axial flow past a finite cylinder ($\textsc{ar}=1$) shows that the flow exhibits chaotic properties for the $Re$ in $420 \leq Re \leq 460$, and the secondary lower frequency ($St_{Re=460}=0.03$) gradually dominates the entire flow field. These results of the transition from a single frequency to a chaotic state is similar to that of the radial flow past a finite cylinder ($\textsc{ar}=1$) as we study here.

\subsection{Global modes} \label{subsec:globalmodes}

The discussions in Sec.~\ref{subsec:nonlinearDNS} pertain to the different flow patterns (P1-P4) one can obtain from the nonlinear DNS when increasing the value of $Re$ and the critical values of $Re$ delimiting the different flow regimes. In this section, we will look into the eigenmodes at certain values of $Re$ and $\textsc{ar}$ for some base states. We report this because we found that when the parameters change, the type of the mode that becomes the most unstable changes. If not stated otherwise, the global linear stability analysis to follow is performed with respect to the mean flow.

\begin{figure}[h]
	\centering\includegraphics[trim=0.0cm 4cm 1.0cm 4.1cm, clip, width=0.99\textwidth]{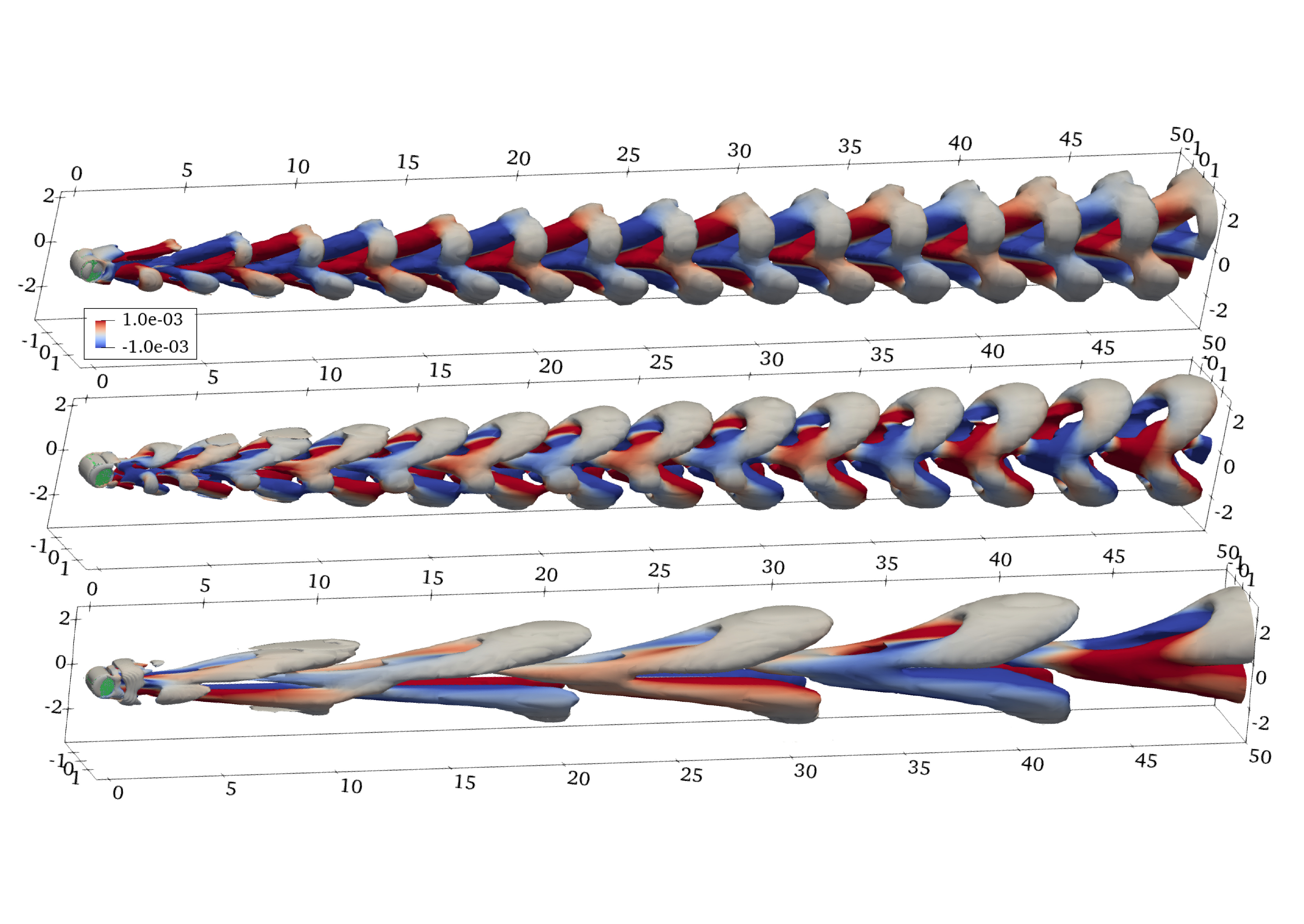}  % in R1, the figname is {Re290eigenmodes.png}
	\put(-385,180){($a$)}
	\put(-385,115){($b$)}
	\put(-385,50){($c$)}
	\caption{The first three eigenmodes at $Re=290$ \& $\textsc{ar}=1$. ($a$) eigenmode A with eigenvalue $\lambda=2.959\times 10^{-3}+0.1403\boldsymbol{i}$ (real part is the growth rate and imaginary part the eigenfrequency divided by $2\pi$), ($b$) eigenmode B with eigenvalue $-3.885\times10^{-2}+0.1258\boldsymbol{i}$, ($c$) eigenmode C with eigenvalue $-7.491\times10^{-2}+0.04345\boldsymbol{i}$. The $Q$-criterion isosurfaces $Q=0$ are colored by the $x$-component of the vorticity ranging from $-10^{-3}$ to $10^{-3}$.}
	\label{fig:eigenmodeAR1}
\end{figure}

Based on the symmetrical and asymmetrical stationary base states, the global eigenmodes obtained by IRAM codes can be divided into two categories. The first type is the eigenmode with only one symmetric plane ($x$-$y$) as shown in {\color{blue}figure~}\ref{fig:eigenmodeAR1}, and it can be found at the cases $Re>Re_{c1}, \textsc{ar}<1.75$.  
The second type is the eigenmode with two mutually perpendicular symmetric planes ($x$-$y$ and $x$-$z$) as shown in {\color{blue}figure~}\ref{fig:eigenmodeAR175} in the parameter ranges $Re<Re_{c1}$ and $\textsc{ar}<1.75$ or $\textsc{ar}\geq 1.75$. 
It can be understood that the global linear eigenmode retains the same structural symmetry as the basic state (either base flow or mean flow), which is consistent with the results of the global LSA of the classical three-dimensional non-rotating and rotating sphere \citep{citro2016}. 

\begin{figure}[h]
	~\centering\includegraphics[trim=0.0cm 0cm 0.0cm 0.0cm,clip,width=0.475\textwidth]{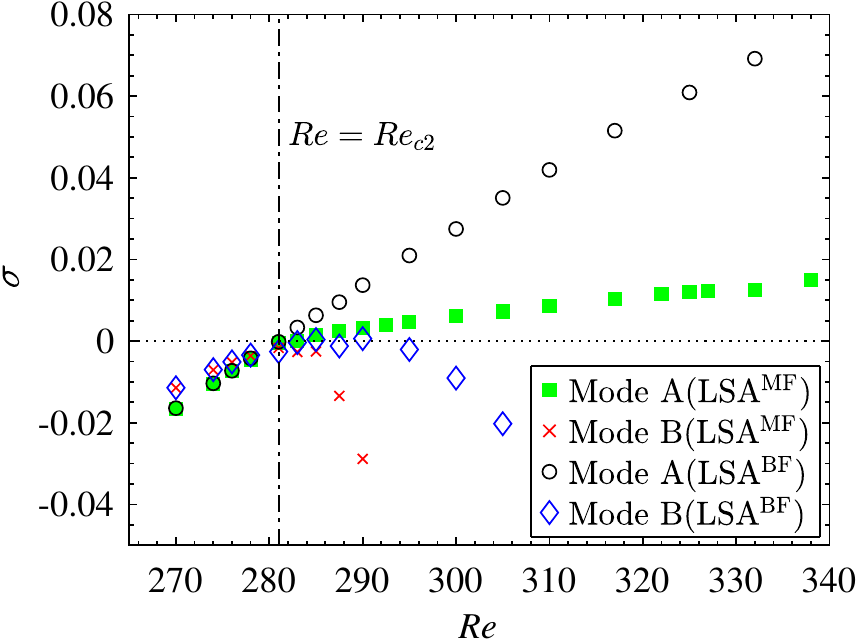}\put(-188,130){($a$)} % in R1, figname is {Re_gr_mean_base.pdf}
	~\centering\includegraphics[trim=0.0cm 0cm 0.0cm 0.0cm,clip,width=0.515\textwidth]{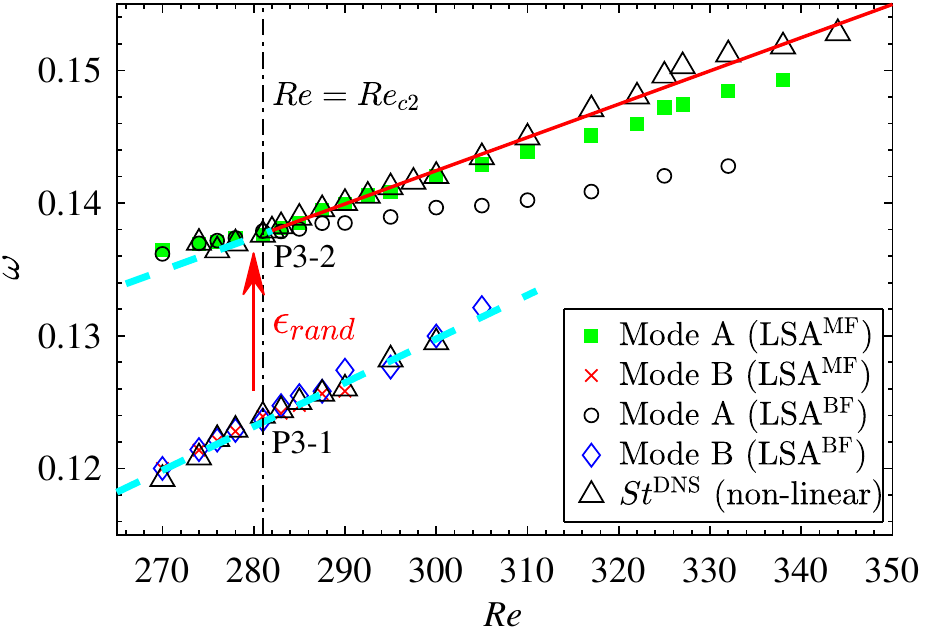}\put(-201,130){($b$)}  % in R1, figname is {Re_st_mean_base.pdf}
	\caption{The eigenvalues and Strouhal numbers as function of $Re$ near the Hopf bifurcation at $\textsc{ar}=1$. The red arrow  '$\uparrow$' indicates that the flow will transition to the wake pattern P3-2 at $Re>Re_{c2}$ under the effect of disturbance. In panel ($b$), the red solid line represents stable solutions and the dashed cyan lines represent transient flows. The specific data is shown in table \ref{tab:BM}.}
	\label{fig:eigenvalue}
\end{figure}

\begin{figure}
\centering\includegraphics[trim=2.5cm 4.0cm 1.3cm 2.7cm, clip, width=1\textwidth]{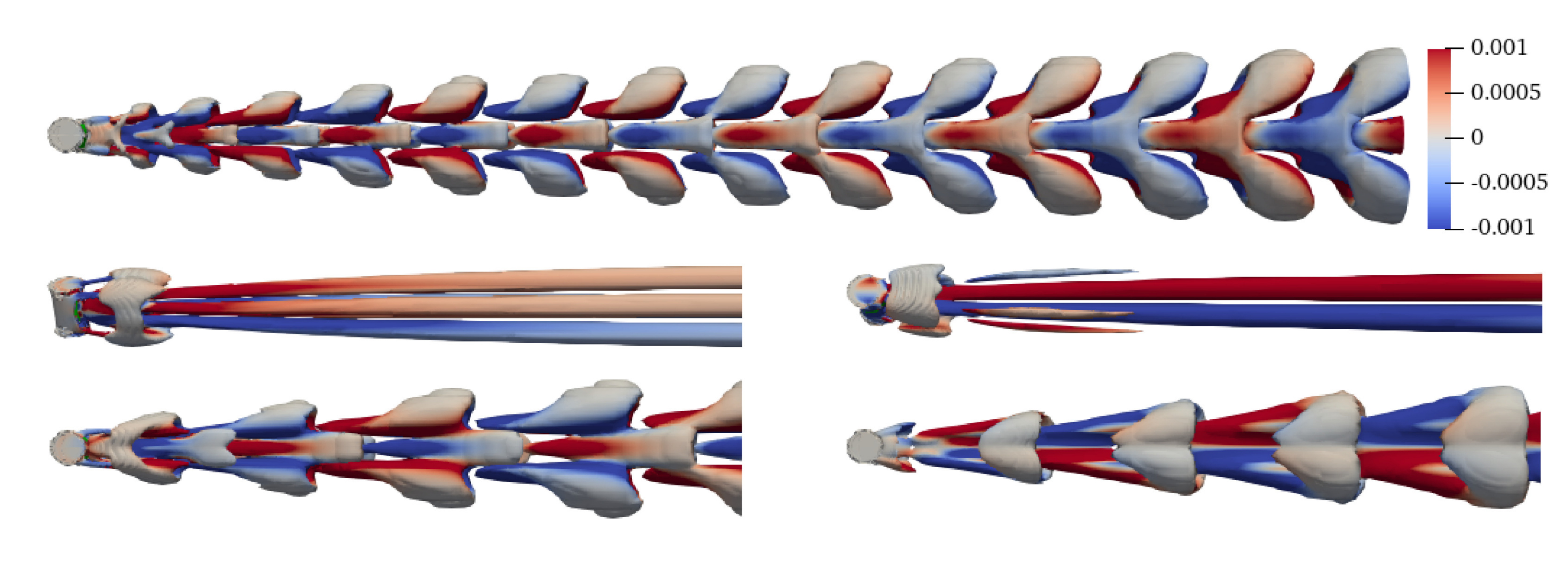} % in R1: the figname is {AR175Re225modes.png}
\put(-385,107){($a$)}
\put(-185,67){($c$)}
\put(-385,67){($b$)}
\put(-385,27){($d$)}
\put(-185,27){($e$)}  
	\caption{The first five eigenmodes I-V based on the mean flow with two symmetric planes at $Re=225$ \& $\textsc{ar}=1.75$. The eigenvalues of ($a$)-($e$) are
$1.432\times10^{-6}+ 0.1201\boldsymbol{i}$, 
$-2.662\times10^{-3}+0\boldsymbol{i}$, 
$-9.895\times10^{-3}+0\boldsymbol{i}$, 
$-2.794\times10^{-2}+0.06716\boldsymbol{i}$ and
$-3.526\times10^{-2}+0.08008\boldsymbol{i}$, respectively. The $Q$-criterion isosurfaces ($Q=0$) are colored by the $x$-component of the vorticity ranging from $-10^{-3}$ to $10^{-3}$.}
	\label{fig:eigenmodeAR175}
\end{figure}

We first discuss the case {$\textsc{ar}=1, Re=290$}, which has the first type of eigenmodes. The typical flow structures of the first three eigenmodes in this case are shown in {\color{blue}figure~}\ref{fig:eigenmodeAR1}, which are ordered by the real part (the growth rate) of the eigenvalues from large to small, and are marked as eigenmodes A, B and C, respectively. 
   The imaginary part of the eigenvalues of the first three eigenmodes are all not equal to zero. Modes A and B appear to have smaller flow structures than mode C, whose frequency is smaller. Interestingly, we found that the eigenmodes A and B can be favourably compared to the DNS results and are highly relevant to, respectively, the P3-2 and P3-1 structures that we have identified and discussed earlier. Following \cite{sansica2018three}, we have also checked that the mode shape and frequency in the dynamic mode decomposition (DMD following \citealt{SCHMID2010}) analysis of the nonlinear flow field snapshots of the P3-1 pattern are almost the same as those of the linear eigenmode B (DMD results are not shown). The same correspondence between the DMD mode of the P3-2 pattern and the global mode A can also be established. 
Based on these strong links, to further understand the Hopf bifurcation, the relationship between the eigenvalues of these two eigenmodes and the Reynolds number in the stability analyses based on the mean flow and the SFD base flow is shown in {\color{blue}figure~}\ref{fig:eigenvalue} (with the specific data in table \ref{tab:BM}). In the two panels, the vertical dash-dot line is $Re=Re_{c2}$, and the dotted horizontal line is the zero growth rate $\sigma=0$. In panel ($b$), the dashed cyan and red solid lines are approximate fittings of the non-dimensional vortex frequency $St$ measured in the nonlinear DNS (the triangles). The red solid line also indicates stable solutions and the dashed cyan ones transient flow states.
   
From panel $a$, one can see that at a subcritical $Re$, the mean flow and the SFD base flow generate the same results for modes A, B.  The decay rates of mode B are both smaller than those of mode A (i.e., mode A is more stable), which is consistent with the result that it is sometimes possible to observe wake pattern P3-1 in nonlinear DNS in this range of $Re$ (note that mode B corresponds to P3-1 pattern). In the supercritical range, the most unstable mode is now eigenmode A (circles based on the SFD base flow and green filled squares based on the mean flow), whose $\sigma$ increases with $Re$. The eigenmode B (red crosses) based on the mean flow is stable. Thus, the amplitude of eigenmode A will increase exponentially with time until the nonlinearity becomes important. This can explain that in the evolution of nonlinear DNS, the frequency information of wake pattern P3-2 (corresponding to  eigenmode A) is observed in a supercritical condition. 

\begin{table}
	\begin{tabular}{p{0.6cm} p{0.8cm} p{0.8cm} p{1.2cm} p{1.2cm} p{3cm} p{3.0cm} p{1.1cm}}
		\hline
		\centering	
	 $Re$ &$x_s^M$&$x_s^B$&$\theta_s^{\rm MF}(^\circ)$&$\theta_s^{\rm BF}(^\circ)$&$\ (\sigma,\omega)^{\rm MF}$  &  $\ (\sigma,\omega)^{\rm BF}$  & $St^{\rm DNS}$\\
%	  274 & 2.639 & 2.635 & 102.885    & 102.825    & (-7.08E-03, 0.1213)    &(-7.00E-03, 0.1214)    &  0.1208\\
	  278 & 2.645 & 2.650 & 102.620    & 102.720    & (-3.61E-03, 0.1228)    &(-3.43E-03, 0.1229)    &  0.1229\\
	  283 & 2.651 & 2.655 & 102.599    & 102.545    & ( 8.43E-05, 0.1381)    &( 3.33E-03, 0.1379)    &  0.1383\\
	  285 & 2.620 & 2.667 & 102.490    & 102.295    & ( 1.40E-03, 0.1385)    &( 6.32E-03, 0.1381)    &  0.1389\\
	  290 & 2.580 & 2.681 & 102.321    & 101.960    & ( 3.23E-03, 0.1400)    &( 1.37E-02, 0.1385)    &  0.1400\\
%	  295 & 2.540 & 2.693 & 102.165    & 101.728    & ( 4.72E-03, 0.1409)    &( 2.10E-02, 0.1390)    &  0.1412\\
	  300 & 2.500 & 2.718 & 102.065    & 101.421    & ( 6.07E-03, 0.1420)    &( 2.75E-02, 0.1397)    &  0.1420\\
	  310 & 2.450 & 2.747 & 101.610    & 100.885    & ( 8.62E-03, 0.1438)    &( 4.19E-02, 0.1402)    &  0.1449\\
\hline		
	\end{tabular}
    \caption{Comparisons between the most unstable eigenvalues and Strouhal numbers near the Hopf bifurcation for the cylinder $\textsc{ar}=1$. The superscripts $^{\rm BF}$ and $^{\rm MF}$ represent SFD base flow and mean flow, respectively.}\label{tab:BM}
\end{table}
   
Now we look at panel $b$ for $\textsc{ar}=1$. One can immediately observe that there are two clusters of eigenfrequencies and they in fact correspond to the P3-1 and P3-2 structures found in the nonlinear DNS around the Hopf bifurcation. By comparison, one can notice that the vortex shedding frequency of the P3-1 wake is consistent with the eigenfrequency of eigenmode B, while the vortex shedding frequency of the P3-2 wake is consistent with the eigenfrequency of eigenmode A. As we have investigated above, when some perturbation is added on the initial condition (see the red arrow in panel $b$), the flow will transition to the P3-2 structure. Besides, the wake P3-2 (eigenmode A) will appear at $Re>Re_{c2}$ as the nonlinearly saturated state. All the flow states superposed by the cyan dashed lines are transient. These results are consistent with the DNS investigations in the end of section \ref{subsec:P3} that the flow at $\textsc{ar}=1$ bifurcates supercritically around the Hopf bifurcation. 
Moreover, in the validation Sec. \S \ref{subsec:valDNS}, we mentioned that the difference of vortex shedding frequency between the present result ($St_{\rm P3-2}=0.142$) and the DNS result in \cite{Inoue2008} ($St=0.127$) at $\textsc{ar}=1, Re=300$ is 10.5\%. We think that the frequency obtained by \cite{Inoue2008} may be the frequency of P3-1 (our $St_{\rm P3-1}=0.1295$, which is closer to their value with $Err=1.9\%$) in present work. According to the results of our nonlinear DNS and LSA, P3-1 is a transitional state.
These results pertain to $\textsc{ar}=1$. In the case of other $\textsc{ar}$ values, the flow bifurcation may be different. In a different but related context, \cite{sheard2004spheres} showed that geometric feature size $\textsc{ar}$ can change the properties of bifurcation in the flow past spheres and circular cylinders (rings). Thus, the bifurcation types of the radial flow at the other values of $\textsc{ar}$ deserve to be investigated further in the future.

When $\textsc{ar}\geq1.75$, no matter whether the flow is steady or unsteady, the basic state obtained by the SFD method and the time-average method has two mutually perpendicular symmetric planes. The eigenmodes obtained based on such symmetrical basic states also retain the symmetry of their basic state, and the typical structures are shown in {\color{blue}figure~}\ref{fig:eigenmodeAR175}. The figure shows the first five eigenmodes for the case $\textsc{ar}=1.75, Re=225$, ordered by the real part of the eigenvalues from large to small from panel $a$ to $e$. The eigenfrequencies (the imaginary parts of the eigenvalues) of the eigenmodes II and III (panel $b$ and $c$) are both zero and the corresponding eigenmodes have flow structures which appear like long strips extending streamwise. On the other hand, the other eigenmodes with non-zero eigenfrequencies have distinct wave structures in the streamwise direction. 

\subsection{$Re_c$-$\textsc{ar}$ diagram} \label{subsec:AReffect}
\begin{figure} % in R1, the figname is {Re-ar.pdf}
	\centering
	\includegraphics[trim=0.3cm 0.2cm 0cm 0cm,clip,width=0.98\textwidth]{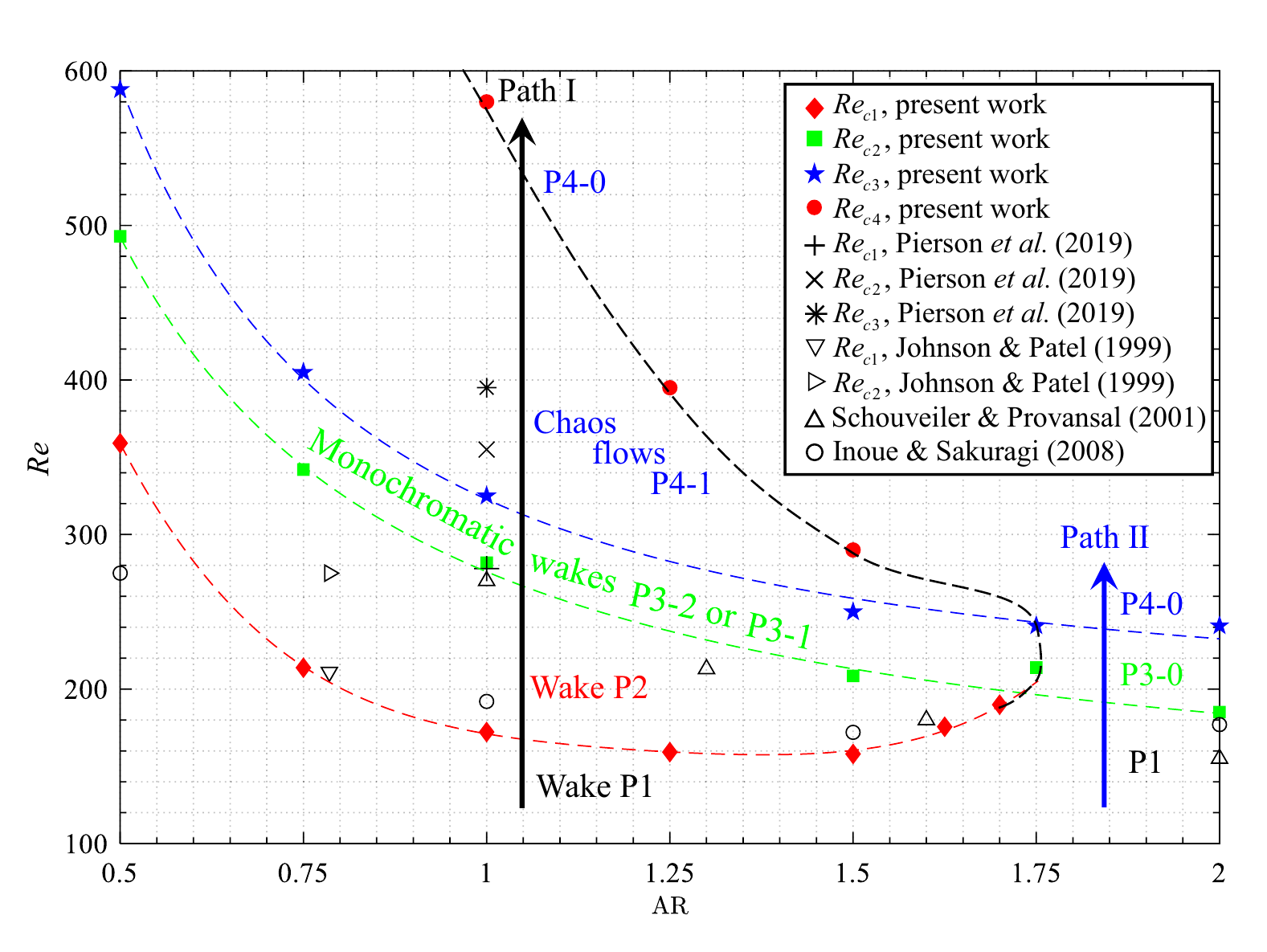} 
	\caption{Different flow regimes in the $Re-\textsc{ar}$ plane. The dashed lines represent the neutral stability. The triangle '$\triangle$' represents the experimental data of the flow past a finite cylinder with hemispherical ends for the transition from a stationary state to a time-dependent flow from \cite{SCHOUVEILER2001}. The circle '$\circ$' represents the DNS of \cite{Inoue2008} (using a finite-difference method) results of the finite cylinder for the transition from steady flow to periodic vortex shedding. Critical Reynolds numbers in the flow past a sphere \citep{Johnson1999} and the axial flow past a cylinder with $\textsc{ar}=1$ \citep{Pierson2019} have also been superposed. The data of sphere \citep{Johnson1999} is placed along the vertical line of $\textsc{ar}=\pi/4\approx0.785$.}
	\label{fig:AR_Rec}
\end{figure}

To summarise this work, we present a $Re_c$-$\textsc{ar}$ diagram in {\color{blue}figure~}\ref{fig:AR_Rec}. The preceding sections \ref{subsec:globalmodes} and \ref{subsec:nonlinearDNS} mainly discussed the various flow patterns and global modes with reference to $\textsc{ar}=1$. A key focus of this work is to study the effect of $\textsc{ar}$ on the flow transition in the flow past a finite-length cylinder (experiencing different flow patterns), which has not been investigated systematically in previous works for this flow. Moreover, combining present works and the previous research on the flow past a finite cylinder with $\textsc{ar}>2$, the transition scenarios of this radial flow past a cylinder with two free ends can be better understood.  

Before embarking on the discussion, we mention that in addition to $Re_{c1}$ (red) and $Re_{c2}$ (green) the definitions of which we have discussed in detail in previous sections, we also define $Re_{c3}$ which marks transition from periodic flow to chaotic flow. More specifically, \cite{sipp2007} and \cite{turton2015mean_flow} have proved theoretically and numerically that if a flow is dominated by a single frequency (i.e. monochromatic wave oscillation), the eigenfrequency of the linearisation operator based on the mean flow is equal to the nonlinear frequency. 
Based on this, $Re_{c3}$ is defined as follows. We increase $Re$ by 5 each simulation from a lower value of $Re$ and keep track of the dominant eigenfrequency as the Reynolds number increases. When the difference between the nonlinear vortex frequency and the dominant eigenfrequency obtained by the linear IRAM code is consecutively greater than 1\%, we define the corresponding $Re$ as the third critical Reynolds number $Re_{c3}$. For example, when $\textsc{ar}=1$, as we increase $Re$, the first time when we have a pair of consecutive $Re$ whose nonlinear and linear frequencies differ by greater than 1\% is $Re=322$ and $327$ and we set $Re_{c3}=322$ for this parameter. The physical meaning of the bifurcation point $Re_{c3}$ is the transition from a single frequency dominated flow to the coexistence of multi-frequency oscillations. As for $Re_{c4}$, it is reminded that this value delimits the two chaos states P4-0 (with $\overline{C}_{ly}=\overline{C}_{lz}=0$) and P4-1 (with $\overline{C}_{ly}\neq 0$, occurs at $\textsc{ar}<1.75$), as we have discussed in the previous section.

In {\color{blue}figure~}\ref{fig:AR_Rec}, one can see that with the increase of $\textsc{ar}$, the critical Reynolds number $Re_{c1}$ first decreases and then increases, while the critical Reynolds numbers $Re_{c2}$ and $Re_{c3}$ decrease monotonically in the range of $\textsc{ar}$ from 0 to 2. The relationships between $Re_{c2}$-$\textsc{ar}$ and $Re_{c3}$-$\textsc{ar}$ approximately conform to power functions, as shown by the green and blue dashed lines in the figure with the corresponding fitting formula being $Re_ {c2}=158.9\textsc{ar}^{-1.241}+117$ and $Re_{c3}=136\textsc{ar}^{-1.572}+186.8$, respectively. Besides, it can be seen from {\color{blue}figure} \ref{fig:Re_cdcly}$(b)$ that as $Re$ increases, $\overline{C}_{ly}$ will turn to zero again, that is, the wake pattern P4-1 of $\overline{C}_{ly}\neq0$ will disappear, and the black-dashed neutral curve is used to fit the critical Reynolds number of this change of wake. Then, the areas where the flows with $\overline{C}_{ly}\neq0$ exist are combined into a thumb-shaped area surrounded by the black and red dashed lines in {\color{blue}figure~}\ref{fig:AR_Rec}.

We have also superposed the results of the critical $Re$ (transitioning from a steady flow to a periodic flow, so should be compared to our $Re_{c2}$) in the radial flow around a short cylinder with double hemispherical free ends in \cite{SCHOUVEILER2001} (experiments) and the double plane free ends in \cite{Inoue2008} (numerical simulations using finite difference method). Their results are supposed to be compared to our green curves and squares. It can be seen that our results are closer to the experimental results of \cite{SCHOUVEILER2001}, and are different from the DNS results of \cite{Inoue2008}. The Hopf bifurcation result of \cite{Inoue2008} appears to be close to our $Re_{c1}$-$\textsc{ar}$ curve obtained. But the regular bifurcation of a radial flow past the cylinder ($\textsc{ar}<1.75$), generating the steady flows with one symmetric plane, found in this paper was not reported in \cite{Inoue2008}. Nevertheless, this regular bifurcation phenomenon has been widely recognised in the flow past small-$\textsc{ar}$ bluff bodies such as a sphere \citep{Johnson1999,TOMBOULIDES2000, thompson2001kinematics, sheard2004spheres}, axial flow around short cylinders with $\textsc{ar}=1$ \citep{Pierson2019} and ellipsoids \citep{sheard2008,Tezuka2006}. Indeed, we have also superposed the $Re_{c1}$ and $Re_{c2}$ results of the flows past a sphere \citep{Johnson1999} in {\color{blue}figure~}\ref{fig:AR_Rec}. Following the discussion in the previous section, we place these data along the vertical line for $\textsc{ar}=\pi/4\approx0.785$.
The $Re_{c1}$ value for the cylinder with $\textsc{ar}=0.75$ is close to that of a sphere in the range of $210<Re_{c1}<212$ (reversed triangle), while the value of $Re_{c2}$ at $\textsc{ar}=0.75$ is larger than that of the sphere's ($270<Re_{c2}<280$), see the black triangle pointing right. Note that this is related to the fact that we place the results of \cite{Johnson1999} at $\textsc{ar}\approx0.785$. Besides, the results of \cite{Pierson2019} on axial flows past an $\textsc{ar}=1$ cylinder have also been superimposed in the figure. The current three critical Reynolds numbers of the radial flow past a cylinder with $\textsc{ar}=1$ are all smaller than those in \cite{Pierson2019}’s DNS results ($Re_{c1}\approx278, Re_{c2}\approx355$ and $Re_{c3}\approx395$\footnote{The definition of $Re_{c3}$ in \cite{Pierson2019} is above which the attractor in the drag coefficient-lift coefficient diagram becomes chaotic. This is in principle different than our definition, but both definitions to some extent measure the onset of chaos in the flow.}, data found in their supplementary material) for the axial flow. This to some extent  indicates that, for the cylinder with $\textsc{ar}=1$, the flow separation due to the curved surface (in our radial flow) is a more efficient destabilising mechanism than the sharp-edge separation due to the front plane end in an axial flow. This may be helpful to the discussion of how to place the finite-length cylinder in a flow to suppress or stimulate flow instability. In this vein, a more general work will be needed to study the flow transition and bifurcation past a cylinder for a series of yaw angles.

\section{Conclusions} \label{sec:Conclusions}

In this paper, we studied systematically the effect of aspect ratio ($0.5\leq \textsc{ar}\leq2$) and Reynolds number ($Re\leq 1000$) on the wake patterns and bifurcations (of regular and Hopf types) in the flow past a finite-length short cylinder by  performing nonlinear DNS and linear stability analyses (based on both time-mean flow and base flow). The detailed transition paths of the radial flow around the short cylinder (shown in {\color{blue}figure~}\ref{fig:AR_Rec}) are obtained, and combined with the previous research on $\textsc{ar}>2$, the present work will help us to better understand the wake transitions of the cylinder with two free ends. 

With the $Re$ increasing from small to large, the first flow pattern is a steady wake P1 with two mutually perpendicular symmetric planes. When $Re>Re_{c1}$, the second flow pattern P2 emerges and it is also a steady wake but with only one symmetric plane (which is perpendicular to the cylinder axis). This P2 pattern was not discovered in the DNS study by \cite{Inoue2008} on the radial flow past a short cylinder.
In the studies of the flows past sphere, ellipsoid and the axial flow past a short cylinder, a similar flow structure with  $\overline{C}_l\neq0$ has also been found in the experimental and numerical results. 
According to our results, this wake pattern P2 only exists in short cylinders with aspect ratio $\textsc{ar}<1.75$; when $\textsc{ar}\gtrsim1.75$, as the $Re$ increases, the wake will directly transition from the pattern P1 to an unsteady wake pattern P3-0 with zero $\overline{C}_l$.

When $Re>Re_{c2}$, vortex shedding happens and the flow becomes periodic and unsteady. Two wake patterns P3-1 and P3-2 have been observed. They have different frequencies and shapes and respectively represent the vortex shedding from the side surface and free ends of the cylinder. 
When $\textsc{ar}=1$, the P3-1 structure is transient and will eventually transition to the nonlinearly stable P3-2 pattern. The transition can take a very long time if $Re$ is close to the critical value and there is no disturbance in the flow field. 
According to our numerical analyses, the bifurcation around $Re_{c2}$ at $\textsc{ar}=1$ is supercritical as we cannot find a nonlinearly stable solution in the subcritical regime when we change the initial conditions. We have not investigated in detail the type of the Hopf bifurcation in this flow at other values of $\textsc{ar}$.
When $Re$ is further increased and is larger than $Re_{c3}$, the flow becomes chaotic. The main message of this work is summarised in the diagram on the relation between the aspect ratio $\textsc{ar}$ and the critical Reynolds numbers. 
The values of $Re_{c2}$, $Re_{c3}$ decreases as $\textsc{ar}$ increases. The critical $Re_{c1}$ first decreases and then increases with the increase of $\textsc{ar}$, and finally intersects the $\textsc{ar}$-$Re_{c2}$ curve at $\textsc{ar}\approx1.75$. 

Based on mean flow and base flow, global linear stability analyses (LSA) have been conducted to analyse the instability and bifurcation in the flow past the short cylinder. By linearly interpolating the real part of the dominant eigenvalue, the aforementioned two critical Reynolds numbers $Re_{c1}$ (regular bifurcation) and $Re_{c2}$ (Hopf bifurcation) obtained from DNS results can also be determined in the LSA. When $Re<Re_{c2}$, the eigenvalues obtained based on mean flow and base flow are the same; when $Re>Re_{c2}$, similar to the 2D cylinder wake flow, the nonlinear vortex frequency is consistent with the most unstable eigenfrequency based on mean flow, but the difference between the nonlinear vortex shedding frequency and the eigenfrequency based on the mean flow gradually becomes larger when $Re$ increases to a non-monochromatic flow state. Finally, around the Hopf bifurcation, the previously-discussed wake patterns P3-1 and P3-2 can be connected to the first two global eigenmodes A and B in the global LSA based on the mean flow. In our numerical analysis, we can find that the frequencies and structures of the first two eigenmodes A, B are very close to those of nonlinear wakes P3-2 and P3-1, respectively. 
Therefore, based on this strong link, by analysing the linear growth rates of these eigenmodes, the evolution of the flow can be better understood and a more accurate reduced-order model of the flow can be proposed in the future. 

For future directions, an interesting topic is to investigate the bifurcation type for other values of $\textsc{ar}$ using a global weakly nonlinear stability analysis or similar stability analysis following Bengana $et ~al.$ (\citeyear{bengana2019}) based on the mean flow past a finite cylinder in LSA to determine the area of hysteresis and the transition conditions of the flow. \\

Declaration of Interests. The authors report no conflict of interest.

\begin{acknowledgments}
The simulations were performed at National Supercomputing Centre, Singapore (NSCC). We acknowledge the financial support from the Ministry of Education, Singapore (a Tier 2 grant with the WBS no. R-265-000-661- 112). 
\end{acknowledgments}

%\bibliographystyle{jfm}
% Note the spaces between the initials jfm
\bibliography{finite_cylinder_ref}
%\printbibliography

\end{document}